\documentclass[a4paper,11pt]{article}
\usepackage{amsfonts,amsthm,amsmath,amssymb,graphicx,float,mathtools}
\usepackage{xcolor}
\usepackage{ulem}
\numberwithin{equation}{section}
\usepackage{subfigure}
\usepackage{color}
\usepackage{appendix}
\usepackage{bbm} 
\usepackage{tensor} 
\usepackage{verbatim} 
\usepackage[
pagebackref=false,
colorlinks=true,
linkcolor=blue,
urlcolor=blue,
filecolor=black,
citecolor=red,
pdfstartview=FitV,
pdftitle={},
pdfauthor={},
pdfsubject={},
pdfkeywords={},
pdfpagemode=None,
bookmarksopen=true
]{hyperref}
\usepackage{ascmac}
\graphicspath{ {Pic/} }
\usepackage{cite}

\begin{document}

\newtheorem{definition}{Definition}[section]
\newcommand{\be}{\begin{equation}}
	\newcommand{\ee}{\end{equation}}
\newcommand{\bea}{\begin{eqnarray}}
	\newcommand{\eea}{\end{eqnarray}}
\newcommand{\LE}{\left[}
\newcommand{\R}{\right]}
\newcommand{\nn}{\nonumber}
\newcommand{\Tr}{\text{Tr}}
\newcommand{\N}{\mathcal{N}}
\newcommand{\G}{\Gamma}
\newcommand{\vf}{\varphi}
\newcommand{\LL}{\mathcal{L}}
\newcommand{\Op}{\mathcal{O}}
\newcommand{\HH}{\mathcal{H}}
\newcommand{\arctanh}{\text{arctanh}}
\newcommand{\up}{\uparrow}
\newcommand{\down}{\downarrow}
\newcommand{\ket}[1]{\left| #1 \right>}
\newcommand{\bra}[1]{\left< #1 \right|}
\newcommand{\ketbra}[1]{\left|#1\right>\left<#1\right|}
\newcommand{\rd}{\partial}
\newcommand{\de}{\partial}
\newcommand{\ba}{\begin{eqnarray}}
	\newcommand{\ea}{\end{eqnarray}}
\newcommand{\db}{\bar{\partial}}
\newcommand{\we}{\wedge}
\newcommand{\ca}{\mathcal}
\newcommand{\lr}{\leftrightarrow}
\newcommand{\f}{\frac}
\newcommand{\s}{\sqrt}
\newcommand{\vp}{\varphi}
\newcommand{\hvp}{\hat{\varphi}}
\newcommand{\tvp}{\tilde{\varphi}}
\newcommand{\tp}{\tilde{\phi}}
\newcommand{\ti}{\tilde}
\newcommand{\ap}{\alpha}
\newcommand{\pr}{\propto}
\newcommand{\mb}{\mathbf}
\newcommand{\ddd}{\cdot\cdot\cdot}
\newcommand{\no}{\nonumber \\}
\newcommand{\la}{\langle}
\newcommand{\lb}{\rangle}
\newcommand{\ep}{\epsilon}
\def\we{\wedge}
\def\lr{\leftrightarrow}
\def\f {\frac}
\def\ti{\tilde}
\def\ap{\alpha}
\def\pr{\propto}
\def\mb{\mathbf}
\def\ddd{\cdot\cdot\cdot}
\def\no{\nonumber \\}
\def\la{\langle}
\def\lb{\rangle}
\def\ep{\epsilon}
\newcommand{\mcl}{\mathcal}
\def\g{\gamma}
\def\Tr{\text{tr}}

\begin{titlepage}
	\thispagestyle{empty}

	\begin{flushright}
		
	\end{flushright}
	\bigskip
	
	\begin{center}
	\noindent{\large \textbf{Entanglement Dynamics by (Non-)Unitary Local Operator Quenches in a 2D Holographic CFT} }
		
	\vspace{2cm}
	\renewcommand\thefootnote{\mbox{$\fnsymbol{footnote}$}}
	Weibo Mao$^{1}$\footnote{maoweibo21@mails.ucas.ac.cn}, 
	Akihiro Miyata$^{2,1}$\footnote{akihiro.miyata@yukawa.kyoto-u.ac.jp}
	Masahiro Nozaki$^{1,3}$\footnote{mnozaki@ucas.ac.cn}
	and Farzad Omidi$^{1}$\footnote{omidi@ucas.ac.cn}
	
	\vspace{1cm}
	${}^{1}${\small \sl Kavli Institute for Theoretical Sciences, University of Chinese Academy of Sciences,
		Beijing 100190, China}\\
	${}^{2}${\small \sl Yukawa Institute for Theoretical Physics, Kyoto University, Kitashirakawa Oiwakecho, Sakyo-ku, Kyoto 606-8502,
		Japan}\\
	${}^{3}${\small \sl RIKEN Interdisciplinary Theoretical and Mathematical Sciences (iTHEMS), \\Wako, Saitama 351-0198, Japan}\\

	\begin{abstract}
		In this paper, we investigate the time evolution of entanglement entropy and mutual information for the spatially-infinite systems where we act with a primary operator on the vacuum state and then time-evolve it with the sequence of the Euclidean and Lorentzian time evolutions. 
		Two-dimensional holographic conformal field theories describe the systems under consideration in this paper.
		The Euclidean time evolution is induced by the Rindler Hamiltonian and behaves as the regulator that tames the divergence induced by the local operator, while the Lorentzian one is induced by the uniform Hamiltonian.  
		Under these time evolutions, we investigate the time ordering effect of the Rindler Euclidean and uniform Lorentzian time evolution operators. 
		Consequently, we find the remarkable differences between those time evolutions are induced by whether those are unitary or non-unitary.
		Especially, we find that the unitary time evolution induces the late-time logarithmic growth of the entanglement entropy, while the non-unitary time evolution induces the late-time constant behavior. 
		Furthermore, we investigate the dual gravity of the systems under consideration.  
		Especially, we investigate the gravity duals of the systems with the insertion of the heavy primary operator and show that it is a black brane with a spacetime-dependent horizon.    
	\end{abstract}
	
	\vspace{1cm}

	\vskip 4em
\end{center}

\end{titlepage} 
\tableofcontents

\section{Introduction}
\label{Sec:intro}

One of the very intriguing and long-standing problems in physics is to understand how systems out of the equilibrium approach to the equilibrium state. In recent years, the non-equilibrium physics has also been investigated extensively in the context of the AdS/CFT correspondence \cite{Maldacena:1997re,Gubser:1998bc,Witten:1998qj}. 
Moreover, it was observed that 
\cite{Wen_2018,Wen:2018agb,Goto:2021sqx,Fan:2019upv,Lapierre:2019rwj,Han:2020kwp,Wen:2020wee,Fan:2020orx,Lapierre:2020ftq,Wen:2021mlv,Goto:2023wai,Jiang:2024hgt} 
the conformal field theories (CFTs) with inhomogeneous Hamiltonians, which are called {\it inhomogeneous} CFTs
\footnote{Inhomogeneous CFTs can also be considered as CFTs living on curved manifolds \cite{Caputa:2020mgb,Jiang:2024hgt,Miyata:2024gvr,Li:2025rzl}. },
have been very fruitful in understanding the phenomenon. This is due to the fact that they are capable of producing very sophisticated time evolutions, and hence, very non-trivial dynamics. 
To reach this goal, various types of inhomogeneous Hamiltonians have been explored such as sine-square deformed (SSD)
\footnote{The SSD Hamiltonian
\cite{Gendiar_2009,Gendiar:2008udd,Hikihara:2011mtb,Gendiar_2011,Katsura:2011ss,Shibata_2011,Katsura:2011zyx,Maruyama:2011njv,Hotta_2012,PhysRevB.87.115128} (see also \cite{Goto:2023yxb})
was historically introduced to reduce the effects of the boundaries in finite-size spin systems with open boundary conditions. 
The corresponding Hamiltonian densities are designed such that they are suppressed at the endpoints of the system. This feature makes it easier to understand the bulk properties of the systems.}
\cite{Katsura:2011ss,Tada:2014kza,Ishibashi:2015jba,Ishibashi:2016bey,Okunishi:2016zat, Wen:2016inm,
Wen_2018,Fan:2019upv,MacCormack:2018rwq,Caputa:2020mgb,Goto:2021sqx,Goto:2023wai,Nozaki:2023fkx,Liu:2023tiq,Mao:2024cnm,Bernamonti:2024fgx}, q-SSD \cite{Fan:2019upv,Jiang:2024hgt,Mao:2025hkp,Bai:2025ysg}, 
M\"{o}bius \cite{Tada:2014kza,Okunishi:2016zat,MacCormack:2018rwq,Goto:2021sqx,Goto:2023wai,Kudler-Flam:2023ahk,Nozaki:2023fkx,Liu:2023tiq,Bernamonti:2024fgx,Mao:2024cnm}, 
q-M\"{o}bius
\cite{Miyata:2024gvr,Bai:2024azk,Li:2025rzl,Bai:2025ysg}
and Rindler Hamiltonians \cite{Kudler-Flam:2023ahk} (see also \cite{Langmann:2018skr,Moosavi:2019fas}).
Furthermore, based on these inhomogeneous Hamiltonians, another interesting type of CFTs called {\it Floquet} CFTs were investigated in which one periodically changes the Hamiltonian in time 
\cite{Berdanier:2017kmd,Wen:2018agb,Fan:2019upv,Lapierre:2019rwj,Wen:2020wee,Han:2020kwp,Lapierre:2020ftq,Fan:2020orx,Wen:2021mlv,Wen:2022pyj,deBoer:2023lrd,Das:2023xaw,Jiang:2024hgt,Lapierre:2024lga,Das:2024lra,Fang:2025rie,Das:2025wjo,2025PhRvB.112j4322L,Lapierre_2020}.

From now on, we focus on the two-dimensional systems, i.e., those we will investigate in this paper.
In two-dimensional CFTs ($2$d CFTs), one starts from the CFT Hamiltonian
\bea
H_0 = \int dx \; h(x),
\label{Hamiltonian-CFT}
\eea 
where $h(x)$ is the Hamiltonian density and the integration is taken over the whole spatial part of the system. Then, one deforms $H_0$ into an inhomogeneous Hamiltonian as
\bea
H= \int dx \; f(x) h(x),
\label{H-inhomogeneous}
\eea 
and study the time evolutions of the states by the new Hamiltonian. Here, $f(x)$ is called the {\it envelope} function and depends on the spatial coordinate $x$
\footnote{For instance, for $H_{q-{\rm SSD}}$ and $H_{q-\text{M\"{o}bius}}$ the envelope functions are defined as
$f_{q{\rm -SSD}} = \sin^2 \left( \frac{q \pi x}{L} \right)$ and
$f_{q-\text{M\"{o}bius}} = 1- \tanh(2 \theta) \cos \left( \frac{2 q \pi x}{L} \right)$,
respectively. Here, $L$ is the total size of the system. Moreover, $q$ and $\theta>0$ are real numbers. For $q=1$, $H_{q-{\rm SSD}}$ and $H_{q-\text{M\"{o}bius}}$ reduce to $H_{{\rm SSD}}$ and $H_{\text{M\"{o}bius}}$, respectively.}.

On the other hand, local operator quench is an interesting type of quench and various aspects of it were investigated extensively both on the CFT and gravity sides with (R\'enyi) entanglement entropy 
\cite{Nozaki:2013wia,Nozaki:2014hna,He:2014mwa,Nozaki:2014uaa,Caputa:2014vaa,Caputa:2014eta,2015JHEP...02..171A,Guo:2015uwa,Chen:2015usa,Nozaki:2015mca,Caputa:2015qbk,Nozaki:2016mcy,David:2016pzn,He:2017lrg,Shimaji:2018czt,Guo:2018lqq,Apolo:2018oqv,Kusuki:2019gjs,He:2019vzf,Kawamoto:2022etl,Bhattacharyya:2019ifi,Doi:2025oma}, 
pseudo (R\'enyi) entanglement entropy 
\cite{Nakata:2020luh, Guo:2022jzs,Guo:2022sfl,Mukherjee:2022jac,Ishiyama:2022odv,He:2023eap,Guo:2023aio,He:2023wko,Omidi:2023env}, time-like entanglement entropy 
\cite{Doi:2023zaf}
and mutual information
\cite{Kudler-Flam:2020xqu}. 
An intriguing phenomenon in the local operator quench was found to be the late-time behavior of the entanglement entropy for the half-line subsystem \cite{Nozaki:2014hna,Nozaki:2014uaa,He:2014mwa,Caputa:2014vaa}.
The $2$d systems, considered in those papers, are defined as
\be
\ket{\Psi(t)}=\f{e^{-it H_1}e^{-\epsilon H_2}\mathcal{O}(x)\ket{0}}{\sqrt{\bra{0}\mathcal{O}^{\dagger}(x)e^{-2\epsilon H_2}\mathcal{O}(x)\ket{0}}}.
\ee
In those papers, two Hamiltonians are set to be the uniform ones, i.e., $H_1=H_2=H_0$. Moreover, $\ket{0}$ and $\mathcal{O}(x)$ denote the vacuum state of $H_0$ and the local operator inserted at the spatial point, $x$.
Note that the Euclidean time evolution is exploited as the regulator taming the divergence induced by the contraction of the local operators.
In the systems described by the non-holographic CFTs, such as free ones, the late-time value of the entanglement entropy is a constant determined by the local operator inserted, while in the ones described by holographic CFTs, the late-time value logarithmically grows in time. 
One possible interpretation of this logarithmic growth is that it suppresses the local operator dependence of the entanglement entropy.
Suppose that we consider this local operator dependence as the dependence on the initial state. 
Under this assumption, the late-time behavior of the entanglement entropy may imply that the holographic systems strongly scramble the initial-state information, and then that information is locally hidden. 

Furthermore, some authors of this paper investigated the late-time behavior of the entanglement entropy in a slightly different setup, which is a spatially-compact system, instead of the infinite one \cite{Mao:2024cnm,Mao:2025hkp}.
Note that in these articles, the Hamiltonian that induces the Euclidean time evolution is different from the one that induces the Lorentzian time evolution, i.e., $H_1 \neq H_2$
\footnote{In ref. \cite{Mao:2024cnm} the Hamiltonians for the Lorentzian  and Euclidean time evolutions were $H_0$ and $H_{\text{M\"{o}bius}}$. 
Moreover, in ref. \cite{Mao:2025hkp} the Hamiltonian for the Lorentzian time evolution was $H_{q{\rm-SSD}}$ with $q=2$, and the Euclidean time evolution was performed by $H_0$ and $H_{\rm SSD}$.
}.
They found that in the systems described by the $2$d holographic CFTs, the CFTs described by gravity, the late-time value of the entanglement entropy depends on the time ordering of the Euclidean and Lorentzian time evolutions: when the holographic systems are evolved by the Euclidean time evolution, and then are evolved by the Lorentzian one, the late-time value of the entanglement entropy logarithmically grows in time, while 
when the holographic systems are evolved by the Lorentzian one, and then are done by the Euclidean one, the late-time value becomes finite.
Note that the time evolution initially triggered by the Euclidean time evolution and then the Lorentzian one can be considered as the unitary process, while the one triggered by the different time ordering of the Euclidean and Lorentzian time evolutions can be considered as the non-unitary process. 

        It should be emphasized that in ref. [40], the Lorentzian time evolution of the holographic CFT system was defined on a {\it curved} background. 
        This raises the question of weather the non-unitary time evolution on a {\it flat} background can suppress the late-time logarithmic growth of the entanglement entropy. To answer this question, we will consider the Lorentzian time evolution of the holographic CFT system on a flat background. More precisely,
we will start from the same system as \cite{Nozaki:2014hna,Caputa:2014vaa,Nozaki:2014uaa,He:2014mwa}, i.e. a spatially-infinite one, with the insertion of the local operator into the vacuum state of $2$d holographic uniform Hamiltonian. Then, we will time-evolve the system with the composite time evolution operator that is constructed of the Euclidean and Lorentzian time evolution operators determined by the $2$d Rindler and uniform holographic Hamiltonians, respectively (for details, see (\ref{eq:systems-under-consideration})).
Subsequently, we will investigate how the time ordering of the Euclidean and Lorentzian time evolutions influences the bipartite entanglement and non-local correlation by exploiting the entanglement entropy and mutual information. 
Furthermore, we will also discuss the gravity dual of the systems considered in this paper.
Note that all parameters and variables in this paper are dimensionless. We take the ultraviolet cutoff, taming the divergence of the entanglement entropy, as the unit of the energy scale.

\subsection*{Summary}

Now, we will report on our main findings of this paper.
As explained above, in the dynamics considered in this paper, we started from the spatially-infinite system in the vacuum state with the insertion of the primary operator, and then time-evolve the system with the composite time evolution operator constructed of the Euclidean and Lorentzian time evolution operators determined by the $2$d Rindler and uniform CFT Hamiltonians, respectively (for details of them, see Section \ref{Sec:Preliminary}).
As the Rindler Hamiltonian does not commute with the uniform one (see Appendix \ref{Sec: Time Evolution of The Primary Operator}), the time ordering of the Euclidean and Lorentzian time evolutions can influence the dynamics of the system.
We investigated how the time ordering of the Euclidean and Lorentzian time evolutions influences the entanglement dynamics by exploiting the energy density, entanglement entropy and mutual information (for the definitions of them, see Section \ref{Sec:Preliminary}).


We introduced an effective description that can qualitatively describe the entanglement dynamics, called the quasiparticle picture.
Furthermore, we investigated how the time ordering of the Euclidean and Lorentzian time evolutions influences the entanglement dynamics, by exploring the formula, elaborated in \cite{Mao:2025hkp}, bridging the bipartite entanglement and the energy-momentum densities.
Then, in the low-energy regime, we found that whether the time evolution is unitary or non-unitary can induce remarkable differences in the energy density, entanglement entropy, and mutual information between those processes.

Especially, we found an intriguing late-time behavior of the entanglement growth for the semi-infinite interval. 
When the system is initially evolved by the Euclidean time evolution, and then is done by the Lorentzian one, the entanglement entropy for the semi-infinite interval logarithmically grows in time, while if the system is initially evolved by the Lorentzian one, and then is done by the Euclidean one, then the entanglement growth for the semi-infinite interval saturates to a certain value depending on the conformal dimension of the primary operator. 
Furthermore, by using the formula that bridges the entanglement entropy and the energy-momentum densities \cite{Mao:2025hkp}, we found that the unitary time evolution can induce the late-time logarithmic growth of the entanglement entropy, while the non-unitary time evolution can induce the late-time constant behavior in time of the entanglement growth (for details, see Section \ref{Sec:QE-EMD}). 
Therefore, the non-unitary time evolution in the holographic CFT system on the flat background suppresses the late-time logarithmic growth of the entanglement entropy, and hence, the strong scrambling property of the holographic CFT system. This behavior is similar to that for the non-unitary time evolution on the curved backgrounds \cite{Mao:2025hkp}.
Finally, we investigated the properties of the gravity dual of the systems under consideration.

\subsection*{Organization of The Paper}

In section \ref{Sec:Preliminary}, we introduce our setup and the necessary tools for our calculations. 
In Section \ref{Sec:Time-dependence-of-ee-and-energy-density}, we calculate the entanglement entropy on the CFT side for finite and semi-infinite intervals and for the cases where the primary operator is inserted into the subsystem or its complement region. We also calculate the energy-momentum tensors of the corresponding states. Next, by applying the energy-momentum tensors, we construct the dual geometry which is a Ba$\tilde{\rm n}$ados geometry. Then, we apply the Ryu-Takayanagi formula \cite{2006PhRvL..96r1602R,2006JHEP...08..045R,2013JHEP...08..090L} in the Ba$\tilde{\rm n}$ados geometry to find the holographic entanglement entropy. Moreover,
we calculate the holographic mutual information by considering the connected and disconnected Ryu-Takayanagi surfaces. In Section \ref{Sec: Revisiting the Gravity Dual}, we write the dual geometry in a new coordinate system which covers the horizon,  and find the shape of the horizon. 
In Section \ref{Sec: Discussions}, we summarize our results and discuss possible future directions.

\section{Preliminary \label{Sec:Preliminary}}
In this section, we will present the details of the systems considered in this paper, define the entanglement entropy and mutual information, and then report on how to analytically calculate them.

\subsection{The Systems under Consideration\label{Sec:Systems-under-consideration}}

The systems considered in this paper are in
\be \label{eq:systems-under-consideration}
\begin{split}
	&| \psi_1(t) \rangle =N_1 e^{-\epsilon H_R} e^{- i H_0 t} \mathcal{O}(x) |0 \rangle = N_1  e^{- \epsilon H_R} e^{- i H_0 t}\mathcal{O}(x) e^{i H_0 t} e^{\epsilon H_R}  |0 \rangle =N_1 \mathcal{O}^H_1 (x,t) | 0 \rangle,\\
	&| \psi_2(t) \rangle =N_2 e^{- i H_0 t} e^{- \epsilon H_R} \mathcal{O}(x)  |0 \rangle = N_2 e^{- i H_0 t} e^{- \epsilon H_R} \mathcal{O}(x) e^{\epsilon H_R} e^{i H_0 t} |0 \rangle= N_2 \mathcal{O}^H_2 (x,t) | 0 \rangle ,\\
\end{split}
\ee
where $\epsilon>0$ and we assume that $|0 \rangle$ is the vacuum state of $H_0$. Moreover, in each line, we exploited the following facts
\footnote{By writing $H_R$ in terms of the Virasoro generators in \eqref{HR-Virasoro}, one can verify that $H_R$ annihilates the vacuum state of $H_0$.},
\be
e^{i H_0 t} e^{\epsilon H_R}  |0 \rangle=|0 \rangle,~~~~~~~ e^{\epsilon H_R}  e^{i H_0 t} |0 \rangle=|0 \rangle.
\ee
Here, the normalization constants, $N_{i=1,2}$, keep the norm of each state to be one. In other words, they are defined as 
\be \label{eq:normalization-constant}
\begin{split}
	N_1=\f{1}{\sqrt{\bra{0}\mathcal{O}^{\dagger}(x)e^{iH_0t}e^{-2\epsilon H_R} e^{-iH_0t}\mathcal{O} (x)\ket{0}}},~~~~~~ N_2=\f{1}{\sqrt{\bra{0}\mathcal{O}^{\dagger}(x)e^{-2\epsilon H_R} \mathcal{O} (x)\ket{0}}}.
\end{split}
\ee
Furthermore, we assume that $\mathcal{O}(x,t)$ is a primary operator with the conformal dimensions $(h_{\mathcal{O} }, h_{\mathcal{O}} )$ 
inserted at $t=0$ and $x >0$. Moreover, we define $\mathcal{O}^H_i (x,t)$ as the operator in the Heisenberg picture,
\be
\begin{split}
	&\mathcal{O}^H_1 (x,t)=e^{-\epsilon H_R} e^{- i H_0 t} \mathcal{O}(x) e^{i H_0 t} e^{\epsilon H_R}, ~~~~~ \mathcal{O}^H_2 (x,t)=e^{- i H_0 t} e^{- \epsilon H_R} \mathcal{O}(x) e^{\epsilon H_R} e^{i H_0 t}.\\
\end{split}
\ee 
Here, we define $H_0$ and $H_{R}$ as the uniform and Rindler Hamiltonians,
\be
H_0 =\int^{\infty}_{-\infty} dx\; h(x), 
~~~~~~~~ H_R=a \int^{\infty}_{-\infty}dx \; x h(x), 
\label{H0-HR}
\ee
where $h(x)$ is the Hamiltonian density and $a$ is a real parameter, which is assumed to be positive.
Note that $H_{R}$ is a simple Hamiltonian that cannot commute with $H_0$. Then, we choose it for the simplicity of the calculations.
Later, we exploit $e^{-\epsilon H_R}$ as the regulator. 
In the positive spatial region, $x>0$, the Hamiltonian density is positive, while in the negative spatial region, $x<0$, that is negative.
This suggests that $e^{-\epsilon H_R}$ may not behave as the regulator in the negative spatial region.
Therefore, we will explore the dynamical properties of the systems in (\ref{eq:systems-under-consideration}) in the positive spatial region, $x>0$
\footnote{However, if we treat $e^{-\epsilon H_R}$ as the evolution operator, there seem to be no problems associated with the negativity of the Hamiltonian density in the negative spatial region. In this paper, to avoid such a subtlety, we consider only the positive spatial region. Another subtlety is that at the origin, $x=0$,  $e^{-\epsilon H_R}$ may not behave as the regulator because of $xh(x)|_{x=0}=0$.}.

We will report the properties of the systems in (\ref{eq:systems-under-consideration}) in the following sections. 
The time evolution operator, acting on $\mathcal{O}(x)$, consists of the Euclidean time evolution operator with $H_{R}$ and the real-time one with $H_0$.
The difference between $i=1$ and $i=2$ is the time ordering of the Euclidean and real-time evolutions.
For $i=1$, the real-time evolution induces the evolution of the primary operator, and then the Euclidean one does.
In contrast, for $i=2$, the Euclidean time evolution induces the operator evolution, and then the real-time one does.
In the following, we refer to the time evolution in $| \psi_{i=1}(t) \rangle$ and $| \psi_{i=2}(t) \rangle$ as the non-unitary and unitary time evolutions, respectively.

\subsection{Entanglement Entropy and Mutual Information \label{Sec:Entanglement-measures}}

Now, we will define the quantum informational quantities, entanglement entropy and mutual information, that will be exploited to investigate the dynamical properties of the systems in (\ref{eq:systems-under-consideration}).
We begin by defining the density operators associated with (\ref{eq:systems-under-consideration}) as
\be \label{eq:lorentzian-density-op}
\rho_{i=1,2} = \ket{\psi_i(t)} \bra{\psi_i(t)}.
\ee
Spatially divide the system into $A$ and $\overline{A}$, the complement spatial region to $A$, and then define the reduced density matrix associated with the subsystem $A$ as 
\be
\rho_{A;i=1,2}= \Tr_{\overline{A}} ( \rho_{i=1,2}) = \sum_{\phi} {}_{\overline{A}} \bra{\phi} \rho_{i=1,2} \ket{\phi}_{\overline{A}},
\ee
where $\ket{\phi}_{\overline{A}}$ are the eigenvectors associated with $\overline{A}$.
Subsequently, we define entanglement entropy, a quantity measuring the bipartite entanglement between adjacent subsystems, as Von Neumann entropy for the reduced density matrix,
\be
S_A=\lim_{n\rightarrow 1}S^{(n)}_A=- \Tr_{A}\left(\rho_A \log{\rho_A}\right).
\label{EE}
\ee
Here, $S^{(n)}_A$ is the $n$-th moment of the R\'enyi entanglement entropy defined as 
\be
S^{(n)}_A =\f{1}{1-n}\log{\left[\Tr_A \left(\rho_A\right)^n\right]},
\label{Renyi Entropy}
\ee
where $n$ is an integer.
Furthermore, we define the mutual information as a quantity measuring the correlation between the distant subsystems.
Spatially divide the system into $A$, $B$, and the complement region to them, and then define the mutual information as
\be \label{eq:def-of-mutual-information}
I_{A,B} =S_A+S_B-S_{A\cup B},
\ee
where $S_A$, $S_B$, and $S_{A\cup B}$ are the entanglement entropies associated with $A$, $B$, and $A\cup B$, respectively.
If $A$ and $B$ are distant spatial regions, $I_{A,B}$ is free from the divergence.

\subsection{Twist Operator Formalism\label{Sec:Twist-operator-formalism}}

In this section, we will explain the analytical method of calculating the entanglement entropy.
As the mutual information is defined as the linear combination of the entanglement entropies, the analytical calculation of the entanglement entropy leads us to the analytical one of the mutual information.
To employ the analytical calculation method, we define the Euclidean density operators as
\be
\begin{split} \label{eq:Euclidean-density-op}
	&\rho^{E}_1=N_{E,1} ^2 e^{- \epsilon H_R} e^{- H_0 \tau_E} \mathcal{O}(x) e^{H_0 \tau_E} e^{ \epsilon H_R} |0 \rangle\bra{0} 
	e^{ \epsilon H_R} e^{- H_0 \tau_E} \mathcal{O}^{\dagger}(x) e^{ H_0 \tau_E} e^{- \epsilon H_R},\\
	&\rho^{E}_2=N_{E,2} ^2 e^{- H_0 \tau_E} e^{- \epsilon H_R} \mathcal{O}(x) e^{ \epsilon H_R} e^{H_0 \tau_E} |0 \rangle\bra{0} 
	e^{- H_0 \tau_E} e^{ \epsilon H_R}  \mathcal{O}^{\dagger}(x) e^{- \epsilon H_R} e^{ H_0 \tau_E},
\end{split}
\ee
where $\tau_E$ and $\epsilon$ are real parameters, and the normalization constants, $N_{E,i=1,2}$, guarantee $\Tr (\rho^E_{i=1,2})=1$.
Then, divide the Euclidean system into $A$ and $\overline{A}$, where the subsystem $A$ is defined as
\be
A=[x_1,x_2].
\ee
In the following, we assume that $0 < x_1 < x_2$. 
Subsequently, define the Euclidean reduced density matrix for $A$ as 
\be
\rho^E_{A;i=1,2}= \Tr_{\overline{A}} (\rho^{E}_{i=1,2}).
\ee
Furthermore, define the $n$-th moment of the Euclidean R\'enyi entanglement entropy as
\be
\begin{split}
	S^{(n)}_{A;i,E}=\f{1}{1-n}\log{\left[\Tr_A \left(\rho^{E}_{A;i}\right)^n\right]}.
\end{split}
\ee
By exploiting the replica trick \cite{Holzhey:1994we,Calabrese:2009qy,Nishioka:2009un,Rangamani:2016dms} and then employing the twist-operator formalism \cite{Calabrese:2009qy,2004JSMTE..06..002C}, 
we obtain the $n$-th moment of the Euclidean R\'enyi entanglement entropy as \cite{Nozaki:2014uaa,
Nozaki:2014hna,Caputa:2014vaa,2015JHEP...02..171A,Mao:2024cnm}
\bea
S^{(n)}_{A;i,E} &= & \frac{1}{1-n} \log \Bigg[ \frac{\langle \mathcal{O}^{H, \dagger}_{n,i}(x) \sigma_n(x_1) \bar{\sigma}_n(x_2) \mathcal{O}^{H}_{n,i}(x) \rangle}{\langle \mathcal{O}^{H, \dagger}_{i}(x) \mathcal{O}^{H}_{i}(x) \rangle^n \;
} \Bigg]
\cr && \cr
& = & \frac{1}{1-n} \log \Bigg[ \frac{\langle \mathcal{O}^{H, \dagger}_{n,i}(z, \bar{z}) \sigma_n(z_{x_1}, \bar{z}_{x_1}) \bar{\sigma}_n(z_{x_2}, \bar{z}_{x_2} ) \mathcal{O}^{H}_{n,i}(z, \bar{z}) \rangle}{\langle \mathcal{O}^{H, \dagger}_{i}(z, \bar{z}) \mathcal{O}^{H}_{i}(z , \bar{z}) \rangle^n 
} \Bigg].
\label{Renyi-Entropy-1}
\eea 
Note that the correlation functions in (\ref{Renyi-Entropy-1}) are calculated on the orbifold $CFT^n / Z^n$.  Moreover, $\sigma_n(x)$ and $\bar{\sigma}_n(x)$ are the twist and anti-twist operators inserted at the endpoints of the subsystem. Furthermore, the complex coordinates $(z, \overline{z})$ are introduced as $(z, \overline{z})=(\tau_E+ix, \tau_E-ix)$, where $(z_{x_i}, \overline{z}_{x_i})=(ix_i,-ix_i)$ and $(z, \overline{z})=(ix,-ix)$.
The twist and anti-twist operators are defined as the primary operators with conformal dimensions $( h_n,  \overline{h}_n)= \left(\frac{c}{24} \left( n - \frac{1}{n} \right),\frac{c}{24} \left( n - \frac{1}{n} \right)\right)$.
The operator, $\mathcal{O}_n$, is defined as the primary operator with conformal dimensions $(nh_{\mathcal{O}}, n h_{\mathcal{O}})$.
For simplicity, we define the primary operators in the Heisenberg picture as
\bea
&& \mathcal{O}^{H}_{n,1}(z, \bar{z}) = e^{- \epsilon H_R} e^{- H_0 \tau_E} \mathcal{O}_{n}(z, \bar{z}) e^{H_0 \tau_E} e^{\epsilon H_R},
\label{O1-Heisneberg}
\\
&& \mathcal{O}^{H}_{n,2}(z, \bar{z}) = e^{- H_0 \tau_E} e^{- \epsilon H_R} \mathcal{O}_{n}(z, \bar{z}) e^{\epsilon H_R} e^{H_0 \tau_E}.
\label{O2-Heisneberg}
\eea 
It should be emphasized that $H_0$ and $H_R$ act on the primary operators as time translation and rotation, respectively. 
Therefore, the time evolution of the operators in \eqref{O1-Heisneberg} and \eqref{O2-Heisneberg} changes the insertion points of the operators to new locations (for more details, refer to Appendix \ref{Sec: Time Evolution of The Primary Operator}).  They are given by \eqref{z1-new-a} and \eqref{z2-new-a} and we rewrite them here
\bea
&& z^{\rm new}_{1, \pm \epsilon} = e^{\pm i a \epsilon} (i x- \tau),
~~~~~~ \bar{z}^{\rm new}_{1, \pm \epsilon} = - e^{\mp i a \epsilon} (i x+ \tau),
\cr && \cr
&& z^{\rm new}_{2, \pm \epsilon} = i x e^{\pm i a \epsilon} - \tau,
~~~~~~~~~ \bar{z}^{\rm new}_{2, \pm \epsilon} = - ( i x e^{\mp i a \epsilon} + \tau).
\label{z-new1-z-new2}
\eea 
Moreover, $a$ is the parameter which appears in the definition of the Rindler Hamiltonian in \eqref{H0-HR}. 
In the following,
we restrict ourselves to $0 \leq a \epsilon \leq 2\pi$, to avoid the insertion points, $z^{\text{new}}_{i=1,2, \pm \epsilon}$ and $\bar{z}^{\text{new}}_{i=1,2, \pm \epsilon}$, not to coincide with each other.
Next, by exploiting \eqref{time evolution-conformal transform-O} and \eqref{time evolution-conformal transform-O-dagger}, one can rewrite \eqref{Renyi-Entropy-1} as 
\bea
S^{(n)}_{A;i,E} = \frac{1}{1-n} \log \Bigg[ \; \frac{\langle \mathcal{O}^{\dagger}_{n}(z^{\rm new}_{i, -\epsilon}, \bar{z}^{\rm new}_{i, -\epsilon}) \sigma_n(z_{x_1}, \bar{z}_{x_1}) \bar{\sigma}_n(z_{x_2}, \bar{z}_{x_2} ) \mathcal{O}_{n}(z^{\rm new}_{i, \epsilon}, \bar{z}^{\rm new}_{i, \epsilon}) \rangle}{ \langle \mathcal{O}^{\dagger}(z^{\rm new}_{i, - \epsilon}, \bar{z}^{\rm new}_{i, - \epsilon}) \mathcal{O}(z^{\rm new}_{i, \epsilon} , \bar{z}^{\rm new}_{i, \epsilon}) \rangle^n 
} \Bigg].
\label{Renyi-Entropy-2}
\eea 
Thus, the conformal factors resulted from the time evolutions of the operators in the numerator and denominator are canceled with each other. 
Subsequently, we take the Von Neumann limit, $n\rightarrow 1$, and then perform the analytic continuation to the real time as $\tau_E=it$.

\section{Entanglement Entropy, Energy-Momentum Densities and Mutual Information}
\label{Sec:Time-dependence-of-ee-and-energy-density}

In this section, we will report the time dependence of the entanglement entropy, 
energy-momentum densities and  mutual information in the systems in (\ref{eq:systems-under-consideration}).
We assume that the systems are described by the $2d$ holographic CFTs.

\subsection{Holographic Entanglement Entropy}
\label{Sec:Holographic-entanglement-entropy}

We will explore the time dependence of the entanglement entropy in $2$d holographic CFTs by exploiting two methods: the first one relies on the holographic conformal blocks \cite{2013arXiv1303.6955H,2014JHEP...08..145F,2015JHEP...02..171A}
\footnote{For similar calculations in the SSD or M\"obius quenches refer to \cite{Mao:2024cnm}. Here, we apply their notations.}
; the second one employs the Ryu-Takayanagi formula \cite{2006PhRvL..96r1602R,2006JHEP...08..045R,2013JHEP...08..090L}.

\subsubsection{Conformal Block Method}
\label{Sec:Conformal-block-method}

We begin with the calculation of the entanglement entropy by exploiting the holographic conformal blocks \cite{2013arXiv1303.6955H,2014JHEP...08..145F,2015JHEP...02..171A}.
To employ the holographic conformal blocks, we will rewrite the holographic entanglement entropy as a function of the cross ratios defined as
\bea
\eta_i = \frac{ \left( z^{\rm new}_{i, - \epsilon} - z_{x_1} \right) \left( z_{x_2} - z^{\rm new}_{i, \epsilon} \right) }{ \left( z_{x_1} - z^{\rm new}_{i, \epsilon} \right) \left( z^{\rm new}_{i, - \epsilon} - z_{x_2} \right)},
~~~~~~~ \bar{\eta}_i = \frac{ \left( \bar{z}^{\rm new}_{i, - \epsilon} - \bar{z}_{x_1} \right) \left( \bar{z}_{x_2} - \bar{z}^{\rm new}_{i, \epsilon} \right) }{ \left( \bar{z}_{x_1} - \bar{z}^{\rm new}_{i, \epsilon} \right) \left( \bar{z}^{\rm new}_{i, - \epsilon} - \bar{z}_{x_2} \right)},
\label{cross-ratios-1}
\eea 
where $i$ labels the systems in (\ref{eq:systems-under-consideration}).
Therefore, we employ the following conformal transformations,
\bea
\tilde{z}_i (z) = \frac{ \left( z^{\rm new}_{i, - \epsilon} - z \right) \left( z_{x_2} - z^{\rm new}_{i, \epsilon} \right) }{ \left( z - z^{\rm new}_{i, \epsilon} \right) \left( z^{\rm new}_{i, - \epsilon} - z_{x_2} \right)},
\;\;\;\;\;\;\;
\bar{\tilde{z}}_i ( \bar{z} ) = \frac{ \left( \bar{z}^{\rm new}_{i, - \epsilon} - \bar{z} \right) \left( \bar{z}_{x_2} - \bar{z}^{\rm new}_{i, \epsilon} \right) }{ \left( \bar{z} - \bar{z}^{\rm new}_{i, \epsilon} \right) \left( \bar{z}^{\rm new}_{i, - \epsilon} - \bar{z}_{x_2} \right)}.
\label{conformal map-EE}
\eea
It is straightforward to verify that under this transformation, the location of the primary operators are changed as
\bea
(z_{i, - \epsilon}, \bar{z}_{i, - \epsilon}) \rightarrow (0, 0),
~~~~~~~~~~ (z_{i, \epsilon}, \bar{z}_{i,  \epsilon}) \rightarrow (\infty, \infty).
\eea 
Moreover, the locations of the twist and anti-twist operators are transformed as
\bea
(z_{x_2}, \bar{z}_{x_2}) \rightarrow (1,1),
~~~~~~~~~~ (z_{x_1}, \bar{z}_{x_1}) \rightarrow (\eta_i, \bar{\eta}_i).
\label{endpoints-entangling regoin}
\eea
In other words, in the new coordinates, the two endpoints of the subsystem are located at $(\tilde{z}, \bar{\tilde{z}})= (1,1)$ and $(\tilde{z}, \bar{\tilde{z}})= (\eta_i, \bar{\eta}_i)$. Furthermore, one has
\bea
\sigma_n(z_{x_1}, \bar{z}_{x_1}) \bar{\sigma}_n(z_{x_2}, \bar{z}_{x_2} ) = \frac{1}{|z_{x_1} - z_{x_2} |^{4 n h_n}} | 1 - \eta_i |^{4 n h_n} \sigma_n(\eta_i, \bar{\eta}_i) \bar{\sigma}_n(1, 1),
\label{twist-operators-conformal-transformation}
\eea 
where $1-\eta_i$ and $1-\bar{\eta}_i$ are given by
\bea
1- \eta_i = \frac{ ( z^{\rm new}_{i, - \epsilon} - z^{\rm new}_{i, \epsilon} ) \left( z_{x_1} - z_{x_2} \right) }{ ( z^{\rm new}_{i, - \epsilon} - z_{x_2} ) ( z_{x_1} - z^{\rm new}_{i, \epsilon} ) },
~~~~~ 1- \bar{\eta}_i = \frac{ ( \bar{z}^{\rm new}_{i, - \epsilon} - \bar{z}^{\rm new}_{i, \epsilon} ) ( \bar{z}_{x_1} - \bar{z}_{x_2} ) }{ ( \bar{z}^{\rm new}_{i, - \epsilon} - \bar{z}_{x_2} ) ( \bar{z}_{x_1} - \bar{z}^{\rm new}_{i, \epsilon} )}.
\label{cross-ratios-2}
\eea 
Putting everything together, we can rewrite $S^{(n)}_{A;i,E}$ in \eqref{Renyi-Entropy-2} as
\bea
S^{(n)}_{A;i,E} = \frac{1}{1-n} \log \Big[ |z_{x_1} - z_{x_2}|^{-4 n h_n} |1 - \eta_i|^{4 n h_n} G_{n} (\eta_i, \bar{\eta}_i) \Big],
\label{Renyi-Entropy-3}
\eea 
where the function depending only on the cross ratios, $G_n(\eta_{i},\bar{\eta}_i)$, is defined as 
\bea
G_{n} (\eta_i, \bar{\eta}_i) = \frac{\langle \mathcal{O}^{\dagger}_{n,i}(0) \sigma_n(\eta_i , \bar{\eta}_i) \bar{\sigma}_n(1, 1) \mathcal{O}_{n,i}(\infty) \rangle}{\langle \mathcal{O}^{\dagger}_{i}(0) \mathcal{O}_{i}(\infty) \rangle^n}.
\label{G-n}
\eea 
To evaluate $G_{n} (\eta_i, \bar{\eta}_i)$, one can expand it in terms of the holomorphic and anti-holomorphic Virasoro conformal blocks. 
In the semi-classical limit, where the central charge is very large and the ratios $h_n/c$ and $n h_{\mathcal{O}}/c$ are held fixed, the Virasoro conformal blocks exponentiate and $G_{n} (\eta_i, \bar{\eta}_i)$ can be written as a sum of exponentials \cite{Zamolodchikov:1987avt,2014JHEP...08..145F}.
Moreover, when the cross ratios are close to one, i.e. $\eta_i \simeq 1, \; \bar{\eta}_i \simeq 1$
\footnote{We will see that the cross ratios for states \eqref{eq:systems-under-consideration} are close to one.}
, the dominant term in the conformal block expansion comes from the identity operator and all of its descendants \cite{2015JHEP...02..171A}. 
Furthermore, in the Von Neumann limit where $n \rightarrow 1$, the conformal dimensions of the twist and anti-twist operators go to zero and they become light operators. Then, for heavy primary operators, one can apply the "light-light-heavy-heavy" approximation and obtain the following expression for $\log G_{n} (\eta_i, \bar{\eta}_i)$ to the leading order in the $(n-1)$ expansion \cite{2014JHEP...08..145F,2015JHEP...02..171A}
\bea
\log G_{n} (\eta_i, \bar{\eta}_i) = \frac{c (1- n)}{6} \log \Bigg[ \frac{\eta_i^{\frac{(1- \alpha_{\mathcal{O}})}{2}} \bar{\eta}_i^{\frac{(1- \alpha_{\mathcal{O}})}{2}} \left( 1 - \eta_i^{\alpha_{\mathcal{O}} } \right) \left( 1 - \bar{\eta}_i^{\alpha_{\mathcal{O}} } \right) }{ \alpha^2_{\mathcal{O}}  } \Bigg] + \mathcal{O} \left( \left( n- 1 \right)^2 \right). \;\;\;\;\;\;
\label{conformal block-von Neumann limit}
\eea 
Here, we defined $\alpha_{\mathcal{O}}$ as
\bea
\alpha_{ \mathcal{O}} = \sqrt{1 - \frac{24 h_{\mathcal{O}}}{c} }.
\label{alpha-alpha-bar}
\eea
It should be emphasized that for $h_{\mathcal{O}} < h_0$, where 
\bea
h_0 = \frac{c}{24},
\label{h0}
\eea 
$\alpha_{\mathcal{O}}$ is real-valued. However, for $h_{\mathcal{O}} > h_0$, it is imaginary. For later convenience, it is better to define $\kappa_{\mathcal{O}}$
as
\bea
\kappa_{\mathcal{O}} = \frac{\sin (\pi \alpha_{\mathcal{O}})}{\alpha_{\mathcal{O}}}.
\label{kappa-kappa-bar}
\eea 
Furthermore, by substituting \eqref{conformal block-von Neumann limit} into \eqref{Renyi-Entropy-3}, one arrives at 
\bea
S^{(n)}_{A;i,E} & \underset{n \approx 1}{\approx} & \frac{c (n+1)}{12} \Bigg[ \log |z_{x_1} - z_{x_2}|^2 
- \log \left( |1 - \eta_i|^2 \right) \Bigg]
\cr && \cr 
&& + \frac{c}{6} \log \Bigg[ \frac{\eta_i^{\frac{(1- \alpha_{\mathcal{O}})}{2}} \bar{\eta}_i^{\frac{(1- \alpha_{\mathcal{O}})}{2}} \left( 1 - \eta_i^{\alpha_{\mathcal{O}} } \right) \left( 1 - \bar{\eta}_i^{\alpha_{\mathcal{O}} } \right) }{ \alpha^2_{\mathcal{O}} } \Bigg]
+ \mathcal{O} \left( \left( n- 1 \right)^2 \right).
\label{Renyi-Entropy-4}
\eea 
Therefore, the entanglement entropy is given by 
\bea \label{eq:Euclidean-EE-CB}
S_{A;i,E} =  \frac{c}{3} \log |x_2 - x_1 |
+ \frac{c}{6} \log \Bigg[ \left( \frac{\eta_i^{\frac{(1- \alpha_{\mathcal{O}})}{2}} \left( 1 - \eta_i^{\alpha_{\mathcal{O}} } \right)  }{ \alpha_{\mathcal{O}} (1 - \eta_i) } \right)
\left( \frac{\bar{\eta}_i^{\frac{(1- \alpha_{\mathcal{O}})}{2}} \left( 1 - \bar{\eta}_i^{\alpha_{\mathcal{O}} } \right)}{\alpha_{\mathcal{O}} (1 - \bar{\eta}_i) } \right)
\Bigg].
\label{EE-2}
\eea 
After obtaining the Euclidean entanglement entropy, we analytically continue to the Lorentzian time as $\tau_E=it$, and then investigate the time dependence of the entanglement entropy, $S_{A;i}$. 
Next, by substituting (\ref{z-new1-z-new2}) into \eqref{cross-ratios-1}, the cross ratios for 
$i=1,2$ result in 
\be
\begin{split}
	&\eta_1  =\frac{(e^{-i a \epsilon} (t -x) + x_1) (t - x + e^{-i a \epsilon} x_2)}{(t - x + e^{-i a \epsilon} x_1) (e^{-i a \epsilon} (t - x) + x_2)},
	~~~~~~ \bar{\eta}_1= \frac{( t + x - e^{-i a \epsilon} x_1) (e^{-i a \epsilon} (t+x) - x_2)}{(e^{-i a \epsilon} (t+x) - x_1) ( t+x - e^{-i a \epsilon} x_2)},\\
	&\eta_2  =\frac{(t+x_1 - e^{-i a \epsilon}  x) (e^{- i a \epsilon} (t+x_2) - x)}{(e^{- i a \epsilon} (t+ x_1) - x) (t+x_2 - e^{- i a \epsilon} x)},
	~~~~~~ \bar{\eta}_2 = \frac{(x+ e^{- i a \epsilon} (t- x_1)) (t-x_2 + e^{- i a \epsilon} x)}{(t- x_1 + e^{- i a \epsilon} x) (x+ e^{- i a \epsilon} (t -x_2))}. 
	\label{eta1-eta2}
\end{split}
\ee 
Subsequently, we take $\epsilon$ to be small, i.e. $a \epsilon \ll 1$, and then by the next-to-leading order in the small $\epsilon$-expansion, the cross ratios are approximated as 
\bea
\eta_i =  1 + i \epsilon f_i (t),
\;\;\;\;\;\;\;\;\;
\bar{\eta}_i =  1 + i \epsilon \bar{f}_i (t),
\eea 
where the imaginary parts of them are given by
\bea
&& f_{1} = - \frac{2 a (t - x) (x_2 - x_1)}{(t - x +x_1) (t - x + x_2)},
~~~~~ \bar{f}_{1} = - \frac{2 a (t+x) (x_2 -x_1)}{(t+x - x_1) (t+x- x_2)},
\cr && \cr
&& f_2 =  \frac{2 a x (x_2 -x_1)}{(t-x +x_1) (t-x+x_2)},
~~~~~~~ \bar{f}_2 = - \frac{2 a x (x_2 - x_1)}{(t+x -x_1) (t+x- x_2)}.
\label{f1-f2}
\eea 
Therefore, one has $\eta_i \simeq 1$ and $\bar{\eta}_i \simeq 1$. 
In the following, by investigating the time dependence of $f_i$ and $\bar{f}_i$, we will report the time dependence of the entanglement entropy for the single interval in Sections \ref{Sec:EE-for-finite-interval} and \ref{Sec:CB-SII}.

\subsection{Time Dependence of Partition Function \label{sec:TDPF}}

Before closely looking at the details of the time dependence of the entanglement entropy for the systems under consideration, we focus on the time dependence of the partition functions.
To this end, we define the Euclidean counterparts of these partition functions as
\be
\mathcal{Z}_{E,i}=\f{1}{N_{E,i}^2}.
\ee
In other words, those counterparts are given as the two-point functions, 
\be
\begin{split}
	\mathcal{Z}_{E,1}=\bra{0}\mathcal{O}^{\dagger}(x)e^{\tau_E H_0}e^{-2\epsilon H_R} e^{-\tau_E H_0}\mathcal{O} (x)\ket{0}, ~~~~~ \mathcal{Z}_{E,2}=\bra{0}\mathcal{O}^{\dagger}(x)e^{-2\epsilon H_R}\mathcal{O} (x)\ket{0}.
\end{split}
\ee
Then, we define the chiral and anti-chiral partition functions as
\be
\begin{split}
	Z_{E,i}(z,\overline{z})=Z_{C,i}(z) Z_{A,i}(\overline{z}), 
\end{split}
\ee
where the chiral and anti-chiral partition functions are defined as the holomorphic and anti-holomorphic functions, respectively.
The expression of $Z_{C,i}(z)$ and $Z_{A,i}(\overline{z})$ 
for each system is given as
\be
\begin{split}
	Z_{C,i}(z) = \frac{1}{(z^{\rm new}_{i,\epsilon} - z^{\rm new}_{i,- \epsilon})^{2h_{\mathcal{O}}}},
	~~~~~~~Z_{A,i}(\overline{z}) = \frac{1}{(\bar{z}^{\rm new}_{i,\epsilon} - \bar{z}^{\rm new}_{i,- \epsilon})^{2h_{\mathcal{O}}}},
\end{split}
\ee
where $z^{\rm new}_{i, \pm \epsilon}$ and $\bar{z}^{\rm new}_{i, \pm \epsilon}$ are given by \eqref{z-new1-z-new2}.
After performing the analytic continuation as $\tau_E=it$, we define the partition functions for the left-moving and right-moving modes $Z_{L,i}$ and  $Z_{R,i}$ as
\be
\begin{split}
	& Z_{L,i} =Z_{C,i}(z)|_{\tau_E=it}=\begin{cases} \frac{1}{(2 (x-t) \sin(a \epsilon))^{2h_{\mathcal{O}}}} ~~~~~~\text{for}~i=1,\\
		\frac{1}{(2 x \sin(a \epsilon))^{2h_{\mathcal{O}}}} ~~~~~~~~~~\text{for}~i=2,\\
	\end{cases}\\
	& Z_{R,i} =Z_{A,i}(\overline{z})|_{\tau_E=it}=\begin{cases} \frac{1}{(2 (x+t) \sin(a \epsilon))^{2h_{\mathcal{O}}}} ~~~~~~\text{for}~i=1,\\
		\frac{1}{(2 x \sin(a \epsilon))^{2h_{\mathcal{O}}}} ~~~~~~~~~~\text{for}~i=2.\\
	\end{cases}
\end{split}
\ee
Thus, the partition function for $i=2$ is independent of time, while that for $i=1$ depends on time.
Then, let us calculate the derivative of the system for $i=2$ with respect to time.
As the normalization constant is independent of time, we obtain it as the Schrödinger equation,
\be
i \partial_t \ket{\psi_2(t)} =H_0\ket{\psi_2(t)}.
\ee
From this Schrödinger equation, we can see that in the system with $i=2$, we start from the vacuum state with the insertion of $e^{-\epsilon H_R} \mathcal{O}(x)$, and {\it unitarily} evolve it in time with $H_0$.
Contrary to this time evolution, the system with $i=1$ is {\it non-unitarily} evolved in time because $e^{-\epsilon H_R} e^{-iH_0 t}$ is not a unitary operator, i.e.,
\be
e^{-\epsilon H_R} e^{-iH_0 t}(e^{-\epsilon H_R} e^{-iH_0 t})^{\dagger} \neq {\bf 1},~~~~~~~ (e^{-\epsilon H_R} e^{-iH_0 t})^{\dagger} e^{-\epsilon H_R} e^{-iH_0 t} \neq {\bf 1}.
\ee
The non-unitarity of the system with $i=1$ induces the time dependence of the partition function for $i=1$.
Finally, we report the asymptotic forms of those partition functions in the small $\epsilon$-expansion, i.e., $a \epsilon \ll 1$.
We expand those partition functions with respect to $a \epsilon \ll 1$, and then let $Z^{0}_{L,i}$ and $Z^{0}_{R,i}$ denote the leading terms in this small $\epsilon$-expansion.
Those leading terms are given as
\be \label{Z0-L-Z0-R}
\begin{split}
	& Z^0_{L,i} =\begin{cases} \frac{1}{(2 a \epsilon (x-t))^{2h_{\mathcal{O}}}} ~~~~~~\text{for}~i=1,\\
		\frac{1}{(2 a \epsilon x)^{2h_{\mathcal{O}}}} ~~~~~~~~~~\text{for}~i=2,\\
	\end{cases}\\
	& Z^0_{R,i} =\begin{cases} \frac{1}{(2 a \epsilon (x+t))^{2h_{\mathcal{O}}}} ~~~~~~\text{for}~i=1,\\
		\frac{1}{(2 a \epsilon x)^{2h_{\mathcal{O}}}} ~~~~~~~~~~\text{for}~i=2.\\
	\end{cases}
\end{split}
\ee

\subsection{Euclidean Time Evolution as the Post-Selecting Projective Measurement \label{eq:ETE-PSPM}}

Here, we will provide an interpretation of the Euclidean time evolution as a unitary time evolution induced by the post-selecting measurement as reported by \cite{2025PhRvB.112j4322L}.
We assume that the system is composed of the $2$d CFT system and a two-level system as
\be
\rho_{\text{Tot.}}=\rho_{\text{s}} \otimes  \rho_{\text{an}},
\ee
where $\rho_{\text{s}}$ and $ \rho_{\text{an}}$ are density operators for the $2$d CFT system and the two-level system, respectively.
Furthermore, we assume that those density operators are 
\be
\begin{split}
\rho_{\text{s}} = \ket {0}\bra{0},~~~~~~~~ \rho_{\text{an}}= \ket{\up}\bra{\up},
\end{split}
\ee
where $\ket{0}$ is the vacuum state of $H_0$. Moreover, $\ket{\up}$ and $\ket{\down}$ are the eigenvectors of $\sigma_z$, the Pauli's spin operator along the $z$ direction, i.e., 
\be
\sigma_z\ket{\up}=\ket{\up},~~~~~~~~ \sigma_z\ket{\down}=-\ket{\down},
\ee
where we assume that $\left \langle p=\up,\down|q=\up,\down\right \rangle=\delta_{p,q}$. Furthermore, we use the following representation
\bea
\ket{\up} = \begin{pmatrix} 
	1 \\ 0
\end{pmatrix}
,~~~~~~~~~~~~~~~~~~ 
\ket{\down} = \begin{pmatrix}
	0 \\ 1
\end{pmatrix}.
\eea 
Note that the eigenvalues of the Rindler Hamiltonian are not bounded from below. 
We define a truncated Hamiltonian $\tilde{H}_R$ and denote its lowest eigenvalue by $E_L$. We assume that $\tilde{H}_R$ reduces to $H_R$ in the limit $E_L \rightarrow - \infty$.
Then, we define a Hermitian Euclidean time evolution operator as
\bea \label{tilde-O}
\tilde{\mathcal{O}}= e^{- \epsilon (\tilde{H}_R - \tilde{E}_L)}.
\eea 
Here, $\tilde{E}_L < E_L$ can be considered as a lower bound on the eigenvalues of $\tilde{H}_R$. It is clear that all of the eigenvalues of $\tilde{O}$ are non-negative and smaller than one. Next, we can construct the following operator
\bea \label{U-tilde}
\tilde{U} = \begin{pmatrix} 
	\tilde{O} & \sqrt{1- \tilde{O}^2} \\
	- \sqrt{1- \tilde{O}^2} & \tilde{O}
\end{pmatrix},
\eea 
which is unitary and non-Hermitian as
\bea
\tilde{U}^\dagger \neq \tilde{U}, ~~~~~~~~~~~~~ \tilde{U}^\dagger \tilde{U} = \tilde{U} \tilde{U}^\dagger= 1.
\eea 
By applying \eqref{U-tilde} on $\ket{\psi} \otimes \ket{\up \down}$ where $| \psi \rangle$ is a state of the system and $\ket{\up \down}$ describes the states in the ancilla, we can obtain
\bea
&&\tilde{U} | \psi \rangle \otimes \ket{\up} = \tilde{O}  \ket{\psi} \otimes \ket{\up}  - \sqrt{1- \tilde{O}^2}  | \psi \rangle \otimes \ket{\down},
\cr && \cr 
&&\tilde{U} \ket{\psi} \otimes \ket{\down} =  \sqrt{1- \tilde{O}^2}  \ket{\psi} \otimes \ket{\up} +   \tilde{O} \ket{\psi} \otimes \ket{\down}.
\eea 
Now, we can rewrite $\tilde{U}$ as an exponential. To do so, we use the fact that all of the eigenvalues of the operator $\tilde{O}$ are non-negative and less than one. Therefore, we can define a new Hermitian operator
\bea
\Theta = \arccos (\mathcal{\tilde{O}}),
\eea 
where $\arccos (\mathcal{\tilde{O}})$ is defined by the following Taylor expansion
\bea
\arccos (\tilde{O}) = \frac{\pi}{2} . {\bf 1} - \sum_{m=0}^{\infty} \frac{(2m)!}{4^m (m!)^2 (2m+1)} \mathcal{\tilde{O}}^{2m+1}.
\eea 
Having said this, we can rewrite the unitary operator $\tilde{U}$ as
\bea
\tilde{U} =   \begin{pmatrix} 
	\cos \Theta & \sin \Theta \\
	- \sin \Theta & \cos \Theta \end{pmatrix}
= e^{- i \Theta \sigma_y}, 
\eea 
where 
\bea
\sigma_y= \begin{pmatrix} 0 & -i \\ i & 0 \end{pmatrix},
\eea 
is the Pauli's spin operator along the $y$ direction. Now, we want to apply $\tilde{U}$ to construct the density operators in \eqref{eq:lorentzian-density-op}. During the non-unitary time evolution corresponding to $i=1$, we insert $\mathcal{O}$ at the spatial location, $x$, evolve the system with $H_0$ that acts only on the $2$d CFT system, and then evolve the system with $\tilde{U}$.
In contrast, during the unitary time evolution corresponding to $i=2$, we insert $\mathcal{O}$ at the spatial location, $x$, evolve the system with $\tilde{U}$, and then evolve the system with $H_0$.
Then, the systems corresponding to $i=1,2$ result in
\be \label{eq:bofore-postselcted}
\begin{split}
&\rho_{\text{Tot},1}\propto e^{-i \Theta \sigma_y} e^{-iH_0 t} \mathcal{O}(x)(\rho_{\text{s}} \otimes  \rho_{\text{an}})\mathcal{O}^{\dagger}(x)e^{iH_0 t}e^{i \Theta \sigma_y},\\
&\rho_{\text{Tot},2}\propto e^{-iH_0 t}e^{-i \Theta \sigma_y}  \mathcal{O}(x)(\rho_{\text{s}} \otimes  \rho_{\text{an}})\mathcal{O}^{\dagger}(x)e^{i \Theta \sigma_y}e^{iH_0 t}.
\end{split}
\ee
Therefore, after projecting the systems on $\ket{\up}$, we obtain
\be
\begin{split}
	&\bra{\up}\rho_{\text{Tot},1}\ket{\up}\propto \tilde{O} e^{-iH_0 t} \mathcal{O}(x)\rho_{\text{s}}\mathcal{O}^{\dagger}(x)e^{iH_0 t} \tilde{O},\\
	&~~~~~~~~~~~~~~~ = e^{- \epsilon (\tilde{H}_R- \tilde{E}_L)} e^{-iH_0 t} \mathcal{O}(x)\rho_{\text{s}}\mathcal{O}^{\dagger}(x)e^{iH_0 t} e^{- \epsilon (\tilde{H}_R- \tilde{E}_L)}, \\
	&\bra{\up}\rho_{\text{Tot},2}\ket{\up}\propto e^{-iH_0 t}\tilde{O} \mathcal{O}(x)\rho_{\text{s}}\mathcal{O}^{\dagger}(x) \tilde{O}e^{iH_0 t}\\
	&~~~~~~~~~~~~~~~ = e^{-iH_0 t} e^{- \epsilon (\tilde{H}_R- \tilde{E}_L)} \mathcal{O}(x)\rho_{\text{s}}\mathcal{O}^{\dagger}(x)  e^{- \epsilon (\tilde{H}_R- \tilde{E}_L)} e^{iH_0 t}.
\end{split}
\ee
Then, we normalize $\bra{\up}\rho_{\text{Tot},i}\ket{\up}$ as
\bea
\rho_{\text{s},i}^{\text{measured}}= \f{\langle \up | \rho_{\text{Tot},i} | \up \rangle}{\text{tr} \left( \langle \up | \rho_{\text{Tot},i} | \up \rangle \right)},
\eea 
and obtain the measured density operators as
\bea \label{rho-measured}
&\rho_{\text{s},1}^{\text{measured}} = \f{e^{- \epsilon \tilde{H}_R} e^{-iH_0 t} \mathcal{O}(x)\rho_{\text{s}}\mathcal{O}^{\dagger}(x)e^{iH_0 t} e^{- \epsilon \tilde{H}_R}}{\text{tr} \left( e^{- \epsilon \tilde{H}_R} e^{-iH_0 t} \mathcal{O}(x)\rho_{\text{s}}\mathcal{O}^{\dagger}(x)e^{iH_0 t} e^{- \epsilon \tilde{H}_R} \right)}, \nonumber \\
&\rho_{\text{s},2}^{\text{measured}} = \f{e^{-iH_0 t} e^{- \epsilon \tilde{H}_R} \mathcal{O}(x)\rho_{\text{s}}\mathcal{O}^{\dagger}(x) e^{- \epsilon \tilde{H}_R} e^{iH_0 t}}{\text{tr} \left( e^{-iH_0 t} e^{- \epsilon \tilde{H}_R} \mathcal{O}(x)\rho_{\text{s}}\mathcal{O}^{\dagger}(x) e^{- \epsilon \tilde{H}_R} e^{iH_0 t} \right)}.
\eea 
Note that the density operators in \eqref{rho-measured} are independent of $\tilde{E}_L$. Therefore, we can send $\tilde{E}_L$ to minus infinity. 
Furthermore, we take the limit $E_L \rightarrow - \infty$. Correspondingly, $\tilde{H}_R$ reduces to $H_R$ in \eqref{rho-measured}. Then, \eqref{rho-measured} becomes the same as (\ref{eq:lorentzian-density-op}).
Thus, we can construct the density operators by post-selecting projection measurement on the ancilla.

Moreover, the norm of the density operators in (\ref{eq:bofore-postselcted}) becomes infinite due to the contraction of the two local operators. 
To keep their norms finite, for example, one may replace $\mathcal{O}(x)$ with $e^{-\tilde{\epsilon}H_{0}}\mathcal{O}(x)$. 
However, after performing post-selecting projection measurement, by setting the trace of the projected density operators to be one, we can obtain the normalized density operators.
Therefore, we would not discuss how to tame the divergence of $\rho_{\text{Tot.},i}$ in this paper.

\subsection{Finite Interval \label{Sec:EE-for-finite-interval}}

In this section, as a warm-up, we will investigate the time dependence of the entanglement entropy for a subsystem which is the finite interval $A \in [x_1 , x_2 ]$ of length $l = x_2 -x_1$. Let $x$ denotes the spatial location of the local operator.
In the following sections, we will consider the setup where the local operator is inserted into the positive spatial region outside of $A$, i.e., $x_2>x_1>x>0$ or $x>x_2>x_1>0$.
We also consider the setup where the local operator is inserted into $A$, i.e., $x_2 > x > x_1 >0$.
Since the time dependence of the entanglement entropy for $x_2>x_1>x>0$ is similar to that for $x_2 > x > x_1> 0$, we postpone the latter one to Appendix \ref{Sec:Insetion-into-A}. 

\subsubsection{Insertion into Positive Spatial Region Outside of A
	\label{Sec:Insertion-Outside}}

In this section, we first investigate and present the time dependence of the entanglement entropy for $x_2>x_1>x>0$.
In this case, 
it is easy to check that the denominator of $f_{i=1,2}$ in \eqref{f1-f2} is non-zero for all times. Therefore, for $a \epsilon \ll 1$, $\eta_{i= 1,2}$ is always very close to one, and we can take the limit $\eta_{i= 1,2} \rightarrow 1$. It is straightforward to verify that for $a \epsilon \ll 1$, one arrives at
\bea
\left( \frac{\eta_i^{\frac{(1- \alpha_{\mathcal{O}})}{2}} \left( 1 - \eta_i^{\alpha_{\mathcal{O}} } \right)  }{ \alpha_{\mathcal{O}} (1 - \eta_i) } \right) \simeq 1.
\label{eta-i-insertion-outside-LHS}
\eea
On the other hand, the denominator of $\bar{f}_{i=1,2}$ becomes zero at $t=x_1 -x$ and $t= x_2 - x$. 
In other words, at these times $\bar{\eta}_{i=1,2}$ becomes infinite, and the sign of its imaginary part changes at these times. 
Therefore, we have to take the correct branch cut of $\bar{\eta}_{i=1,2}$. In this case, one has
\be
\begin{split}
	\overline{f}_{i} =\begin{cases}
		\text{negative}~&~\text{for}~ x_1-x>t>0,\\
		\text{positive}~&~\text{for}~ x_2-x>t>x_1-x,\\
		\text{negative} ~&~\text{for}~ t>x_2-x.
	\end{cases}
\end{split}
\ee
For the early-time region, $x_1 -x>t>0$, $\bar{\eta}_{i=1,2}$ is close to one and we can take the limit $\bar{\eta}_{i=1,2} \rightarrow 1$. Therefore, in this limit, the entanglement entropy is approximated as the vacuum one,
\bea
S_{A;i} = S^{\rm Vac}_{A} = \frac{c}{3} \log \left( l \right).
\label{EE-Vacuum}
\eea 
For the middle time interval, $x_1 - x < t < x_2 -x$, we need to take another branch different from the one for the early time. Therefore, $\bar{\eta}_{i}$ obtains the phase factor as $\bar{\eta}_i \rightarrow e^{-2 \pi i} \bar{\eta}_i$. Then, when we take the limit, $\bar{\eta}_i \rightarrow 1$, at the leading order, we arrive at
\be
\begin{split}
	\left( \frac{\bar{\eta}_i^{\frac{(1- \alpha_{\mathcal{O}})}{2}} \left( 1 - \bar{\eta}_i^{\alpha_{\mathcal{O}} } \right)}{\alpha_{\mathcal{O}} (1 - \bar{\eta}_i) } \right) \simeq  
	\frac{2 \kappa_{\mathcal{O}} }{\epsilon \bar{f}_i(t)} = 
	\begin{cases}
		\frac{\kappa_{\mathcal{O}} (t+x -x_1)(x_2-x -t)}{a \epsilon l (t+x)} > 0, ~~~&~~~\text{for}~i=1,\\
		\frac{\kappa_{\mathcal{O}} (t+x -x_1)(x_2-x -t)}{a \epsilon l x} > 0, ~~~&~~~\text{for}~i=2.
	\end{cases}
\end{split}
\label{eta1-etab-1-EE-2x bigger}
\ee
Next, by plugging \eqref{eta-i-insertion-outside-LHS} and \eqref{eta1-etab-1-EE-2x bigger} into \eqref{EE-2}, we obtain
\be
\begin{split}
	S_{A;i} &= 
	S_A^{\rm Vac} + \frac{c}{6} \log { \left[\frac{2 \kappa_{\mathcal{O}} }{ \epsilon \bar{f}_{i=1,2} (t)}\right]}\\
	&=S_A^{\rm Vac} +\begin{cases}
		\frac{c}{6} \log {\left[ \frac{\kappa_{\mathcal{O}} (t+x -x_1) (x_2- x -t)}{a \epsilon l (t+ x)}\right]},~~~&~~~\text{for}~i=1,
		\vspace{1mm}\\
		\frac{c}{6} \log {\left[ \frac{\kappa_{\mathcal{O}} (t+x -x_1) (x_2- x -t)}{a \epsilon l x}\right]},~~~&~~~\text{for}~i=2.
	\end{cases}
\end{split}
\ee
For the late-time region, $t >x_2 -x$, the sign of $\bar{f}_i$ is the same as that for the time $t < x_1 -x$. This suggests that we should take the same branch for the early-time region. Therefore, $\bar{\eta}_i$ does not obtain the additional phase factor.
Consequently, in the small $\epsilon$ limit, the entanglement entropy reduces to the vacuum one.
Putting everything together, one can write
\be
\begin{split}
	&S_{A;1}= S_A^{\rm Vac} + 
	\begin{cases}
		0, &~~ 0 < t < x_1 -x, \\
		\frac{c}{6} \log \left[ \frac{\kappa_{\mathcal{O}} (t+x -x_1) (x_2- x -t)}{a \epsilon l (t+ x)}
		\right], &~~  x_1-x < t < x_2 -x, \\
		0 &~~ t>  x_2- x,
	\end{cases}\\
	&S_{A;2}= S_A^{\rm Vac} +
	\begin{cases}
		0, &~~ 0 < t < x_1 -x, \\
		\frac{c}{6} \log \left[ \frac{\kappa_{\mathcal{O}} (t+x -x_1) (x_2- x -t)}{a \epsilon l x}
		\right], &~~  x_1-x < t < x_2 -x, \\
		0, &~~ t>  x_2- x.
	\end{cases}
	\label{SA-i-finite-interval-outside-ER-LHS}
\end{split}
\ee
Thus, in the intermediate time interval, $x_1-x<t<x_2-x$, $S_{A;i}$ is larger than the vacuum entanglement entropy. 
Let us compare $S_{A;1}$ to $S_{A;2}$ for $x_1-x<t<x_2-x$.
By subtracting $S_{A;1}$ from $S_{A;2}$, we obtain the difference, i.e. $\delta S_{A}$, between them as 
\be \label{eq:deltaSA1}
\delta S_{A}=S_{A;2}-S_{A;1}=\f{c}{6}\log{\left(\f{x+t}{x}\right)} >0, ~~~~~ \text{for}~~~ x_1-x<t<x_2-x,
\ee
where $(t+x)/x>1$.

Moreover, when the operator is inserted into the right region to $A$, i.e., $0<x_1<x_2<x$, the entanglement entropy results in
\be
\begin{split}
	&S_{A;1} = S_A^{\rm Vac} +
	\begin{cases}
		0,  &~~ 0 < t < x -x_2, \\
		\frac{c}{6} \log \left[ \frac{\kappa_{\mathcal{O}} (x- t -x_1) (t - x +x_2)}{a \epsilon l (x - t)}
		\right], &~~  x -x_2 < t < x -x_1, \\
		0, &~~ t>  x- x_1,
	\end{cases}\\
	&S_{A;2}= S_A^{\rm Vac} +
	\begin{cases}
		0, &~~ 0 < t < x -x_2, \\
		\frac{c}{6} \log \left[ \frac{\kappa_{\mathcal{O}} (x -t -x_1) (t -x +x_2)}{a \epsilon l x}
		\right], &~~  x-x_2 < t < x -x_1, \\
		0, &~~ t>  x- x_1.
	\end{cases}
\end{split}
\label{SA-i-finite-interval-outside-ER-RHS}
\ee
Thus, similar to the previous case, in the intermediate time interval, $x_2-x<t<x-x_1$, $S_{A;i}$ is larger than the vacuum entanglement entropy.
In contrast to the previous case, $\delta S_A$ in this time interval is negative as in
\be \label{eq:deltaSA2}
\delta S_{A}=S_{A;2}-S_{A;1}=\f{c}{6}\log{\left(\f{x-t}{x}\right)} <0, ~~~~~ \text{for}~~~x_2-x<t<x-x_1,
\ee
where $(x-t)/x<1$.
In Section \ref{sec:TE-EMT}, we will explain what induces the difference in entanglement entropy, reported in (\ref{eq:deltaSA1}) and (\ref{eq:deltaSA2}).

\subsection{Semi-Infinite Interval \label{Sec:CB-SII}}

Now, we move on to the main target of this paper, i.e., the research on the time dependence of the entanglement entropy for the semi-infinite interval $A \in [x_1 , x_2 = \infty]$.
In the spatially-infinite system in the vacuum state, the entanglement entropy for the finite single-interval, $A$, with the subsystem size of $l$ behaves as 
\be
S^{\rm Vac}_A\approx\f{c}{3}\log{ \left( l \right)}.
\ee
Correspondingly, the entanglement entropy for the semi-infinite interval ($l \rightarrow \infty$) is infinite.
Therefore, in this section, we will report the time dependence of the growth of the entanglement entropy, instead of the entanglement entropy.
To this end, we define the growth of the entanglement entropy as
\be
\Delta S_{A;i} =S_{A;i} -S^{\text{Vac}}_{A},
\ee
where $S_{A;i}$ is the the entanglement entropy for (\ref{eq:systems-under-consideration}), while $S^{\text{Vac}}_A$ is the one for the vacuum state. 
By applying \eqref{conformal map-EE}, \eqref{twist-operators-conformal-transformation} and \eqref{G-n} and then performing analytic continuation as $\tau_E=it$, we obtain $\Delta S_{A;i}$ as
\bea
\Delta S_{A;i} 
&=& \lim_{n \rightarrow 1} \frac{1}{1-n} \log \Bigg[ \frac{G_{n} (\eta_i, \bar{\eta}_i) }{\langle \sigma_n(\eta_i, \bar{\eta}_i) \bar{\sigma}_n(1, 1)  \rangle} \Bigg] =  \lim_{n \rightarrow 1} \frac{1}{1-n} \log \Bigg[ \frac{G_{n} (\eta_i, \bar{\eta}_i) }{|1 - \eta_i|^{-4 n h_n}} \Bigg]
\cr && \cr 
&=& \frac{c}{6} \log \Bigg[ \left( \frac{\eta_i^{\frac{(1- \alpha_{\mathcal{O}})}{2}} \left( 1 - \eta_i^{\alpha_{\mathcal{O}} } \right)  }{ \alpha_{\mathcal{O}} (1 - \eta_i) } \right)
\left( \frac{\bar{\eta}_i^{\frac{(1- \alpha_{\mathcal{O}})}{2}} \left( 1 - \bar{\eta}_i^{\alpha_{\mathcal{O}} } \right)}{\alpha_{\mathcal{O}} (1 - \bar{\eta}_i) } \right)
\Bigg],
\label{Delta-S-3}
\eea 
where in the last line we applied \eqref{conformal block-von Neumann limit}. 
In the following, we will report the time dependence of $\Delta S_{A;i}$ when the local operator is inserted into the spatial open set of $(0,x_1)$, while in Appendix \ref{Sec:Insetion-into-A}, we will report that of $\Delta S_{A;i}$ when the local operator is inserted into $A$.

\subsubsection{Insertion into Spatial Region, $0<x<x_1$ 
\label{Sec:Insertion into Spatial Region, $0<x<x_1$}}

In this section, we will investigate the time dependence of $\Delta S_{A;i}$ for $i=1,2$ when the local operator is inserted into $0<x<x_1$.
We closely look at the cross ratios by the next-to-leading order in the small $\epsilon$ limit.
In this expansion, the leading terms of the cross ratios are one, while the coefficients of $\epsilon$ are given by 
\be
\begin{split}
	&f_1 = - \frac{2 a (t- x)}{t-x+x_1},~~~~~ \bar{f}_1 =\frac{2 a (t+x)}{t+x -x_1},
	\\
	&f_2 = \frac{2 a x}{t-x +x_1},~~~~~~~ \bar{f}_2 = \frac{2 a x}{t+x- x_1}.
\end{split}
\label{f1-fb1-semi-infinite}
\ee
It is easy to verify that for $t>0$, $f_{i=1,2}$ is always finite, and hence, $\eta_{i=1,2}$ is close to one. In this case, \eqref{eta-i-insertion-outside-LHS} is still valid.
On the other hand, in the small $\epsilon$ limit, the denominator of $\bar{f}_{i=1,2}$ becomes zero at $t = x_1 -x$, and one has
\be
\begin{split}
	\overline{f}_{i=1,2}=\begin{cases}
		\text{negative}~&~\text{for}~ 0<t<x_1-x,\\
		\text{positive}~&~\text{for}~ x_1-x<t.\\
	\end{cases}
\end{split}
\ee
For the early time region, $0 < t < x_1 -x$, $\bar{\eta}_i $ does not receive additional phase factor. 
Therefore, in the small $\epsilon$ limit, the time dependence of $S_{A;i}$ under consideration is the same as the vacuum one.
Moreover, for the late time region, $t> x_1 -x$, $\bar{\eta}_i$ obtains an additional phase factor as $\bar{\eta}_i \rightarrow e^{-2 \pi i} \bar{\eta}_i$.
Then, in the small $\epsilon$ limit, we obtain
\be
\begin{split}
	\left( \frac{\bar{\eta}_i^{\frac{(1- \alpha_{\mathcal{O}})}{2}} \left( 1 - \bar{\eta}_i^{\alpha_{\mathcal{O}} } \right)}{\alpha_{\mathcal{O}} (1 - \bar{\eta}_i) } \right) \simeq  \frac{2 \kappa_{\mathcal{O}} }{ \epsilon \bar{f}_i(t)}=
	\begin{cases}
		\frac{\kappa_{\mathcal{O}} (t+x -x_1)}{a \epsilon (t+x)} > 0 ~&~\text{for}~ i=1,\\
		\frac{\kappa_{\mathcal{O}} (t+x -x_1)}{a \epsilon x} > 0 ~&~\text{for}~ i=2.\\
	\end{cases}
\end{split}
\label{eta1-etab-1-EE-2x bigger-semi-infinite}
\ee
Subsequently, by substituting \eqref{eta-i-insertion-outside-LHS} and \eqref{eta1-etab-1-EE-2x bigger-semi-infinite} into \eqref{Delta-S-3}, we obtain the time dependence of $\Delta S_{A;i}$ as
\be
\begin{split}
	& \Delta S_{A;1}=  
	\begin{cases}
		0, ~&~\text{for}~ 0 < t < x_1 -x, \\
		\frac{c}{6} \log \left[ \frac{\kappa_{\mathcal{O}} (t+x -x_1)}{a \epsilon (t+ x)} \right], ~&~\text{for}~  x_1-x < t.
	\end{cases}\\
	& \Delta S_{A;2}= 
	\begin{cases}
		0, ~&~\text{for}~ 0 < t < x_1 -x, \\
		\frac{c}{6} \log \left[ \frac{\kappa_{\mathcal{O}} (t+x -x_1)}{a \epsilon x} \right], ~&~\text{for}~  x_1-x < t.
	\end{cases}
\end{split}
\label{SA-1-semi-infinite-interval-outside-ER}
\ee
Then, to closely look at the late-time behavior, we take the large time limit, $t\gg x_1 -x$, and then expand the function with respect to $t$.
At the leading order of the large time expansion, $\Delta S_{A;i}$ is approximated as
\be
\begin{split}
	\Delta S_{A;i} \underset{t \gg x_1-x}{\approx} \begin{cases}
		\frac{c}{6} \log {\left( \frac{\kappa_{\mathcal{O}}}{a \epsilon }\right)},~&~\text{for}~ i=1,\\
		\frac{c}{6} \log {\left( \frac{\kappa_{\mathcal{O}} t}{a \epsilon x}\right)},~&~\text{for}~ i=2.\\
	\end{cases}
\end{split}
\label{SA-1-late time-1}
\ee
Thus, these late-time approximations suggest that in the large time limit, $\Delta S_{A,i=1}$ saturates to a certain value independent of the subsystem size and its endpoints, while $\Delta S_{A;i=2}$ logarithmically grows with time. 
In Section \ref{Sec:physical-interpretaion}, we will investigate how this difference between $i=1$ and $i=2$ emerges.

\subsection{Time Evolution of Energy-Momentum Tensors \label{sec:TE-EMT}}

Now, before moving on to the report on the entanglement dynamics investigated by the different computational methods, we will present the time dependence of the expectation values of the chiral and anti-chiral energy-momentum tensors. 
We begin with the Euclidean calculation of the chiral and anti-chiral energy-momentum densities, and then analytically continue to the real time.
In the Euclidean counterparts of the systems under consideration, the expectation values of the chiral and anti-chiral energy-momentum tensors are given as 
\be
\begin{split}
	&\langle T(z_X) \rangle_{i}=\langle \psi_i(t) | T(z_X) | \psi_i(t) \rangle 
	= \frac{\langle 0| \mathcal{O}^{\dagger} (z^{\rm new}_{i, - \epsilon}, \bar{z}^{\rm new}_{i, - \epsilon}) T(z_X) \mathcal{O} (z^{\rm new}_{i, \epsilon}, \bar{z}^{\rm new}_{i, \epsilon}) |0 \rangle}{\langle 0| \mathcal{O}^{\dagger} (z^{\rm new}_{i, - \epsilon}, \bar{z}^{\rm new}_{i, - \epsilon}) \mathcal{O} (z^{\rm new}_{i, \epsilon}, \bar{z}^{\rm new}_{i, \epsilon}) |0 \rangle},
	\\
	&\langle \bar{T}(\bar{z}_X) \rangle_{i} = \langle \psi_i(t) | \bar{T}(\bar{z}_X) | \psi_i(t) \rangle = \frac{\langle 0| \mathcal{O}^{\dagger} (z^{\rm new}_{i, - \epsilon}, \bar{z}^{\rm new}_{i, - \epsilon}) \bar{T}(\bar{z}_X) \mathcal{O} (z^{\rm new}_{i, \epsilon}, \bar{z}^{\rm new}_{i, \epsilon}) |0 \rangle}{\langle 0| \mathcal{O}^{\dagger} (z^{\rm new}_{i, - \epsilon}, \bar{z}^{\rm new}_{i, - \epsilon}) \mathcal{O} (z^{\rm new}_{i, \epsilon}, \bar{z}^{\rm new}_{i, \epsilon}) |0 \rangle},
\end{split}
\ee
where $z_X= i X$. Then, by exploiting the Ward-Takahashi identities as \cite{Ginsparg:1988ui}
\bea
&& \langle 0 | T(z) \mathcal{O} (z_1, \bar{z}_1) \mathcal{O} (z_2, \bar{z}_2) | 0\rangle = \sum_{j=1}^{2} 
\Bigg[ \frac{h_{\mathcal{O}}}{(z- z_j)^2} + \frac{1}{(z- z_j) } \partial_{z_j} \Bigg] \langle \mathcal{O} (z_1, \bar{z}_1) \mathcal{O} (z_2, \bar{z}_2) \rangle,
\cr && \cr
&& \langle 0 | \bar{T}( \bar{z}) \mathcal{O} (z_1, \bar{z}_1) \mathcal{O} (z_2, \bar{z}_2) | 0\rangle = \sum_{j=1}^{2} 
\Bigg[ \frac{\bar{h}_{\mathcal{O}}}{(\bar{z}- \bar{z}_j)^2} + \frac{1}{(\bar{z}- \bar{z}_j) } \partial_{\bar{z}_j} \Bigg] \langle \mathcal{O} (z_1, \bar{z}_1) \mathcal{O} (z_2, \bar{z}_2) \rangle,
\nonumber
\\
\eea 
we obtain
\bea
&& \langle T(z_X) \rangle_{i} = h_{\mathcal{O}} \Bigg[ \frac{1}{(z_X - z^{\rm new}_{i, - \epsilon})} - \frac{1}{(z_X - z^{\rm new}_{i, \epsilon})} \Bigg]^2,
\cr && \cr 
&& \langle \bar{T}(\bar{z}_X) \rangle_{i} = \bar{h}_{\mathcal{O}} \Bigg[ \frac{1}{(\bar{z}_X - \bar{z}^{\rm new}_{i, - \epsilon})} - \frac{1}{(\bar{z}_X - \bar{z}^{\rm new}_{i, \epsilon})} \Bigg]^2.
\label{T-holo-antiholo-1-2}
\eea 
Subsequently, by exploiting the analytic continuation, $\tau_E=it$, we obtain the time dependence of the chiral and anti-chiral energy-momentum densities as
{\small
	\be \label{eq:time-dependence-of-EM}
	\begin{split}
		&\langle T(z_X) \rangle_{1} =\frac{4 h_{\mathcal{O}} (t-x)^2 \sin^2 (a \epsilon)}{\Big[ X^2 + (t-x)^2 + 2 X (t-x) \cos (a \epsilon) \Big]^2},~
		\langle \bar{T}(\bar{z}_X) \rangle_{1} = \frac{4 \bar{h}_{\mathcal{O}} (t+x)^2 \sin^2 (a \epsilon) }{\Big[  X^2 + (t+x)^2 - 2 X (t+x) \cos (a \epsilon) \Big]^2},\\
		&\langle T(z_X) \rangle_{2} = \frac{4 h_{\mathcal{O}} x^2 \sin^2 (a \epsilon)}{\Big[ x^2 + (t+X)^2 - 2 x (t+X) \cos (a \epsilon) \Big]^2},~
		\langle \bar{T}(\bar{z}_X) \rangle_{2} = \frac{4 \bar{h}_{\mathcal{O}} x^2 \sin^2 (a \epsilon) }{\Big[ x^2 + (t-X)^2 + 2 x (t-X) \cos (a \epsilon) \Big]^2}.
	\end{split}
	\ee}
Note that $X$ and $x$ are the spatial locations of the energy-momentum tensor and primary operator, respectively. 
Moreover, if we take $\epsilon$ to be zero, then these expectation values vanish. 
This suggests that the growth, induced by the insertion of the local operators, of these energy-momentum densities in the Heisenberg picture needs the Euclidean process that tames the divergence induced by the contraction of the local operators.
To investigate the fine-grained properties of the systems under consideration, as a first step, we closely look at the leading order of $\left \langle T(z_X)\right \rangle_{i}$ and $\left \langle \bar{T}(\bar{z}_X)\right \rangle_{i}$ in the large time expansion, $t \gg x, X$, and then obtain them as
\bea
&& \langle T(z_X) \rangle_{1} \propto \frac{1}{t^2}, \;\;\;\;\;\;\;\; \langle \bar{T}(\bar{z}_X) \rangle_{1} \propto \frac{1}{t^2},
\cr && \cr
&& \langle T(z_X) \rangle_{2} \propto \frac{1}{t^4}, \;\;\;\;\;\;\;\; \langle \bar{T}(\bar{z}_X) \rangle_{2}  \propto \frac{1}{t^4}.
\label{T-holo-antiholo-late-times}
\eea 
Note that we omitted the details of the coefficients of the leading order term.
Thus, the time ordering of the Euclidean and Lorentzian processes affects the asymptotic behaviors of the energy-momentum densities: the energy-momentum densities for $i=2$ decay faster than those for $i=1$.
Moreover, at $t=0$, both energy-momentum densities are equal to each other as we expected.
We can easily check that $\left \langle T(z_X)\right \rangle_{i}$ and $\left \langle \bar{T}(\bar{z}_X)\right \rangle_{i}$ are finite since the denominators of these expectation values are positive. 
To simplify and investigate the time dependence of the energy-momentum densities, we consider the small $\epsilon$-expansion, $a \epsilon \ll 1$.
At the leading order in the small $\epsilon$-expansion, these expectation values reduce to 
\be \label{eq:time-EM-smallepsilon}
\begin{split}
	& \langle T(z_X) \rangle_1 =   \frac{ 4 h_{\mathcal{O}}(a \epsilon)^2 (t-x)^2 }{\left( X+ t -x \right)^4}= \f{\mathcal{F}_L(t,x,X)}{(Z^0_{L,1})^{\f{1}{h_{\mathcal{O}}}}},
	~~ \langle \bar{T}(\bar{z}_X) \rangle_1 = \frac{ 4 h_{\mathcal{O}} (a \epsilon)^2 (t+x)^2 }{\left( X- t -x \right)^4} =\f{\mathcal{F}_R(t,x,X)}{(Z^0_{R,1})^{\f{1}{h_{\mathcal{O}}}}},\\
	& \langle T(z_X) \rangle_2 =   \frac{ 4 h_{\mathcal{O}}(a \epsilon)^2 x^2 }{\left( X+ t -x \right)^4} =\f{\mathcal{F}_L(t,x,X)}{(Z^0_{L,2})^{\f{1}{h_{\mathcal{O}}}}},
	~~~~~~~ \langle \bar{T}(\bar{z}_X) \rangle_2 =   \frac{ 4 h_{\mathcal{O}} (a \epsilon)^2 x^2 }{\left( X- t -x \right)^4} =\f{\mathcal{F}_R(t,x,X)}{(Z^0_{R,2})^{\f{1}{h_{\mathcal{O}}}}},\\
\end{split}
\ee
where the functions, $\mathcal{F}_{\alpha=R,L}(t,x,X)$, are defined as
\be
\mathcal{F}_L(t,x,X)=\f{h_{\mathcal{O}}}{(X+t-x)^4}, ~~~~~~ \mathcal{F}_R(t,x,X)=\f{h_{\mathcal{O}}}{(X-t-x)^4}.
\ee
Thus, the difference between the expectation values of the energy-momentum densities for $i=1$ and that for $i=2$ is induced by the difference between the partition functions for $i=1$ and that for $i=2$.
Since the denominators of the energy-momentum densities in (\ref{eq:time-EM-smallepsilon}) become zero at $t=\pm (X-x)$, the time dependence of  (\ref{eq:time-EM-smallepsilon}) suggests that local excitations emerge at the insertion point of the local operator, and then propagate to the left and right at the speed of light, so that the energy-momentum densities at the spatial point, $X$, become large.

\begin{figure}
	\begin{center}
		\includegraphics[scale=0.48]{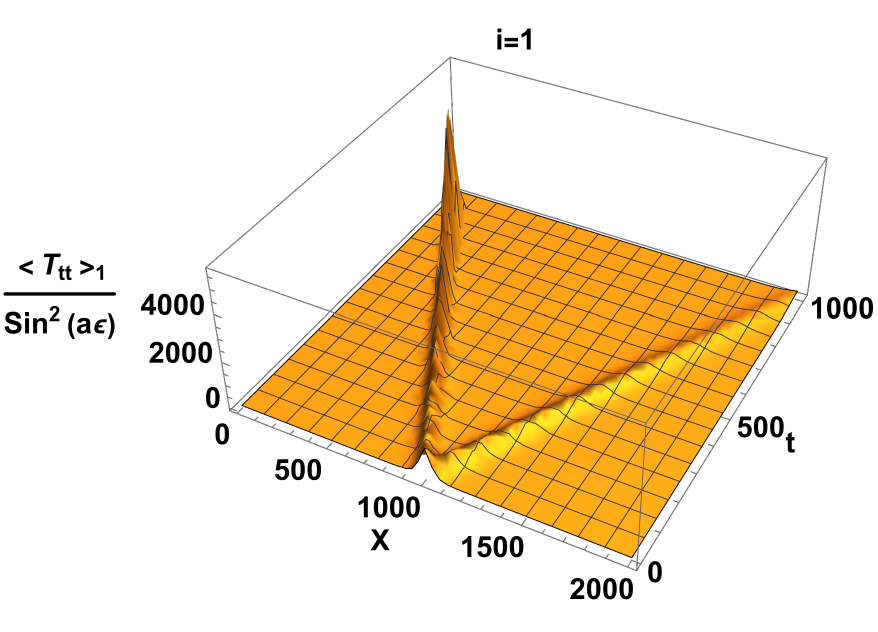}
		\hspace{0.3cm}
		\includegraphics[scale=0.48]{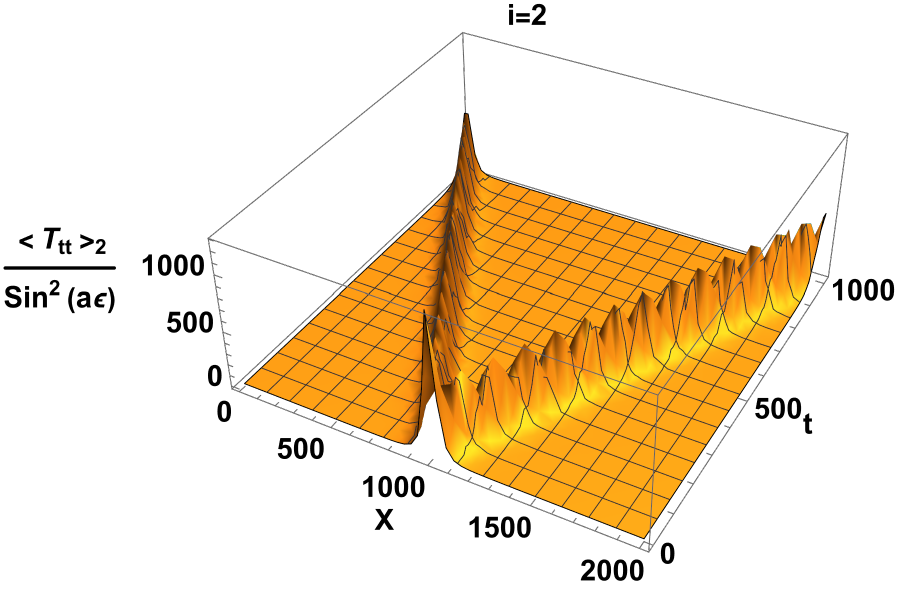}
	\end{center}
	\caption{$\frac{\langle T_{tt} \rangle_{i}}{\sin^2(a \epsilon)}$ as a function of $X$ and $t$ when the primary operator is inserted at $x= 10^3$. 
		{\it Left}) For the non-unitary time evolution where $i=1$. {\it Right}) For the unitary time evolution where $i=2$. At the insertion point of the primary operator, two energy excitations are created and they propagate along two null directions.
		We set $h_{\mathcal{O}}= \bar{h}_{\mathcal{O}}=10^3$, $a=10^{-3}$ and $\epsilon=50$.
	}
	\label{fig: T-x}
\end{figure}

Now, we closely look at the time dependence of the energy density, which is defined as $\langle T_{tt} \rangle_i=\langle T(z_X) \rangle_i+\langle \bar{T}(\bar{z}_X) \rangle_i$.
Combining the time dependence of the chiral and anti-chiral energy-momentum densities in (\ref{eq:time-dependence-of-EM}), we obtain that of the energy density as
{\small
	\bea
	&& \langle T_{tt} \rangle_1 = 4 \sin^2(a \epsilon)h_{\mathcal{O}} \cdot \Bigg[ \frac{ (t-x)^2}{\big[ X^2 + (t-x)^2 + 2X (t-x) \cos (a \epsilon) \big]^2}
	+ \frac{ (t+x)^2}{\big[ X^2 + (t+x)^2 - 2X (t+x) \cos (a \epsilon) \big]^2}\Bigg],
	\cr && \cr 
	&& \langle T_{tt} \rangle_2 = 4 x^2 \sin^2(a \epsilon)h_{\mathcal{O}}  \cdot \Bigg[ \frac{1}{ \big[ x^2 + (t+X)^2 - 2 x (t+X) \cos(a \epsilon) \big]^2} + 
	\frac{1 }{ \big[ x^2 + (t-X)^2 + 2 x (t-X) \cos(a \epsilon) \big]^2} \Bigg].
	\nonumber
	\\
	\label{T-tt-psi-1-psi-2}
	\eea 
}
Next, we take the small $\epsilon$ limit, i.e., $a \epsilon \ll 1$, and then at the leading order in this limit, we obtain the time dependence of $\langle T_{tt} \rangle_i$ as
\bea
&& \langle T_{tt} \rangle_1 =  4 (a \epsilon)^2 h_{\mathcal{O}}  \cdot \Bigg[ \frac{ (t-x)^2 }{\left( X+ t -x \right)^4}
+ \frac{(t+x)^2 }{\left( X - x-t \right)^4}
\Bigg],
\cr && \cr 
&& \langle T_{tt} \rangle_2 = 4 x^2 (a \epsilon)^2 h_{\mathcal{O}} \cdot \Bigg[ \frac{1 }{\left( X- x +t \right)^4} 
+ \frac{ 1 }{\left( x- X+t \right)^4} \Bigg].
\label{T-tt-psi1-psi2-small epsilon}
\eea 
In Fig. \ref{fig: T-x}, we plot the energy densities as functions of $X$ and $t$.
From the figures, one can see that the energy densities are localized along two null directions which are described by $t=\pm (X-x)$ when $ a \epsilon \ll 1$. As mentioned before, these peaks can be interpreted as the propagation of two energy excitations.

\subsection{Entanglement Entropy as Geodesic Length}
\label{Sec:Geodesic}

In the previous sections \ref{Sec:Holographic-entanglement-entropy}, \ref{Sec:EE-for-finite-interval}, and \ref{Sec:CB-SII}, we calculated the entanglement entropy by exploiting the method of conformal blocks in the large central charge expansion.
However, only when $\eta_{i}$ and $\overline{\eta}_{i}$ are certain values, the analytic form of the conformal blocks is known.
Therefore, we can analytically investigate the time dependence of the entanglement entropy only for the restricted parameter regions. Furthermore, we applied the "light-light-heavy-heavy" approximation in the conformal block method which is valid when the primary operators are heavy. To avoid these shortcomings and as a consistency check for our previous results, we apply the so-called Ryu-Takayanagi formula \cite{2006PhRvL..96r1602R,2006JHEP...08..045R,2013JHEP...08..090L} to calculate the entanglement entropy in this section. This method is valid both for light and heavy primary operators. The celebrated Ryu-Takayanagi formula
states that the entanglement entropy for the subsystem, $A$, is holographically given as
\bea
S_A = \frac{\mathcal{L}(\gamma_A)}{4 G_N},
\label{RT-formula}
\eea 
where $\mathcal{L}(\gamma_A)$ is the length of the geodesic $\gamma_A$ anchored at the boundary of the subsystem $A$, and $G_{N}$ is the Newton constant. 
By exploiting (\ref{RT-formula}), we can analytically investigate the time dependence of the entanglement entropy for the parameter regime beyond that for the calculation in terms of the conformal blocks.
To this end, in the following sections,  we will begin with the three-dimensional Euclidean gravity ($3$d Euclidean gravity) with the negative cosmological constant, a counterpart of the Lorentzian gravity under consideration. Next, we calculate the Euclidean entanglement entropy as the length of a geodesic, perform the analytic continuation to real time, and then investigate the time dependence of the entanglement entropy. 

\subsubsection{Gravity Dual}
\label{Sec: Gravity Dual}

In this section, we will present the gravitational dual of the systems considered in the previous sections.
To this end, we follow the prescription applied in refs. \cite{2015JHEP...02..171A,Banados:1998gg}, and derive the spacetime-dependent background corresponding to the systems under consideration. 
It is well known that the most general solution to vacuum  Einstein's equations for the $3$d Euclidean gravity with negative cosmological constant is given by the Ba$\tilde{\rm n}$ados metric as \cite{Banados:1998gg,Roberts_2012}
\bea
ds^2 = L_{AdS}^2 \Bigg[ \frac{dy^2}{y^2} + \frac{L(z)}{2} dz^ 2 + \frac{\bar{L}(\bar{z})}{2} d\bar{z}^2 + \left( \frac{1}{y^2} + \frac{y^2}{4} L(z) \bar{L}(\bar{z}) \right) dz d\bar{z} \Bigg].  
\label{metric-Banados-complex coordinates}
\eea 
Here, $y$ is the radial coordinate and $(z, \bar{z})$ are the complex coordinates on the asymptotic boundary. Moreover, near the boundary, i.e. $y\approx 0$, the geometry reduces to a Euclidean $AdS_3$ spacetime in the Poincar\'e coordinates \cite{Banados:1998gg}. 
Furthermore, 
the functions $L(z)$ and $\bar{L}(\bar{z})$ are related to the expectation values of the chiral and anti-chiral parts of the energy-momentum tensor under consideration as \cite{Roberts_2012}
\bea
L(z) = - \frac{12}{c} \langle T(z) \rangle_i, \;\;\;\;\;\;\;\;\;\;\;\;\; \bar{L}(\bar{z}) = - \frac{12}{c} \langle \bar{T}(\bar{z}) \rangle_i.
\label{L-T}
\eea 
Thus, the information about the systems under consideration is encoded into (\ref{metric-Banados-complex coordinates}) as the functions, $L(z)$ and $\bar{L}(\bar{z})$ \cite{2015JHEP...02..171A,Roberts_2012,2013arXiv1311.2562U,2016arXiv160407830M,Mao:2025hkp}.
Here, $\langle T(z) \rangle_i$ and $\langle \bar{T}(\bar{z}) \rangle_i$ are given by \eqref{T-holo-antiholo-1-2}.
As a consequence, we obtain 
\bea
&& L(z) = -\frac{12h_{\mathcal{O}}}{c} \left[ \frac{1}{(z- z^{\rm new}_{i, - \epsilon})} - \frac{1}{(z- z^{\rm new}_{i, \epsilon})} \right]^2,
\cr && \cr
&& \bar{L}(\bar{z}) = -\frac{12h_{\mathcal{O}}}{c} \left[ \frac{1}{(\bar{z}- \bar{z}^{\rm new}_{i, - \epsilon})} - \frac{1}{(\bar{z}- \bar{z}^{\rm new}_{i, \epsilon})} \right]^2,
\label{eq:detail-of-EMD}
\eea 
where $z=z_X = iX$. Therefore, from the above expressions one can easily verify that the metric in \eqref{metric-Banados-complex coordinates} is singular at the insertion points $z= z^{\rm new}_{i, \pm \epsilon}$ of the primary operators. This will be important when we calculate the holographic entanglement entropy in the following section.

\subsubsection{Geodesic Length}
\label{Sec: Geodesic Length}

\begin{figure}[h!]
	\begin{center}
		\includegraphics[scale=0.35]{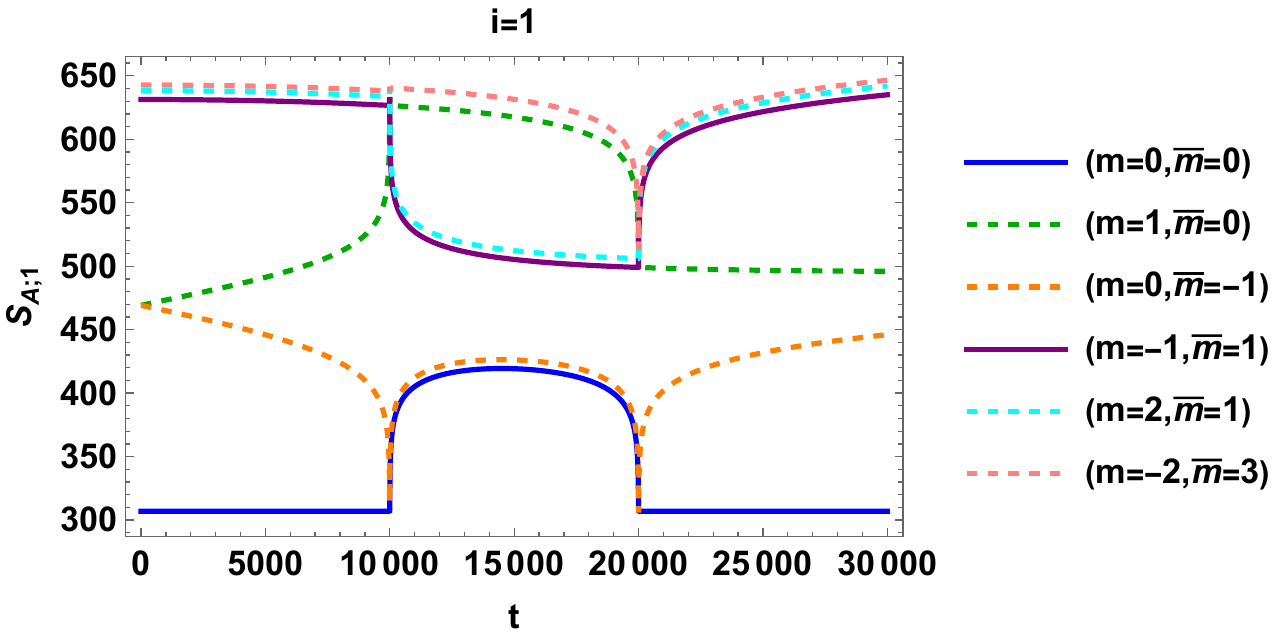}
		\hspace{0.2cm}
		\includegraphics[scale=0.32]{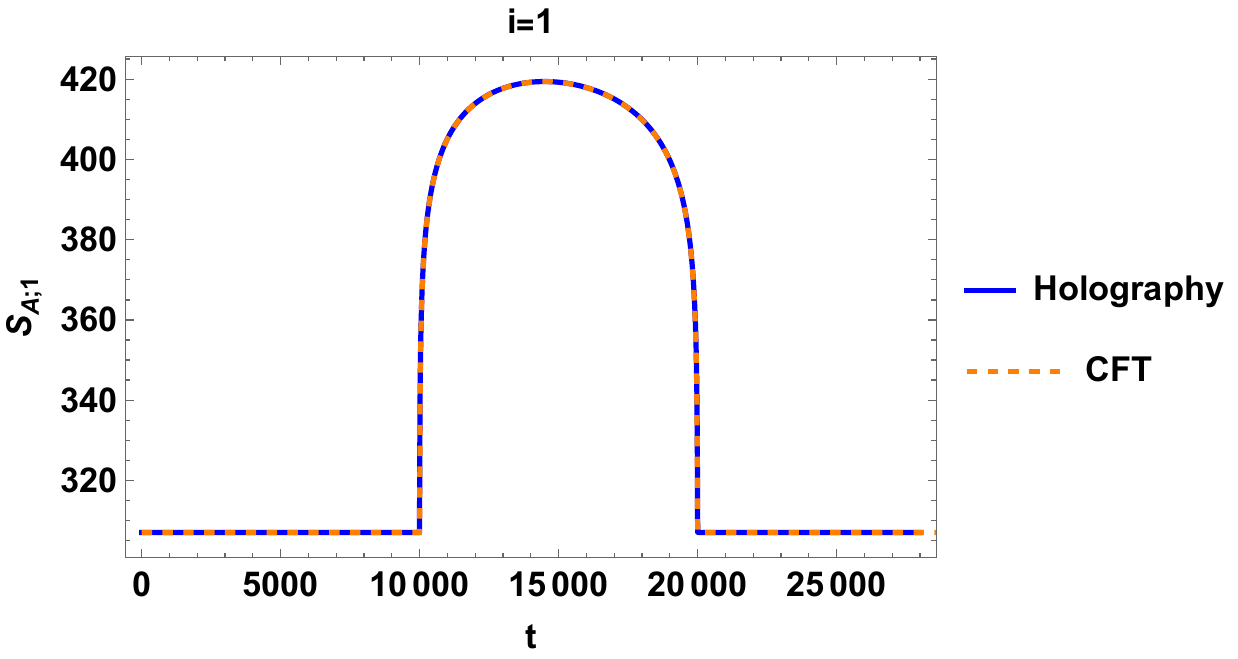}
	\end{center}
	\caption{$S_{A;1}$ as a function of time when the primary operator is inserted outside and on the left hand side of the subsystem. 
		{\it Left)} The contributions of different geodesics with winding numbers $(m, \overline{m})$. 
		{\it Right)} Comparison of the entanglement entropy on the CFT and gravity sides. Here, we set $x=10^{4}$, $x_1= 2 \times 10^4$, $x_2= 3 \times 10^4$, $a= 10^{-5}$, $c= 100$, $\epsilon = 10$ and $h_{\mathcal{O}}= 0.4 h_{0}$ where $h_{0} = \frac{c}{24}$. 
	}
	\label{fig: HEE-psi1-2-outside-LHS}
\end{figure}

To calculate the geodesic length, we will use two facts that the solutions for the three-dimensional Euclidean Einstein Hilbert action with the negative cosmological constant can be locally mapped to a three-dimensional pure Anti-de Sitter geometry ($AdS_3$), and the geodesic length is invariant under the coordinate transformations.
We locally map from the Ba$\tilde{\rm n}$ados geometry to the pure $AdS_3$ as
\bea
ds^2 = \frac{L_{AdS}^2}{u^2} \left( dw d\bar{w} + du^2 \right),
\label{AdS-Poincare-metric}
\eea 
where $w$ is the complex coordinate on the boundary and $u$ is the radial coordinate. 
The map from the Ba$\tilde{\rm n}$ados geometry to the pure $AdS_3$ is given by \cite{Banados:1998gg,Roberts_2012,Shimaji:2018czt} 
\bea
&& w = f(z) - \frac{2 y^2 f'(z)^2 \bar{f}''(\bar{z})}{ 4 f'(z) \bar{f}'(\bar{z}) + y^2 f''
	(z) \bar{f}''(\bar{z})},
\cr && \cr
&& \bar{w} = \bar{f}(\bar{z}) - \frac{2 y^2 \bar{f}'(\bar{z})^2 f''(z)}{ 4 \bar{f}'(\bar{z}) f'(z) + y^2 \bar{f}''
	(\bar{z}) f''(z)},
\cr && \cr
&& u = y \left( \frac{4 (f'(z) \bar{f}'(\bar{z}) )^{\frac{3}{2}} }{4 f'(z) \bar{f}'(\bar{z}) + y^2 f''(z) \bar{f}''(\bar{z})} \right).
\label{Banados map}
\eea
Here, $F^{\overbrace{\prime \cdots \prime}^n}(z)$ and $\bar{F}^{\overbrace{\prime \cdots \prime}^n}(\bar{z})$ denote $\f{d^nF(z)}{dz^n}$ and $\f{d^n\bar{F}(\bar{z})}{d\bar{z}^n}$, where $F(z)$ and $\bar{F}(\bar{z})$ are arbitrary holomorphic and anti-holomorphic functions. 
Thus, $w$ and $\bar{w}$ depend on both $f(z)$ and $\bar{f}(\bar{z})$.
In other words, those functions, $w$ and $\bar{w}$, are neither holomorphic nor anti-holomorphic functions of $z$.
In contrast, near the asymptotic boundary, i.e., $y \approx 0$, the 
maps are determined by $f(z)$, $\bar{f}(\bar{z})$ and their first derivatives \cite{Banados:1998gg,Roberts_2012}
\be
\begin{split}
	w(z)  \approx  f(z), ~~~~~ \bar{w}(\bar{z})  \approx  \bar{f}(\bar{z}),~~~~~ u  \approx  y \sqrt{f'(z) \bar{f}'(\bar{z})}.
\end{split}
\label{Banados map-asymptotic form}
\ee
Thus, the boundary maps, $f(z)$ and $\bar{f}(\bar{z})$, are holomorphic and anti-holomorphic functions, respectively.
These functions map the systems to the vacuum one, and the expectation values of the chiral and anti-chiral energy-momentum densities are determined by
\be \label{eq:Equation-for-fandbf}
\begin{split}
	&\langle T(z) \rangle = \left( \frac{df(z)}{d z} \right)^2 \langle T(f(z)) \rangle + \frac{c}{12} \{f(z), z \} = \frac{c}{12} \{f(z), z \}, \\
	&\langle \bar{T}(\bar{z}) \rangle = \left( \frac{d\bar{f}(\bar{z})}{d \bar{z}} \right)^2 \langle \bar{T}(\bar{f}(\bar{z})) \rangle + \frac{c}{12} \{\bar{f}(\bar{z}), \bar{z} \} = \frac{c}{12} \{\bar{f}(\bar{z}), \bar{z} \}.
\end{split}
\ee
Note that $\langle T(f(z)) \rangle$ and $\langle \bar{T}(\bar{f}(\bar{z})) \rangle$ are the vacuum energy-momentum densities which vanish, and the Schwarzian derivatives are defined as
\be \label{beta}
\begin{split}
	\{f(z), z \}=\f{f'''(z)}{f'(z)}-\f{3}{2}\left(\f{f''(z)}{f'(z)}\right)^2,
	~~~~~ \{\bar{f}(\bar{z}), \bar{z} \}=\f{\bar{f}'''(\bar{z})}{\bar{f}'(\bar{z})}-\f{3}{2}\left(\f{\bar{f}''(\bar{z})}{\bar{f}'(\bar{z})}\right)^2.
\end{split}
\ee
First, we solve (\ref{eq:Equation-for-fandbf}) with respect to $f(z)$ and $\bar{f}(\bar{z})$, and obtain them as
\bea
f(z) = 
\frac{c_2}{2 c c_1 \alpha_{\mathcal{O}}^2 (z^{\rm new}_{i, - \epsilon} -z^{\rm new}_{i, \epsilon})^2} \left( \frac{ 1 \pm c_1 \alpha_{\mathcal{O}}  (z^{\rm new}_{i, - \epsilon} - z^{\rm new}_{i, \epsilon}) \left( \frac{z- z^{\rm new}_{i, - \epsilon}}{z- z^{\rm new}_{i, \epsilon}} \right)^{\pm \alpha_{\mathcal{O}}} }{ -1 \pm c_1 \alpha_{\mathcal{O}} (z^{\rm new}_{i, - \epsilon} - z^{\rm new}_{i, \epsilon}) \left( \frac{z-z^{\rm new}_{i, - \epsilon}}{z - z^{\rm new}_{i, \epsilon}} \right)^{\pm \alpha_{\mathcal{O}}} } \right) + c_3,
\label{boundary-map-f}
\eea
\bea
\bar{f}(\bar{z}) = \frac{\bar{c}_2}{2 c \bar{c}_1 \alpha_{\mathcal{O}}^2 (\bar{z}^{\rm new}_{i, - \epsilon} -\bar{z}^{\rm new}_{i, \epsilon})^2} \left( \frac{ 1 \pm \bar{c}_1 \alpha_{\mathcal{O}}  (\bar{z}^{\rm new}_{i, - \epsilon} - \bar{z}^{\rm new}_{i, \epsilon}) \left( \frac{\bar{z}- \bar{z}^{\rm new}_{i, - \epsilon}}{\bar{z}- \bar{z}^{\rm new}_{i, \epsilon}} \right)^{\pm \alpha_{\mathcal{O}}} }{ -1 \pm \bar{c}_1 \alpha_{\mathcal{O}} (\bar{z}^{\rm new}_{i, - \epsilon} - \bar{z}^{\rm new}_{i, \epsilon}) \left( \frac{\bar{z}-\bar{z}^{\rm new}_{i, - \epsilon}}{\bar{z} - \bar{z}^{\rm new}_{i, \epsilon}} \right)^{\pm \alpha_{\mathcal{O}}} } \right) + \bar{c}_3.
\label{boundary-map-f-bar}
\eea 
Here, $c$ is the central charge and $\alpha_{\mathcal{O}}$ is defined as \eqref{alpha-alpha-bar}. Moreover, $c_{i=1,2,3}$  and $\bar{c}_{i=1,2,3}$ are arbitrary constants which are not fixed. Furthermore, $\bar{c}_i$ is the complex conjugate of $c_i$.
For later convenience, we first choose the coefficients $c_{2,3}$ as 
\bea
&& c_2 = \mp c \alpha_{\mathcal{O}} \left( z^{\rm new}_{i, - \epsilon} - z^{\rm new}_{i, \epsilon} \right), ~~~~~~~~~~~ \bar{c}_2 = \mp c \alpha_{\mathcal{O}} \left( \bar{z}^{\rm new}_{i, - \epsilon} - \bar{z}^{\rm new}_{i, \epsilon} \right),
\cr && \cr 
&& c_3 = \mp \frac{1}{2 c_1 \alpha_{\mathcal{O}} \left( z^{\rm new}_{i, - \epsilon} - z^{\rm new}_{i, \epsilon} \right)}, ~~~~~~~~  \bar{c}_3 = \mp \frac{1}{2 c_1 \alpha_{\mathcal{O}} \left( \bar{z}^{\rm new}_{i, - \epsilon} - \bar{z}^{\rm new}_{i, \epsilon} \right)},
\eea 
and then set $c_1= \bar{c}_1 =0$. 
With these choices, one can simplify the holomorphic and anti-holomporphic boundary maps as
\bea
f(z) = \left( \frac{z- z^{\rm new}_{i, - \epsilon}}{z- z^{\rm new}_{i, \epsilon}}\right)^{a \alpha_{\mathcal{O}}}, \;\;\;\;\;\;\;\;\;\; \bar{f}(\bar{z}) = \left( \frac{\bar{z}- \bar{z}^{\rm new}_{i, - \epsilon}}{\bar{z}- \bar{z}^{\rm new}_{i, \epsilon}}\right)^{\bar{a} \alpha_{\mathcal{O}}},
\label{boundary-map-2}
\eea 
where $(a, \bar{a})= (\pm 1, \pm1)$. Therefore, there are four choices for the signs in $f(z)$ and $\bar{f}(\bar{z})$. To fix the signs, one should note that in the Euclidean signature, the coordinates $z$ and $\bar{z}$ are complex conjugate of each other. Therefore, to have a real-valued radial coordinate $u$ in \eqref{Banados map}, one has to impose the following constraint
\bea
\left( f(z) \right)^* = \bar{f}(\bar{z}).
\label{fstar=fbar}
\eea
For $h_\mathcal{O}  < h_0= \frac{c}{24}$, the geometry is a conical $AdS_3$ \cite{2015JHEP...02..171A}. As mentioned below  \eqref{alpha-alpha-bar}, in this case, $\alpha_\mathcal{O}$ is a real constant. 
Then, if one chooses the signs as 
\bea
(a, \bar{a}) = (+1 , +1),\;\;\;\;\;\;\;\;\; \text{or} \;\;\;\;\;\;\;\;\; (a, \bar{a}) = (-1 , -1),
\label{a-a-bar}
\eea 
the condition in \eqref{fstar=fbar} is satisfied. It should be pointed out that the two choices in \eqref{a-a-bar} are equivalent to each other and they give the same expressions for the entanglement entropy. This is due to the fact that they are related to each other by $\alpha_{\mathcal{O}} \rightarrow - \alpha_{\mathcal{O}}$, and it is straightforward to see that \eqref{eq:Euclidean-EE-CB} is invariant under this transformation. Having said this, we choose the signs in \eqref{boundary-map-2} as
\bea
f(z) = \left( \frac{z- z^{\rm new}_{i, - \epsilon}}{z- z^{\rm new}_{i, \epsilon}}\right)^{\alpha_{\mathcal{O}}}, \;\;\;\;\;\;\;\;\;\; \bar{f}(\bar{z}) = \left( \frac{\bar{z}- \bar{z}^{\rm new}_{i, - \epsilon}}{\bar{z}- \bar{z}^{\rm new}_{i, \epsilon}}\right)^{ \alpha_{\mathcal{O}}}.
\label{boundary-map-conical-AdS}
\eea 
On the other hand, for $h_\mathcal{O}  > h_0$ the dual geometry is a non-rotating BTZ black brane \cite{2015JHEP...02..171A}. From \eqref{alpha-alpha-bar}, one can see $\alpha_\mathcal{O}$ is pure imaginary.
Moreover, the inverse temperature $\beta$ of the black brane is related to the conformal dimension of the primary operator as \cite{2015JHEP...02..171A}
\bea
\beta = \frac{2 \pi L_{AdS}^2}{r_h} = \frac{2 \pi L_{AdS}}{\sqrt{ \frac{24 h_{\mathcal{O}}}{c} -1} },
\label{T-h}
\eea 
where $r_h$ is the horizon radius. Then, by combining \eqref{alpha-alpha-bar} and \eqref{T-h}, one obtains the relation between $\alpha_{\mathcal{O}}$ and the horizon radius as
\bea
\alpha_{\mathcal{O}} = - \frac{i r_h}{L_{AdS}}.
\label{rh-alpha}
\eea 
In this case, one should choose 
\bea
(a, \bar{a}) = (+1 , -1),\;\;\;\;\;\;\;\;\; \text{or} \;\;\;\;\;\;\;\;\; (a, \bar{a}) = (-1 , +1),
\eea 
to satisfy the constraint in \eqref{fstar=fbar}. Since the two choices are again equivalent to each other, we choose the following signs
\footnote{In refs. \cite{Abajian:2023bqv,Mao:2025hkp}, the signs were chosen as $(a, \bar{a}) = (-1 , +1)$.}
\bea
f(z) = \left( \frac{z- z^{\rm new}_{i, - \epsilon}}{z- z^{\rm new}_{i, \epsilon}}\right)^{\alpha_{\mathcal{O}}}, \;\;\;\;\;\;\;\;\;\; \bar{f}(\bar{z}) = \left( \frac{\bar{z}- \bar{z}^{\rm new}_{i, - \epsilon}}{\bar{z}- \bar{z}^{\rm new}_{i, \epsilon}}\right)^{- \alpha_{\mathcal{O}}}.
\label{boundary-map-BTZ}
\eea 
Now, let us consider the length of a geodesic, $\gamma'_1$, in pure $AdS_3$.
Assume that the endpoints of the geodesic are given by
\bea
(u, w) = (u_{\infty,+}, w_{\infty,+}, \bar{w}_{\infty,+}), \;\;\;\;\;\;\;\; (u, w) = (u_{\infty,-}, w_{\infty, -}, \bar{w}_{\infty,-}).
\eea 
Then, these endpoints express the geodesic length as 
\bea
\mathcal{L}(\gamma'_1) = L_{AdS} \log \Bigg[ \frac{\left( w_{\infty,+} - w_{\infty, -} \right) \left( \bar{w}_{\infty,+} - \bar{w}_{\infty, -} \right)}{ u_{\infty,+} u_{\infty, -}} \Bigg].
\label{geodesic lenght-AdS}
\eea 
Next, we pull it back to another geodesic $\gamma_1$ in the Ba$\tilde{\rm n}$ados geometry and find its length. It should be pointed out that this method was also applied in refs. \cite{Roberts_2012,2013arXiv1311.2562U,2015JHEP...02..171A,Shimaji:2018czt}. 
We assume that $(u_{\infty,\pm},w_{\infty,\pm}, \bar{w}_{\infty,\pm})$, the endpoints of $\gamma'_{1}$, are mapped to
$(1 ,z_{x_{i=1,2}}, \bar{z}_{x_{i=1,2}})$
\footnote{In the Ba$\tilde{\rm n}$ados geometry, we choose the same radial cutoff at both endpoints, i.e. 
$y(z_{x_1}) = y(z_{x_2}) = 1$. }
as
\be
(u_{\infty,-},w_{\infty,+}, \bar{w}_{\infty,+}) \rightarrow (1,z_{x_{i=1}}, \bar{z}_{x_{i=1}}),~~~~~(u_{\infty,-},w_{\infty,-}, \bar{w}_{\infty,-}) \rightarrow (1,z_{x_{i=2}}, \bar{z}_{x_{i=2}}).
\ee
As mentioned in Section \ref{Sec:intro}, by dividing the dimensional parameters and variables by the ultraviolet cutoff, we defined all dimensionless ones. 
As the location, along $u$, of the surface, where the geodesic ends corresponds to the ultraviolet cutoff, we set it to be one.
By plugging \eqref{Banados map-asymptotic form} and \eqref{geodesic lenght-AdS} into \eqref{RT-formula}, and exploiting the fact that $c= \frac{3 L_{AdS}}{2 G_N}$ \cite{Brown:1986nw}, 
one has
\bea
S_{A;i} (\gamma_1) &=& \frac{c}{6} \log \Bigg[ \frac{\left( w (z_{x_1}) - w(z_{x_2}) \right) \left( \bar{w}(\bar{z}_{x_1}) - \bar{w}(\bar{z}_{x_2}) \right)}{ u(z_{x_1}, \bar{z}_{x_1}) u(z_{x_2}, \bar{z}_{x_2})} \Bigg]
\cr && \cr
&=& \frac{c}{6} \log \Bigg[ \frac{\left( f(z_{x_1}) - f(z_{x_2}) \right) \left( \bar{f}(\bar{z}_{x_1}) - \bar{f}(\bar{z}_{x_2}) \right)}{
	\left(f'(z_{x_1}) \bar{f}'(\bar{z}_{x_1}) f'(z_{x_2}) \bar{f}'(\bar{z}_{x_2}) \right)^{\frac{1}{2}}} \Bigg].
\label{L-gamma-1}
\eea
Then, when we exploit the map in \eqref{boundary-map-conical-AdS} for the conical AdS case and exploit that in \eqref{boundary-map-BTZ} for the BTZ case, \eqref{L-gamma-1} results in (\ref{eq:Euclidean-EE-CB}).
Therefore, the entanglement entropy obtained as the geodesic length with certain values of the parameters is exactly equal to the one obtained by the method of conformal blocks. 
Moreover, as mentioned above, the geometry in \eqref{metric-Banados-complex coordinates} is singular at two points, i.e. $z= z^{\rm new}_{i, \pm \epsilon}$ and $\bar{z}= \bar{z}^{\rm new}_{i, \pm \epsilon}$. Thus, a geodesics can wrap around each singularity. 
On the other hand, in $(\tilde{z}, \bar{\tilde{z}})$, the coordinates introduced in \eqref{conformal map-EE}, we have only a singularity at $\tilde{z}= \bar{\tilde{z}} =0$.
Therefore, in the $(\tilde{z}, \bar{\tilde{z}})$ coordinates, the geodesics wrap around only one singularity, and the calculations are easier in these coordinates.  
Thus, to calculate the geodesic length easily, we exploit the conformal transformation \eqref{conformal map-EE}, and then move on to the $(\tilde{z}, \bar{\tilde{z}})$ coordinates. 
Then, we write the Ba$\tilde{\rm n}$ados geometry by applying the energy-momentum tensor in these coordinates. Next, we find the geodesic length in this geometry. In the $(\tilde{z}, \bar{\tilde{z}})$ coordinates, the energy-momentum tensor is simply given by
\bea
\langle T (\tilde{z}) \rangle_i &=& \left( \frac{d\tilde{z}}{dz} \right)^{-2} \left[ \langle T(z) \rangle_i - \frac{c}{12} \{ \tilde{z} ; z \} \right]
\cr && \cr
&=& \frac{h_{\mathcal{O}}(z- z^{\rm new}_{i, \epsilon})^2 (z^{\rm new}_{i, -\epsilon} - z_{x_2})^2 }{(z^{\rm new}_{i, -\epsilon} - z)^2 (z_{x_2} - z^{\rm new}_{i, -\epsilon})^2}
= \frac{h_{\mathcal{O}}}{\tilde{z}^2}.
\label{T-znew}
\eea 
Here, we used \eqref{T-holo-antiholo-1-2} and the fact that for the conformal transformation $z \rightarrow \tilde{z}(z)$, the Schwarzian derivative is zero. Similarly, one obtains
\bea
\langle \bar{T} (\bar{\tilde{z}}) \rangle_i = \frac{h_{\mathcal{O}}}{\bar{\tilde{z}}^2}.
\label{Tbar-znew}
\eea
Next, by plugging \eqref{T-znew} and \eqref{Tbar-znew} into \eqref{metric-Banados-complex coordinates}, one has \cite{2015JHEP...02..171A}
\bea
d \tilde{s}^2 = L_{AdS}^2 \Bigg[ \frac{d \tilde{y}^2}{\tilde{y}^2} - \frac{6 h_{\mathcal{O}}}{\tilde{z}^2} d\tilde{z}^ 2 - \frac{6 h_{\mathcal{O}}}{\bar{\tilde{z}}^2} d\bar{\tilde{z}}^2 + \left( \frac{1}{\tilde{y}^2} + \frac{36 h^2_{\mathcal{O}} \ \tilde{y}^2 }{c^2 \tilde{z}^2 \bar{\tilde{z}}^2 } \right) d \tilde{z} d\bar{\tilde{z}} \Bigg].
\label{metric-Banados-complex coordinates-z-tilde}
\eea 
As mentioned in ref. \cite{2015JHEP...02..171A}, this geometry has a singularity at $\tilde{z} = \bar{\tilde{z}} =0$. Note that on the boundary, one of the primary operators is located at $\tilde{z} =0$ and the other one is located at $\tilde{z} = \infty$. 
Now, by applying the aforementioned method, one can map this geometry to the $AdS_3$ Poincar\'e spacetime. The boundary maps are as
\bea
&& \tilde{f}(\tilde{z}) = \xi \; \tilde{z}^{\alpha_{\mathcal{O}}}, \;\;\;\;\;\;\;\;\;\;\;\; \bar{\tilde{f}}(\bar{\tilde{z}}) =  \bar{\xi} \; \bar{\tilde{z}}^{ \alpha_{\mathcal{O}}}, \;\;\;\;\;\;\; \text{for $h_{\mathcal{O}} < \frac{c}{24}$},
\cr && \cr
&& \tilde{f}(\tilde{z}) = \xi \; \tilde{z}^{\alpha_{\mathcal{O}}}, \;\;\;\;\;\;\;\;\;\;\;\; \bar{\tilde{f}}(\bar{\tilde{z}}) =  \bar{\xi} \; \bar{\tilde{z}}^{- \alpha_{\mathcal{O}}}, \;\;\;\;\; \text{for $h_{\mathcal{O}} > \frac{c}{24}$},
\label{boundary-map-BTZ-z-tilde}
\eea 
where $\xi$ and $\bar{\xi}$ are two arbitrary constants. For later convenience, we set 
\bea 
\xi = \left( \frac{z_{x_2} - z^{\rm new}_{i, - \epsilon} }{z_{x_2} - z^{\rm new}_{i, \epsilon}} \right)^{\alpha_{\mathcal{O}}},
\;\;\;\;\;\;\;\;\;\;\;\;
\bar{\xi} = \left( \frac{\bar{z}_{x_2} - \bar{z}^{\rm new}_{i, - \epsilon} }{\bar{z}_{x_2} - \bar{z}^{\rm new}_{i, \epsilon}} \right)^{\alpha_{\mathcal{O}}}.
\eea 
Then, one has
\bea
\tilde{f} (\tilde{z}) = f(z), \;\;\;\;\;\;\;\;\;\;\;\; \bar{\tilde{f}}(\bar{\tilde{z}}) = \bar{f}(\bar{z}),
\label{tilde-f-f}
\eea 
where $f(z)$ and $\bar{f}(\bar{z})$ are given by \eqref{boundary-map-BTZ}.
The geodesic length is simply given by (see also \cite{2015JHEP...02..171A})
\footnote{It should be emphasized that the length of the geodesic in this geometry was calculated before in ref. \cite{2015JHEP...02..171A}. It was shown that it is related to the semiclassical conformal block, i.e.
$\lim_{n \rightarrow 1} \frac{1}{(1-n)} \log G_n (\eta_i, \bar{\eta}_i)$. 
Moreover, the authors considered the same radial cutoffs at the two endpoints. 
Here, we use two different radial cutoffs at the endpoints.}
\bea 
\mathcal{L}(\tilde{\gamma}) = \log \Bigg[ \frac{\eta_i^{\frac{(1- \alpha_{\mathcal{O}})}{2}} \bar{\eta}_i^{\frac{(1- \alpha_{\mathcal{O}})}{2}} \left( 1 - \eta_i^{\alpha_{\mathcal{O}} } \right) \left( 1 - \bar{\eta}_i^{\alpha_{\mathcal{O}} } \right) }{ \alpha^2_{\mathcal{O}} \; \tilde{y}(\tilde{z}_{x_1}) \tilde{y}(\tilde{z}_{x_2})}  \Bigg].
\label{L-gamma-tilde-1}
\eea 
Note that $\tilde{z}_{x_1}= \tilde{z}(z_{x_1}) = \eta_i $ and $\tilde{z}_{x_2}= \tilde{z}(z_{x_2}) = 1 $ are the two endpoints of the subsystem. Moreover, we choose two different radial cutoffs $\tilde{y}(\tilde{z}_{x_{1,2} })$ at the two endpoints. To find these cutoffs,
we apply \eqref{Banados map-asymptotic form} and obtain the connection between the radial coordinates $y$ and $\tilde{y}$ of the two Bana$\tilde{\text{n}}$dos geometries as 
\footnote{We have two Bana$\tilde{\text{n}}$dos geometries with coordinates $(y, z, \bar{z})$ and $(\tilde{y}, \tilde{z}, \bar{\tilde{z}})$. We map both of them to the same $AdS_3$ Poincar\'e spacetime with coordinates $(u, w, \bar{w})$. Then, we apply \eqref{Banados map} and find \eqref{tilde-y-y} which is valid near those boundaries.}
\bea
\tilde{y} \approx y \left( \frac{f'(z) \bar{f}'(\bar{z})}{\tilde{f}'(\tilde{z}) \bar{\tilde{f}}'(\bar{\tilde{z}}) } \right)^{\frac{1}{2}} 
= \left( \frac{d\tilde{z}}{dz} \frac{d \bar{\tilde{z}}}{d \bar{z}} \right)^{\frac{1}{2}}.
\label{tilde-y-y}
\eea 
Next, by applying the above expression, one arrives at
\bea
\tilde{y}(\tilde{z}_{x_1}) \tilde{y}(\tilde{z}_{x_2}) \approx y(z_{x_1}) y(z_{x_2}) \bigg{|} \frac{1- \eta_i }{z_{x_1} - z_{x_2} } \bigg{|}^2 
\approx 
\bigg{|} \frac{1- \eta_i }{z_{x_1} - z_{x_2} } \bigg{|}^2,
\label{cutoffs-tilde-y-and-y-coordinates}
\eea 
where in the last equality we applied 
$y(z_{x_1})  = y(z_{x_2}) = 1$.
Then, by plugging \eqref{cutoffs-tilde-y-and-y-coordinates} into \eqref{L-gamma-tilde-1}, one obtains 
\bea
S_{A;i} (\tilde{\gamma})  =  
\frac{c}{3} \log |z_{x_1} - z_{x_2}| 
+ \frac{c}{6}  \log \Bigg[ \frac{\eta_i^{\frac{(1- \alpha_{\mathcal{O}})}{2}} \bar{\eta}_i^{\frac{(1- \alpha_{\mathcal{O}})}{2}} \left( 1 - \eta_i^{\alpha_{\mathcal{O}} } \right) \left( 1 - \bar{\eta}_i^{\alpha_{\mathcal{O}} } \right) }{ \alpha^2_{\mathcal{O}}  |1 - \eta_i|^2 } \Bigg] \Bigg],
\label{S-A-gamma-tilde-1}
\eea
which is the same as \eqref{EE-2} that was obtained by the method of conformal blocks. 
It should be pointed out that for both the conical AdS and BTZ geometries, one can consider some other geodesics $\tilde{\gamma}_{m , \overline{m}}$ which anchored at the same endpoints $\tilde{z}_{x_1}$ and $\tilde{z}_{x_2}$. However, they have different numbers of winding around the singularity at $\tilde{z} = 0$. As mentioned in ref. \cite{2015JHEP...02..171A}, if one changes 
\bea
\tilde{z}(z_{x_1}) \rightarrow e^{2 \pi i m} \tilde{z}(z_{x_1}),
\label{zt-to-E-zt}
\eea 
both are the same points. On the other hand, after mapping the Ba$\tilde{\rm n}$ados geometry in \eqref{metric-Banados-complex coordinates-z-tilde} to the Poincar\'e $AdS_3$ spacetime, they are different points corresponding to $w_{- \infty}$ and $e^{2 \pi i m \alpha_{\mathcal{O}}} w_{- \infty}$, respectively. Similarly, one can apply the following transformation 
\bea
\bar{\tilde{z}}(\bar{z}_{x_1}) \rightarrow e^{2 \pi i \overline{m}} \bar{\tilde{z}}(\bar{z}_{x_1}),
\label{zbt-to-E-zbt}
\eea 
for the anti-holomorphic part. 
Having said this and using the fact that $(\eta_i, \bar{\eta}_i )= (\tilde{z} (z_{x_1}), \bar{\tilde{z}} (\bar{z}_{x_1}) )$ and \eqref{S-A-gamma-tilde-1}, one can write the entanglement entropy for the geodesics $\tilde{\gamma}_{m,\overline{m}}$ as 
\bea
S_{A;i} (\tilde{\gamma}_{m , \overline{m}}) &=& \frac{c}{3} \log \left( | z_{x_1} - z_{x_2} | \right) 
\cr && \cr
&& +  \frac{c}{6} \log \Bigg[ \frac{E^{\frac{(1- \alpha_{\mathcal{O}})}{2}} \eta_i^{\frac{(1- \alpha_{\mathcal{O}})}{2}} \bar{E}^{\frac{(1- \alpha_{\mathcal{O}})}{2}} \bar{\eta}_i^{\frac{(1 - \alpha_{\mathcal{O}})}{2}} \left( 1 - E^{\alpha_{\mathcal{O}} } \eta_i^{\alpha_{\mathcal{O}} } \right) \left( 1 - \bar{E}^{ \alpha_{\mathcal{O}} } \bar{\eta}_i^{ \alpha_{\mathcal{O}} } \right) }{ \alpha^2_{\mathcal{O}} (1- \eta_i) (1- \bar{\eta}_i) } \Bigg]. \;\;\;\;\;\;\;\;\;\;\;
\label{S-A-gamma-tilde-2}
\eea
Here, we defined $E= e^{2 \pi i m}$ and $\bar{E} = e^{2 \pi i \overline{m}}$. 
For later convenience, we rename $\tilde{\gamma}_1$ as $\tilde{\gamma}_{0,0}$.
It should be emphasized that all of the geodesics $\tilde{\gamma}_{m,\overline{m}}$ anchored at the endpoints of the subsystem which are located at $\tilde{z}_{x_1}$ and $\tilde{z}_{x_2}$. However, they have different winding numbers $(m, \overline{m})$ around the singularity \cite{2015JHEP...02..171A}.
Furthermore, the transformations in \eqref{zt-to-E-zt} and \eqref{zbt-to-E-zbt} can be interpreted as taking different branches for the cross ratios \cite{2015JHEP...02..171A}. In other words, choosing geodesics with different winding numbers corresponds to choosing different branches for the cross ratios on the CFT side.
Since the holographic Euclidean entanglement entropy is given as the minimal geodesic length, we obtain it as \cite{2015JHEP...02..171A}
\be
S_{A;i,E}=\f{1}{4G_N}\text{Min}\left[\left\{ \mathcal{L}(\gamma_{m,\overline{m}})\right\}\right],
\ee
where $\left\{ \mathcal{L}(\gamma_{m,\overline{m}})\right\}$ denotes the set of the geodesic lengths associated with $\gamma_{m,\overline{m}}$.
After performing the analytic continuation as $\tau_E=it$, we investigate the time dependence of the entanglement entropy as
\be
S_{A;i}=\f{1}{4G_N}\text{Min}\left[\left\{ \mathcal{L}(\gamma_{m,\overline{m}})\right\}\right].
\label{HEE-gama-m-m-bar}
\ee
Moreover, in the left panel of Fig. \ref{fig: HEE-psi1-2-outside-LHS}, we show the contributions of the geodesics with different winding numbers $(m, \overline{m})$ to the entanglement entropy for the case where the primary operator is inserted outside and on the left hand side of the subsystem $A$.
From this panel, we can observe that the geodesic with $m= \overline{m}= 0$ has the minimum length. 
Note that by increasing the winding numbers, a geodesic can wind around the singularity several times and it might have a larger length.
In the right panel of Fig. \ref{fig: HEE-psi1-2-outside-LHS}, we show the time dependence of the entanglement entropy by calculating the minimum value of the geodesic lengths. 
We observe that the results from the CFT and gravity sides are consistent with each other.

\begin{figure}[h!]
	\begin{center}
		\includegraphics[scale=0.35]{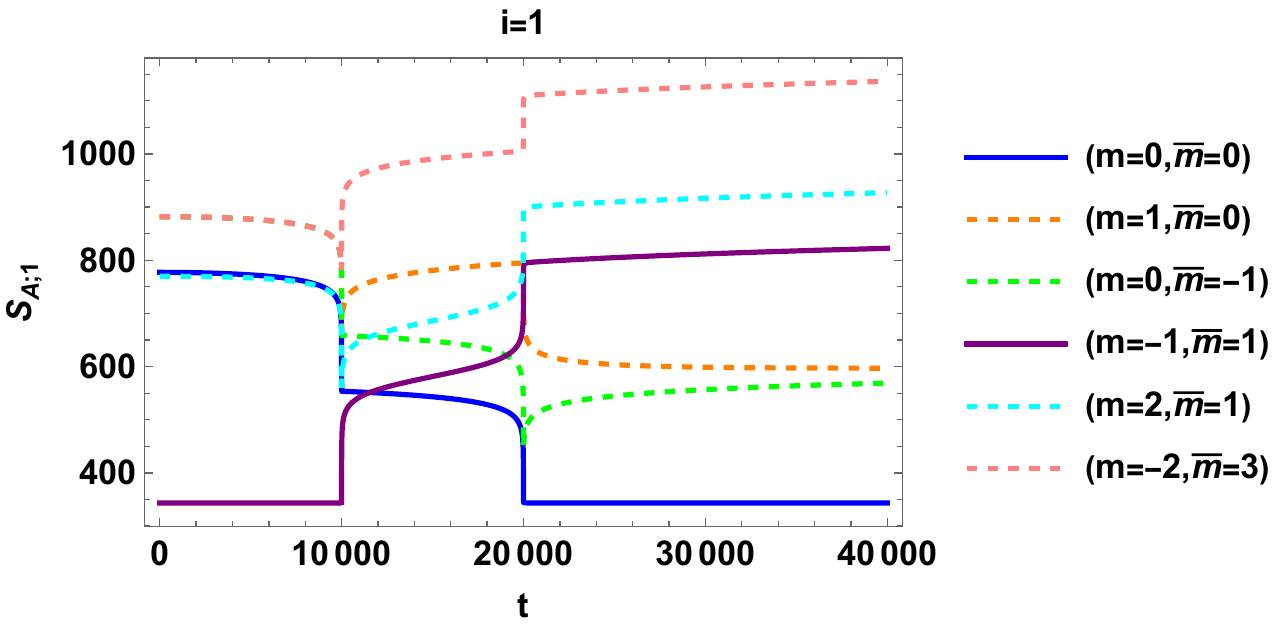}
		\hspace{0.2cm}
		\includegraphics[scale=0.32]{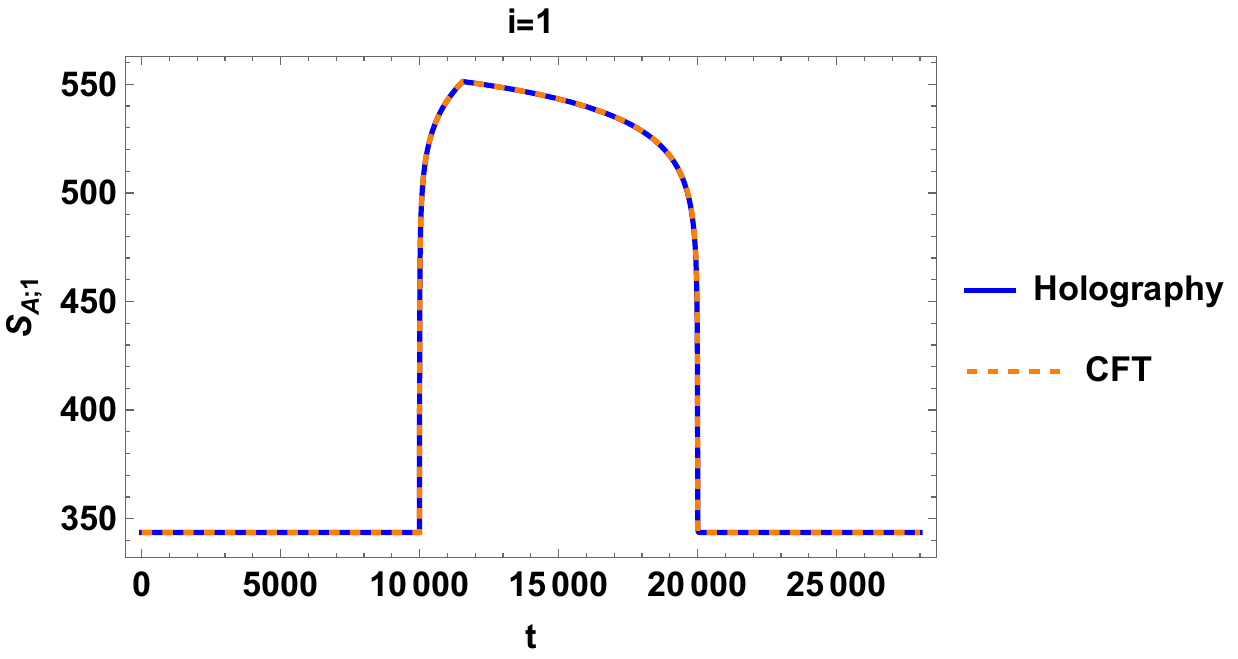}
	\end{center}
	\caption{$S_{A;1}$ as a function of time when the primary operator is inserted inside the subsystem. 
		{\it Left)} The contributions of different geodesics with winding numbers $(m, \overline{m})$. 
		{\it Right)} Comparison of the entanglement entropy on the CFT and gravity sides. 
	Here, we set $x=2 \times 10^4$, $x_1= 10^4$, $x_2= 4 \times 10^4$, $a= 10^{-5}$, $c= 100$, $\epsilon = 10$ and $h_{\mathcal{O}}= 5 h_{0}$.
}
\label{fig: HEE-psi1-2-inside}
\end{figure}

Furthermore, in Fig. \ref{fig: HEE-psi1-2-inside}, we show the contributions of different geodesics to the entanglement entropy for the case where the primary operator is inserted inside the subsystem $A$. We can observe that at early times, the geodesic with $(m=-1 , \overline{m}=1)$ has the minimum length. On the other hand, at late times, the geodesic with $(m= 0, \overline{m}= 0)$ has the minimum length. In this case, the entanglement entropy is given by the combination of these two geodesics.
Since the behavior of $S_{A;i}$ for $i=2$ is similar to $i=1$,  we only report the results for $i=1$ in Figs. \ref{fig: HEE-psi1-2-outside-LHS} and \ref{fig: HEE-psi1-2-inside}.

\subsection{Holographic Mutual Information}
\label{Sec: Holographic Mutual Information}

\begin{figure}
\begin{center}
	\includegraphics[scale=0.62]{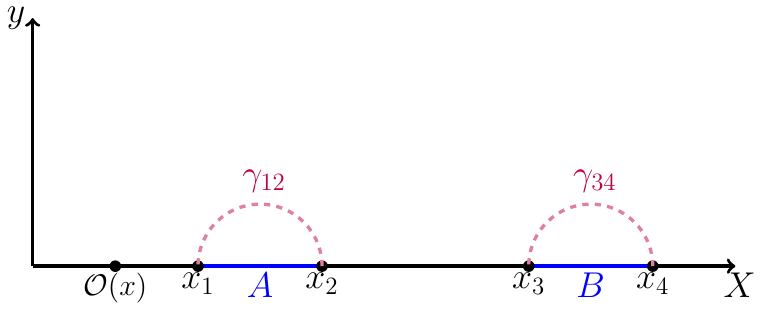}
	\hspace{0.1cm}
	\includegraphics[scale=0.62]{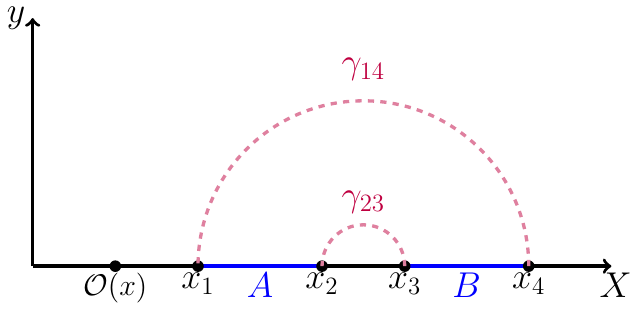}
\end{center}
\caption{The disconnected $\gamma_{12} \cup \gamma_{34}$ and connected $\gamma_{14} \cup \gamma_{23}$ Ryu-Takayanagi surfaces for the subsystem $A \cup B$. Here, $A \in [x_1, x_2]$ and $B \in [x_3, x_4]$ and they are separated by the distance $d$. The primary operator $\mathcal{O}$ is inserted at $x$ on the left hand side of the subsystem $A$.
}
\label{fig: RT-con-dis}
\end{figure}

\begin{figure}
\begin{center}
\includegraphics[scale=0.6]{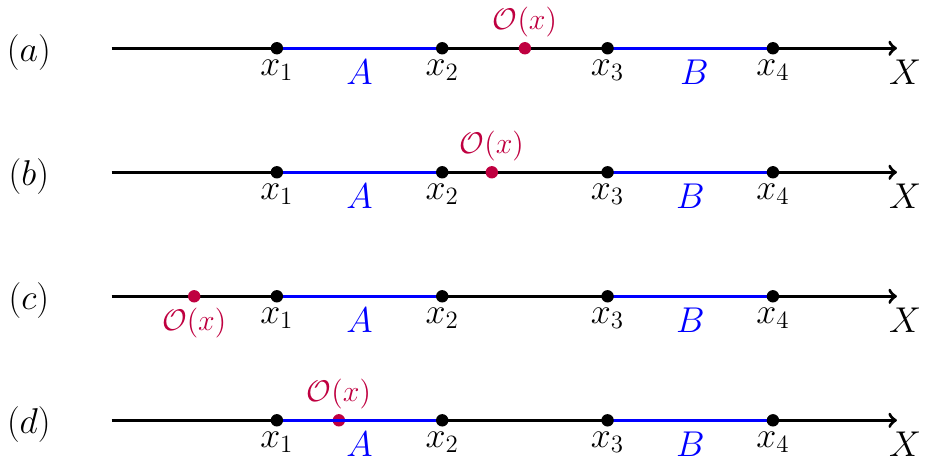}
\end{center}
\caption{Different configurations for the calculation of the holographic mutual information: 
({\it a}) symmetric configuration where the operator is inserted between the subsystems $A$ and $B$ and at equal distances from them.
({\it b}) asymmetric configuration where the operator is inserted between the subsystems $A$ and $B$ and closer to $A$.
({\it c}) the operator is located on the left hand side of the subsystems $A$ and $B$.
({\it d}) the operator is inside the subsystem $A$.
}
\label{fig: HMI-configurations}
\end{figure}

In the previous sections, by exploiting the entanglement entropy, we investigated how the time ordering of the Euclidean and Lorentzian time evolutions influences the bipartite entanglement.
Now, let us move on to this time-ordering effect on the non-local correlations in this section.
Here, we will exploit the mutual information defined as (\ref{eq:def-of-mutual-information}), and investigate how the time ordering of the Euclidean and Lorentzian time evolutions influences the non-local correlations.
To make the mutual information free from the divergence, we consider the union of two disjoint intervals, $A\cup B$ as the subsystem.
We calculated the entanglement entropies for the single intervals, $S_{A;i}$ and $S_{B;i}$, in the previous sections.
Therefore, from here, we closely look at the entanglement entropy for the union of the two intervals, $S_{A\cup B}$.
We will calculate $S_{A\cup B}$ as the geodesic length.
Since the two disjoint intervals have four endpoints, we have two configurations \cite{2010PhRvD..82l6010H} that can dominantly contribute to $S_{A\cup B}$ as in Fig. \ref {fig: RT-con-dis}.
These configurations are called "connected" and "disconnected".
In the connected configuration, the geodesic length is determined by that of two geodesics connecting an endpoint of $A$ with that of $B$.
Here, we assume that these two geodesics do not intersect with each other.
In the disconnected configuration, the geodesic length is determined by the geodesics anchored at $A$ and $B$.
Let $S_{\text{dis}}$ and $S_{\text{con}}$ denote the geodesic length in the disconnected and connected configurations, respectively.
The entanglement entropy for $A\cup B$ is given by the minimal geodesic of $S_{\text{dis}}$ and $S_{\text{con}}$ as \cite{2010PhRvD..82l6010H} 
\bea
S_{A \cup B} = \text{Min} \left[ S_{\rm con} , S_{\rm dis} \right].
\eea 
We choose $A \in [x_1, x_2]$ and $B \in [x_3, x_4]$. Moreover, let $l_A$, $l_B$ and $d$ denote the sizes of $A$, $B$ and the distance between $A$ and $B$, respectively.
Furthermore, let $S_{ij}$ denotes the length of the geodesic anchored at the spatial interval $[x_i, x_j]$.
Then, $S_{\rm dis}$ and $S_{\rm con}$ are given by
\bea
&& S_{\rm dis} = S_A + S_B = S_{12} + S_{34},
\cr && \cr 
&& S_{\rm con} = S_{14} + S_{23}.
\label{S-con-dis}
\eea
Consequently, $I_{A,B}$ results in 
\be
I_{A, B} = S_{A} + S_{B} - \text{Min} \{ S_{\rm con} , S_{\rm dis} \}= S_{\rm dis} - \text{Min} \{ S_{\rm con} , S_{\rm dis} \}.
\label{MI-2}
\ee
For example, for the vacuum state and for subsystems with equal lengths $l_A = l_B = l$, one has 
\bea
I_{A,B}^{\rm Vac} =  
\begin{cases}
I_0,~~~~ &~\text{for}~ d < d_{\rm crit}^{\rm Vac},  \\
0,~~~~ &~\text{for}~ d > d_{\rm crit}^{\rm Vac}, 
\end{cases}
\label{I-Vac}
\eea 
where 
\bea
I_0 = \frac{c}{3} \log \left( \frac{l^2}{(2l+d) d} \right).
\label{I0}
\eea 
Moreover, the critical distance $d_{\rm crit}^{\rm Vac}$
is given by
\bea
d_{\rm crit}^{\rm Vac} = l (\sqrt{2} -1).
\label{d-crit-vac}
\eea 
Note that this is a distance at which $S_{\rm con} = S_{\rm dis}$.
Now, we closely look at the time dependence of the mutual information in the systems under consideration.
We consider four different configurations shown in Fig. \ref{fig: HMI-configurations}.
For simplicity, we choose the union of two equal length intervals
as $A\cup B$.  
In this section, we will report on the time dependence of the mutual information for the symmetric configuration in Fig.  \ref{fig: HMI-configurations}(a), while that of the mutual information for the rest of the configurations will be reported in Appendix \ref{sec:R-TE-MI}.

\subsubsection{Symmetric Configuration}
\label{Sec: Operator Between the Regions A and B: Symmetric Configuration}

\begin{figure}
\begin{center}
\includegraphics[scale=0.41]{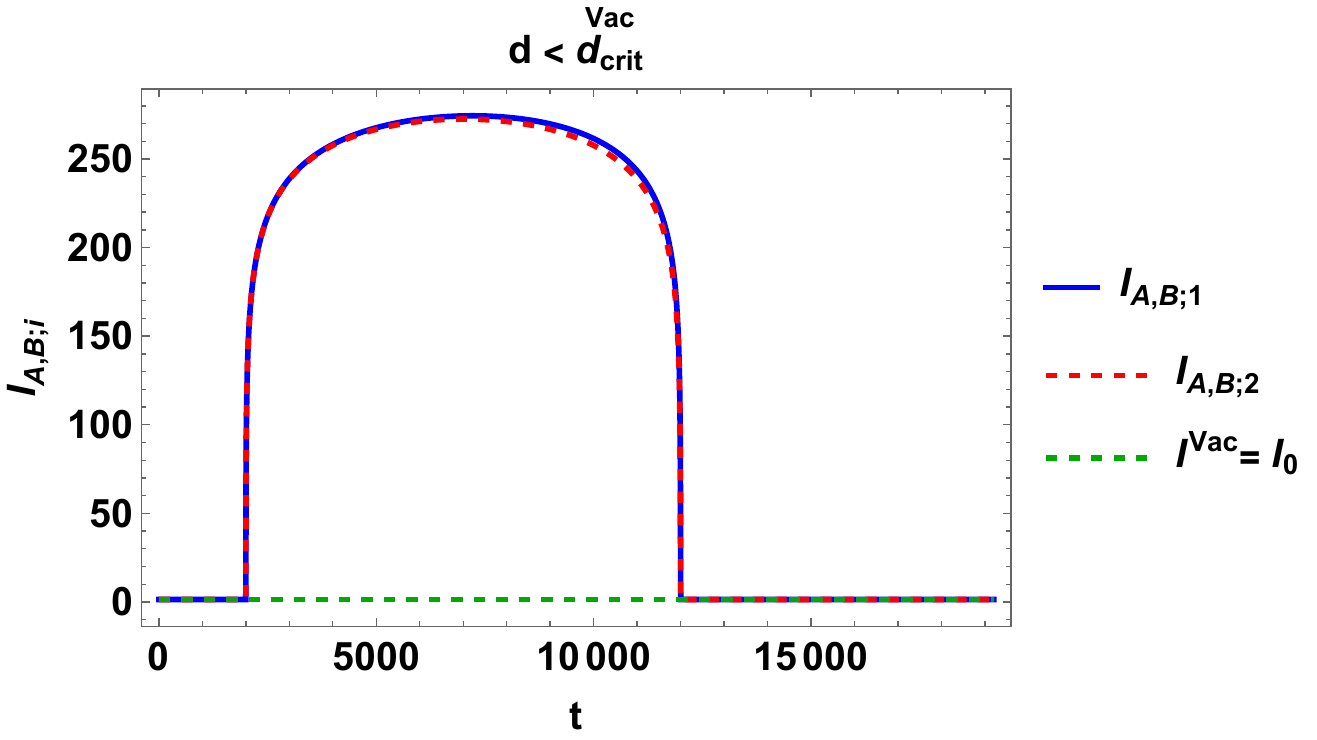}
\end{center}
\caption{  
$I_{A,B;i}$ as a function of time for the symmetric configuration where the primary operator is inserted 
between the subsystems $A$ and $B$ and at equal distances from them.
Here, $d < d_{\rm crit}^{\rm Vac}$. Moreover, we set $x_1= 10^4$, $x_2= 2 \times 10^4$, $x= 2.2 \times 10^4$, $x_3 = 2.4 \times 10^4$, $x_4= 3.4 \times 10^4$, $a= 10^{-4}$, $\epsilon = 10$, $c= 100$ and $h_{\mathcal{O}} =  3 h_{0}$.
}
\label{fig: HMI-symmetric}
\end{figure}

Here, we will present the time dependence of the holographic mutual information for the symmetric configuration where the operator is inserted into the middle of $A$ and $B$, i.e., $0 < x_1 < x_2 < x < x_3 < x_4$ (see Fig. \ref{fig: HMI-configurations}(a)).
Then, one has
\bea
x - x_2 = x_3 - x = \frac{d}{2}, \;\;\;\;\;\;\; 
x - x_1 = x_4 - x = \frac{d}{2} + l.
\eea 
Note that the operator is located on the right hand side of $[x_1 , x_2]$, and on the left hand side of $[x_3,x_4]$. Moreover, it is inside $[x_1, x_4]$ and $[x_2, x_3]$. 
In this case, by using \eqref{SA-i-finite-interval-outside-ER-LHS}, \eqref{SA-i-finite-interval-outside-ER-RHS}, and \eqref{SA-1-finite-interval-inside-ER}, we obtain the time dependence of the disconnected pieces as
\be
\begin{split}
S_{\rm dis;1} = \frac{c}{3} \log \left( l^2 \right) + 
\begin{cases}
0, &~~ 0 < t < \frac{d}{2},  \\
\frac{c}{3} \log \Big[ \frac{\kappa_{\mathcal{O}} (\frac{d}{2} + l - t) (t- \frac{d}{2})}{a \epsilon l \sqrt{x^2- t^2}}
\Big], &~~  \frac{d}{2} < t < \frac{d}{2} + l, \;\;\;\; \\
0, & ~~ t >  \frac{d}{2} + l,
\end{cases}\\
S_{\rm dis;2} =  \frac{c}{3} \log \left( l^2 \right) + 
\begin{cases}
0, &~~ 0 < t < \frac{d}{2},  \\
\frac{c}{3} \log \Big[ \frac{\kappa_{\mathcal{O}} (\frac{d}{2} + l - t) (t- \frac{d}{2})}{a \epsilon l x}
\Big], &~~  \frac{d}{2} < t < \frac{d}{2} + l, \;\;\;\; \\
0, &~~ t >  \frac{d}{2} + l,
\end{cases}
\label{S-disconnected-outside-ER-between-A-B-symmteric}
\end{split}
\ee
while we obtain the time dependence of the connected pieces as
\be
\begin{split}
&S_{\rm con;1} = S_{\rm con;2} = \frac{c}{3} \log \left( (2l + d) d \right).\\
\end{split}
\label{S-connected-outside-ER-between-A-B-symmteric}
\ee
Thus, the connected pieces are independent of the time ordering of the Euclidean and Lorentzian time evolutions and time. Then, by comparison of \eqref{S-disconnected-outside-ER-between-A-B-symmteric} and \eqref{S-connected-outside-ER-between-A-B-symmteric} in the time intervals $t \in (0, \frac{d}{2} )$ and $t \in (\frac{d}{2} + l, \infty)$, regardless of the time ordering of the Euclidean and Lorentzian time evolutions, one can easily verify that for $d > d_{\rm crit}^{\rm vac}$, one has $S_{{\rm dis};i} < S_{{\rm con};i}$. Therefore, the holographic mutual information is zero. 
On the other hand, by comparing $S_{{\rm dis};i}$ and $S_{{\rm con};i}$ for $t \in (\frac{d}{2}, \frac{d}{2} +l)$, one can verify that there is a phase transition at $d_{{\rm crit};i}$ between $S_{{\rm dis};i}$ and $S_{{\rm con};i}$, when the following equations are satisfied
\be
\begin{split}
& \frac{\kappa_{\mathcal{O}}  \; l \; (\frac{d}{2} + l - t) (t- \frac{d}{2})}{a \epsilon \sqrt{x^2 - t^2} (2l +d) d} = 1, ~~~~~ \text{for}~~~i=1, \\
&\frac{\kappa_{\mathcal{O}} \; l \; (\frac{d}{2} + l - t) (t- \frac{d}{2})}{a \epsilon x (2l +d) d} = 1, ~~~~~\text{for}~~~ i=2.
\end{split}
\ee
Therefore, in the time interval, $t \in (\frac{d}{2}, \frac{d}{2} +l)$, when the distance between the two intervals is larger than $d_{{\rm crit};i}$, the holographic mutual information is zero.
Having said this, one can show that for $d < d_{\rm crit}^{\rm Vac}$, one has
\bea
&I_{A,B;1} =  
\begin{cases}
I_0, &~~ 0 < t < \frac{d}{2},  \\
\frac{c}{3} \log \Big[ \frac{\kappa_{\mathcal{O}} \; l \; (\frac{d}{2} + l - t) (t- \frac{d}{2})}{a \epsilon \sqrt{x^2- t^2} (2l+d) d}
\Big], &~~  \frac{d}{2} < t < \frac{d}{2} + l, \;\;\;\; \\
I_0, &~~ t >  \frac{d}{2} + l,
\end{cases} \nonumber \\
& I_{A,B;2} = 
\begin{cases}
I_0, &~~ 0 < t < \frac{d}{2},  \\
\frac{c}{3} \log \Big[ \frac{\kappa_{\mathcal{O}} \; l \; (\frac{d}{2} + l - t) (t- \frac{d}{2})}{a \epsilon x (2l+d) d}
\Big], &~~  \frac{d}{2} < t < \frac{d}{2} + l, \;\;\;\; \\
I_0, &~~ t >  \frac{d}{2} + l,
\end{cases}
\label{I-outside-ER-between-A-B-symmetric-d-smaller-dcrit-vac}
\eea 
where $I_0$ is defined as \eqref{I0}.
On the other hand, for $d_{\rm crit}^{\rm Vac} < d < d_{{\rm crit};i}$, one has
\bea
&I_{A,B;1} = 
\begin{cases}
0, &~~ 0 < t < \frac{d}{2},  \\
\frac{c}{3} \log \Big[ \frac{ \kappa_{\mathcal{O}} \; l \; (\frac{d}{2} + l - t) (t- \frac{d}{2})}{a \epsilon \sqrt{x^2- t^2} (2l+d) d}
\Big], &~~  \frac{d}{2} < t < \frac{d}{2} + l, \;\;\;\; \nonumber \\
0, &~~ t >  \frac{d}{2} + l,
\end{cases}\\
& I_{A,B;2} = 
\begin{cases}
0, &~~ 0 < t < \frac{d}{2},  \\
\frac{c}{3} \log \Big[ \frac{\kappa_{\mathcal{O}}\; l \; (\frac{d}{2} + l - t) (t- \frac{d}{2})}{a \epsilon x (2l+d) d}
\Big], &~~  \frac{d}{2} < t < \frac{d}{2} + l, \;\;\;\; \\
0, &~~ t >  \frac{d}{2} + l.
\end{cases}
\label{I-outside-ER-between-A-B-symmetric-d-larger-dcrit-vac}
\eea 
Moreover, when $d> d_{{\rm crit};i}$, the holographic mutual information is zero for all times.
Comparison of \eqref{I-outside-ER-between-A-B-symmetric-d-smaller-dcrit-vac} and \eqref{I-outside-ER-between-A-B-symmetric-d-larger-dcrit-vac} shows that for $d < d_{{\rm crit};i}$, the early and late time behaviors of $I_{A,B;i=1,2}$ are the same and determined by $I^{\rm Vac}_{A,B}$. 
Furthermore, since the profile of $I_{A,B;i}$ for $d< d_{\rm crit}^{\rm Vac}$ and $d_{\rm crit}^{\rm Vac} < d < d_{{\rm crit};i}$ are very similar to each other, we only plot $I_{A,B;i}$ for the former one in Fig. \ref{fig: HMI-symmetric}. 

\subsection{Mechanism Inducing the Entanglement Dynamics}
\label{Sec:physical-interpretaion}

In this section, we will investigate the mechanism inducing the dynamics of the systems under consideration.
To investigate it, we will begin with the quasiparticle picture \cite{Caputa:2014vaa,
Nozaki:2014hna,
Nozaki:2014uaa,2005JSMTE..04..010C,Calabrese:2007mtj},
the one describing the entanglement dynamics qualitatively, and then quantitatively investigate how the entanglement dynamics is induced by exploiting the relation between the entanglement entropy and energy-momentum densities \cite{Mao:2025hkp}.

\subsubsection{Quasiparticle Picture \label{Sec:quasiperticle-picture}}

Now, we introduce an effective description of the entanglement dynamics induced by the insertion of the local operator in the Heisenberg picture.
We assume that in the small $\epsilon$-expansion, this picture can qualitatively describe the time dependence of the entanglement entropy and mutual information at the order of $\mathcal{O}(\log{\left(1/a \epsilon\right)})$.
For example, this cannot describe them at $\mathcal{O}(1)$, such as the vacuum entanglement entropy.
This picture is an effective description that can describe when the entanglement entropy and mutual information grow to $\mathcal{O}(\log{\left(1/a \epsilon\right)})$ and when they decrease to  $\mathcal{O}(1)$.
Note that this effective description for the holographic systems is different from that for the non-holographic systems proposed in \cite{Nozaki:2014hna,Caputa:2014vaa,Nozaki:2014uaa,He:2014mwa} because the effective description in those papers can describe the behavior of the entanglement entropy at the order of $\mathcal{O}(1)$.
In this picture, the entangled pair emerges at the insertion point of the local operator.
This entangled pair consists of two quasiparticles: one of them propagates left and another propagates right.
Since the Hamiltonian $H_0$, inducing the real-time evolution, is uniform, the propagation speed is that of light
\footnote{For $2$d CFT Hamiltonians on the curved background, the propagation speed is determined by that of the massless particles on the curved background \cite{Mao:2024cnm,2019PhRvL.122w1302C,Fan:2020orx}.}.
The quasiparticles, constructing the entangled pair, are entangled with each other. 
In other words, if only one quasiparticle is in $A$, the subsystem under consideration, quantum entanglement between them can contribute to $S_{A}$, so that the value of $S_{A}$ grows in the time region, where only a single quasiparticle is in $A$. 
In contrast, only when the entangled pair is in the subsystem under consideration, the quantum entanglement can contribute to the mutual information associated to this subsystem.
Consider $A \cup B$, the union of the distant spatial intervals, $A$ and $B$.
In this case, only when one of the quasiparticles of the entangled pair is in $A$, while the other is in $B$, the quantum entanglement can contribute to $I_{A,B}$, such that its value may grow.
However, this cannot describe how large $S_A$ and $I_{A, B}$ become due to the quantum entanglement of the entangled pair.
Furthermore, this cannot describe how the time ordering of the Euclidean and Lorentzian time evolutions affects the time growth in $S_A$ and $I_{A,B}$.
In the next section, we will deeply investigate why the time ordering affects the entanglement dynamics by exploiting the relation between the entanglement entropy and the energy-momentum densities.

\subsubsection{Relation between Quantum Entanglement and Energy-Momentum Densities \label{Sec:QE-EMD}}

In this section, we will investigate the mechanism that can quantitatively describe how the time ordering of the Euclidean and Lorentzian time evolutions influences the entanglement dynamics.
Here, we assume that $a\epsilon \ll 1$.
As reported in \cite{Mao:2025hkp}, the Euclidean holographic entanglement entropy under consideration can be expressed as
\begin{equation}
\begin{aligned}\label{eq:S-T-formula-1}
S^{\text{Euc.}}_A\left(z_{x_1}, z_{x_2}\right)=\min & \left\{ \frac{c}{6} \log \left[\frac{4 \left|\sinh \left(\frac{1}{2} \int_A \mathcal{T}(z) d z\right)\right|\left|\sinh \left(\frac{1}{2} \int_{A} \bar{\mathcal{T}}(\bar{z}) d\bar{z}\right)\right|}{
	\left[\mathcal{T}\left(z_{x_1}\right) \bar{\mathcal{T}}\left(\bar{z}_{x_1}\right)\right]^{1 / 2}\left[\mathcal{T}\left(z_{x_2}\right) \bar{\mathcal{T}}\left(\bar{z}_{x_2}\right)\right]^{1 / 2}}\right], \right. \\
& \left. \vphantom{\frac{c}{6} \log \left[\frac{4 \left|\sinh \left(\frac{1}{2} \int_A \mathcal{T}(z) d z\right)\right|}{
	\left[\mathcal{T}\left(z_{x_1}\right) \right]^{1 / 2}}\right]} 
\frac{c}{6} \log \left[\frac{4 \left|\sinh \left(\frac{1}{2} \int_{\bar{A}} \mathcal{T}(z) d z\right)\right|\left|\sinh \left(\frac{1}{2} \int_{\bar{A}} \bar{\mathcal{T}}(\bar{z}) d\bar{z}\right)\right|}{
	\left[\mathcal{T}\left(z_{x_1}\right) \bar{\mathcal{T}}\left(\bar{z}_{x_1}\right)\right]^{1 / 2}\left[\mathcal{T}\left(z_{x_2}\right) \bar{\mathcal{T}}\left(\bar{z}_{x_2}\right)\right]^{1 / 2}}\right]
\right\},
\end{aligned}
\end{equation}
where $z_{x_{i=1,2}}$ are the locations of the endpoints of the subsystem $A \in [x_1, x_2]$ on the complex plane, and $\bar{A}$ is the region complement to $A$. Moreover, $\mathcal{T}$ and $\bar{\mathcal{T}}$ are functions of the chiral and anti-chiral energy-momentum tensors, respectively.
In the systems under consideration, the functions, $\mathcal{T}(z)$ and $\bar{\mathcal{T}}(\bar{z})$, are defined as
\be \label{eq:condition-for-simplification}
\mathcal{T}(z)= C \sqrt{\langle T(z) \rangle} = \frac{f'(z)}{f(z)},~~~~~~\bar{\mathcal{T}}(\bar{z}) = C^* \sqrt{\langle \bar{T}(\bar{z}) \rangle}=\frac{\bar{f}'(\bar{z})}{\bar{f}(\bar{z})}.
\ee 
Furthermore, the complex constants, $C$ and $C^*$, and the holomorphic and anti-holomorphic functions are given as
\bea \label{eq:parameters-and-functions}
&& C = \frac{\alpha_{\mathcal{O}}}{\sqrt{h_{\mathcal{O}}} }\;\;\;\;\;\;\;\;\;\; 
\text{for}\;\;\;\;\;\; 
f(z) = \left( \frac{z- z^{\rm new}_{i, - \epsilon}}{z- z^{\rm new}_{i, \epsilon}}\right)^{ \alpha_{\mathcal{O}}}, 
\cr && \cr 
&& C^* = \pm \frac{\bar{\alpha}_{\mathcal{O}}}{\sqrt{ \bar{h}_{\mathcal{O}}} }\;\;\;\;\;
\text{for}
\;\;\;\;\;\; \bar{f}(\bar{z}) = \left( \frac{\bar{z}- \bar{z}^{\rm new}_{i, - \epsilon}}{\bar{z}- \bar{z}^{\rm new}_{i, \epsilon}}\right)^{\pm \bar{\alpha}_{\mathcal{O}}}.
\label{C}
\eea 
In the systems under consideration, by exploiting (\ref{eq:condition-for-simplification}) and (\ref{eq:parameters-and-functions}), we can simplify (\ref{eq:S-T-formula-1}) as
\begin{equation}
\begin{aligned}\label{eq:S-T-formula-2}
S^{\text{Euc.}}_A\left(z_{x_1}, z_{x_2}\right)=\min & \left\{ \frac{c}{6} \log \left[\frac{4 \left|\sinh \left(\frac{C}{2} \int_A \sqrt{T(z)} d z\right)\right|\left|\sinh \left(\frac{C^*}{2} \int_{A} \sqrt{\bar{T}(\bar{z})} d\bar{z}\right)\right|}{
	|C|^2\left[T\left(z_{x_1}\right) \bar{T}\left(\bar{z}_{x_1}\right)\right]^{1 / 4}\left[T\left(z_{x_2}\right) \bar{T}\left(\bar{z}_{x_2}\right)\right]^{1 / 4}}\right] \right. 
    \vphantom{\frac{c}{6} \log \left[\frac{4 \left|\sinh \left(\frac{C}{2} \int_A \sqrt{T(z)} d z\right)\right|}{
	|C|^2\left[T\left(z_{x_1}\right) \right]^{1 / 4}}\right]}, \\
& \left. \frac{c}{6} \log \left[\frac{4\left|\sinh \left(\frac{C}{2} \int_{\bar{A}} \sqrt{T(z)} d z\right)\right|\left|\sinh \left(\frac{C^*}{2} \int_{\bar{A}} \sqrt{\bar{T}(\bar{z})} d\bar{z}\right)\right|}{
	|C|^2\left[T\left(z_{x_1}\right) \bar{T}\left(\bar{z}_{x_1}\right)\right]^{1 / 4}\left[T\left(z_{x_2}\right) \bar{T}\left(\bar{z}_{x_2}\right)\right]^{1 / 4}}\right] \right\}
    \vphantom{\frac{c}{6} \log \left[\frac{4 \left|\sinh \left(\frac{C}{2} \int_A \sqrt{T(z)} d z\right)\right|}{
	|C|^2\left[T\left(z_{x_1}\right) \right]^{1 / 4}}\right]}.
\end{aligned}
\end{equation}
As reported in (\ref{eq:time-EM-smallepsilon}), when we perform the analytic continuation and consider the small $\epsilon$-expansion, $a\epsilon \ll 1$, then at the leading order in this expansion, the system-dependence of the expectation values of the energy-momentum densities is induced by the partition functions.
If we calculate the sine-hyperbolic functions in the numerator and expand them with respect to $a\epsilon \ll 1$, then the leading terms in this limit are given as
\footnote{Note that if we first expand $\langle T(z) \rangle_i$ and $\langle \bar{T}(\bar{z}) \rangle_i$ with respect to $a\epsilon \ll 1$, at the leading order in this expansion, $\int\sqrt{\langle T(z) \rangle_i} \; dz$ and $\int\sqrt{ \langle \bar{T}(\bar{z}) \rangle}_i \; d\bar{z}$ become infinite, because $\mathcal{F}_{s=L,R}(t,x,X)$ are infinite at $X=t-x$ or $X=t+x$.}
\bea \label{eq:sinh-functions}
&& \sinh \left(\frac{C}{2} \int_A \sqrt{\langle T(z) \rangle_i } \; d z\right) \approx \sin \left( \left( {Z^{0}_{L,i}} \right)^{- \frac{1}{2 h_{\mathcal{O}}}} \tilde{\mathcal{F}}_L \right) 
\approx \left( {Z^{0}_{L,i}} \right)^{- \frac{1}{2 h_{\mathcal{O}}}} \tilde{\mathcal{F}}_L,
\cr && \cr
&& \sinh \left(\frac{C^*}{2} \int_A \sqrt{ \langle \bar{T}(\bar{z}) \rangle_i } \; d \bar{z}\right) \approx \mp \sin \left( \pi \alpha_{\mathcal{O}} - (Z^0_{R,i})^{- \frac{1}{2 h_{\mathcal{O}}}} \tilde{\mathcal{F}}_R \right) 
\approx \mp \sin \left( \pi \alpha_{\mathcal{O}} \right),
\cr && \cr
&& \sinh \left(\frac{C}{2} \int_{\bar{A}} \sqrt{ \langle T(z) \rangle_i } \; d z\right) \approx \sin \left( \pi \alpha_{\mathcal{O}} - \left( Z^0_{L,i} \right)^{- \frac{1}{2 h_{\mathcal{O}}} } \tilde{\mathcal{F}}_L \right) 
\approx \sin \left( \pi \alpha_{\mathcal{O}} \right),
\cr && \cr 
&& \sinh \left(\frac{C^*}{2} \int_{\bar{A}} \sqrt{ \langle \bar{T}(\bar{z}) \rangle_i } \; d \bar{z}\right) \approx \mp \sin \left( \left( Z^0_{R,i} \right)^{- \frac{1}{2 h_{\mathcal{O}}}} \tilde{\mathcal{F}}_R \right)
\approx \mp \left( Z^0_{R,i} \right)^{- \frac{1}{2 h_{\mathcal{O}}}} \tilde{\mathcal{F}}_R.
\eea
Here, the system-independent function $\tilde{\mathcal{F}}_{\alpha=L,R}$ are defined as
\bea \label{F-tilde-L-R}
\tilde{\mathcal{F}}_L = \frac{\alpha_{\mathcal{O}} (x_2-x_1)}{2(t-x+x_1) (t-x+x_2)},
~~~~~~
\tilde{\mathcal{F}}_R = \frac{\alpha_{\mathcal{O}} (x_2-x_1)}{2 (t+x-x_1) (x_2-t-x)}.
\eea 
Therefore, the system-dependence of (\ref{eq:sinh-functions}) is also determined by the partition functions. Consequently, at the leading order in the small $\epsilon$-expansion, (\ref{eq:S-T-formula-2}) is simplified as 
\begin{equation}
\begin{aligned} \label{eq:EE-for-finite}
S_{A;i} = \min & \left\{ \f{c}{6} \log \left[ \f{ \tilde{\mathcal{F}}_L \sin (\pi \alpha_{\mathcal{O}}) \left( Z^0_{R,i} \right)^{\f{1}{2 h_{\mathcal{O}}}} }{
	|C|^2 \left[ \mathcal{F}_L(t,x,x_1) \mathcal{F}_R(t,x,x_1) \mathcal{F}_L(t,x,x_2) \mathcal{F}_R(t,x,x_2) \right]^{\f{1}{4}}}
\right] \right. \vphantom{\f{c}{6} \log \left[ \f{ \tilde{\mathcal{F}}_L \sin (\pi \alpha_{\mathcal{O}}) \left( Z^0_{R,i} \right)^{\f{1}{2 h_{\mathcal{O}}}} }{
	|C|^2 \left[ \dots \right]^{\f{1}{4}}} \right]}, \\
& \left. \f{c}{6} \log \left[ \f{ \tilde{\mathcal{F}}_R \sin (\pi \alpha_{\mathcal{O}}) \left( Z^0_{L,i} \right)^{\f{1}{2 h_{\mathcal{O}}}} }{
	|C|^2 \left[ \mathcal{F}_L(t,x,x_1) \mathcal{F}_R(t,x,x_1) \mathcal{F}_L(t,x,x_2) \mathcal{F}_R(t,x,x_2) \right]^{\f{1}{4}}}
\right] \right\} \vphantom{\f{c}{6} \log \left[ \f{ \tilde{\mathcal{F}}_L \sin (\pi \alpha_{\mathcal{O}}) \left( Z^0_{R,i} \right)^{\f{1}{2 h_{\mathcal{O}}}} }{
	|C|^2 \left[ \dots \right]^{\f{1}{4}}} \right]}.
\end{aligned}
\end{equation}
Thus, in this low energy limit, $a\epsilon \ll 1$, the system-dependence of the single-interval entanglement entropy emerges as its partition-function-dependence.
As the mutual information is defined as the linear combination of the entanglement entropies, in this low energy limit, $a\epsilon \ll 1$, the system-dependence of the mutual information is also determined by the partition function.  
Furthermore, from (\ref{eq:EE-for-finite}), we can see $\delta S_A$, reported in (\ref{eq:deltaSA1}) and (\ref{eq:deltaSA2}), is induced by the difference between the partition functions as
\be
\delta S_{A} = \f{c}{6}\log{\left[\left(\f{Z^{0}_{\alpha,i=2}}{Z^{0}_{\alpha,i=1}}\right)^{\f{1}{2h_{\mathcal{O}}}}\right]}.
\ee
We close this section with the description of the mechanism inducing the remarkable difference between $i=1$ and $i=2$, which is reported in Section \ref{Sec:CB-SII}.
More precisely, we explain how the remarkable difference between the late time behaviors of $\Delta S_{A;1}$ and $\Delta S_{A;2}$ appears by using (\ref{eq:S-T-formula-1}) and (\ref{eq:S-T-formula-2}).
To this end, we will begin with the Euclidean counterpart of $\Delta S_{A;i}$ for the finite interval, $A=[x_1,x_2]$, take $x_2$ to be large but finite, i.e., $x_2\gg 1$, and then perform an analytic continuation as $\tau_E=it$.
In this way, we will investigate what induces the difference in the late-time behavior of $\Delta S_{A;i}$.
The single-interval entanglement entropy for (\ref{eq:systems-under-consideration}) can be obtained as the analytic continued counterpart of the simplified formula in (\ref{eq:S-T-formula-2}) because the expectation values of the energy-momentum densities satisfy (\ref{eq:condition-for-simplification}), while the single-interval vacuum entanglement entropy should be obtained as (\ref{eq:S-T-formula-1}) because the expectation values of the energy-momentum densities do not. 
As the entanglement entropy for the vacuum state is independent of time, we will closely look at the late-time behavior of $S_{A;i}$.
To this end, we first focus on the leading terms in the small $\epsilon$-expansion, $a\epsilon \ll 1$. 
As reported in (\ref{eq:time-EM-smallepsilon}), in this limit, the system-dependence of the expectation value of the energy-momentum densities are induced by the partition functions.
Then, define the late-time limit for the half space, i.e. $x_2 \gg x_1$, as $t\gg x$ and $t \gg x_1-x$, where we initially take $x_2$ to be large, and take a large $t$ limit.
At the leading order in this limit, the time dependence of the expectation values of the energy-momentum densities are given as
\be
\begin{split}
\left\langle T (z_{x_1})\right\rangle_i \approx\left\langle \bar{T} (\bar{z}_{x_1})\right\rangle_i \approx \f{h_{\mathcal{O}}}{t^4\left(Z^{0,0}_{i}\right)^{\f{1}{h_{\mathcal{O}}}}},~~~  \left \langle T(z_{x_2}) \right \rangle_{i} \approx \left \langle \bar{T}(\bar{z}_{x_2}) \right \rangle_{i} \approx\f{h_{\mathcal{O}}}{x_2^4\left(Z^{0,0}_{i}\right)^{\f{1}{h_{\mathcal{O}}}}}, 
\end{split}
\ee
where the late-time partition functions for the left and right movers, $Z^{0,0}_{i}$, are defined as
\be \label{eq:late-time-partition-functions}
\begin{split}
Z^{0,0}_{i}=\begin{cases}
\f{1}{\left(2a \epsilon t\right)^{2h_{\mathcal{O}}}}~&~\text{for}~i=1\\
\f{1}{\left(2a \epsilon x\right)^{2h_{\mathcal{O}}}}~&~\text{for}~i=2\\
\end{cases}.
\end{split}
\ee
Thus, in the late-time limit for the half space, the partition function for the left and right movers becomes independent of those movers. 
Furthermore, we obtain $\tilde{\mathcal{F}}_{\alpha=L,R}$ in this limit as
\be
\tilde{\mathcal{F}}_{\alpha=L,R} \approx \tilde{\mathcal{F}}=\f{\alpha_{\mathcal{O}}}{2t}.
\ee
Consequently, in the late-time limit for the half space, we obtain $\Delta S_{A;i}$ as
\bea
\label{eq:Delta-S-T-formula}
\Delta S_{A;i} &\approx&  \frac{c}{6} \log \left[ \f{4 t^2  \tilde{\mathcal{F}} \sin (\pi \alpha_{\mathcal{O}} )  \left( Z^{0,0}_i \right)^{\f{1}{2 h_{\mathcal{O}}}} }{\alpha^2_{\mathcal{O}}} \right]
\cr && \cr 
&\approx& \frac{c}{6} \log \left[ 2 t \kappa_{\mathcal{O}} \right]+\frac{c}{6} \log \left[  \left( Z^{0,0}_i \right)^{\f{1}{2 h_{\mathcal{O}}}} \right].
\eea
Thus, in the late-time limit for the half space, the system dependence of $\Delta S_{A;i}$ is induced by $Z^{0,0}_i$. As in (\ref{eq:late-time-partition-functions}), $Z^{0,0}_{i=1}$ decays in time as $t^{-2h_{\mathcal{O}}}$, while $Z^{0,0}_{i=2}$ is independent of time. For $i=1$, the late-time dependence of the partition function cancels with that in the first term in \eqref{eq:Delta-S-T-formula}, while for $i=2$, that is time-independent and does not. Consequently, we obtain the late-time value of $S_{A;i}$ as \eqref{SA-1-late time-1}. As reported in Section \ref{sec:TE-EMT}, the time dependence of $Z^{0,0}_{i=1}$ is induced by the non-unitary time evolution, while the time-independent behavior of $Z^{0,0}_{i=2}$ is induced by the unitary time evolution. Therefore, we can see that the system-dependence of $\Delta S_{A;i}$, i.e., where $\Delta S_{A;i}$ does depend on time is induced by whether the time evolution is unitary or non-unitary.

\section{Revisiting the Gravity Dual for $h_{\mathcal{O}} = \bar{h}_{\mathcal{O}} > \frac{c}{24}$}
\label{Sec: Revisiting the Gravity Dual}

In this section, we will present the details of the gravitational dual of the systems with the insertion of the local operator with large conformal dimensions, i.e., $h_{\mathcal{O}}=\overline{h}_{\mathcal{O}}>\f{c}{24}$.

\subsection{Dual Geometry \label{sec:Map-from-BN-to-One}}

As discussed in Section \ref{Sec: Gravity Dual}, the dual geometry of the systems under consideration is expressed as a Ba$\tilde{\rm n}$ados geometry given in \eqref{metric-Banados-complex coordinates}.
It was shown in ref. \cite{Abajian:2023jye} (see also \cite{Tian:2024fmo}) that this geometry is written in a coordinate system which cannot cover the regions deep inside the bulk spacetime. 
In particular, these coordinates cannot reach the horizon 
\footnote{More precisely, the Ba$\tilde{\rm n}$ados geometry can be mapped \cite{Abajian:2023bqv} to a geometry which is called the {\it cone} geometry \cite{Abajian:2023jye}. The latter has a hypersurface deep inside the bulk spacetime which is called {\it wall}. On the wall the determinant of the metric is zero and the coordinates cannot cover the region behind the wall. Moreover, the horizon is located inside the wall \cite{Abajian:2023jye}. Therefore, in the coordinate systems used in the cone and Ba$\tilde{\rm n}$ados geometries, one cannot reach the horizon of the black hole.}.
Therefore, to investigate the deep region of the gravity dual under consideration, we need to exploit the coordinates to cover that region.
To this end, we begin with the Ba$\tilde{\rm n}$ados geometry, and then map from this one to the pure $AdS_3$ by exploiting the coordinate transformations in (\ref{Banados map}).
Then, map from the pure $AdS_3$ in (\ref{AdS-Poincare-metric}) to the BTZ geometry by exploiting the coordinate transformations defined as 
\be
\begin{split}
r = r_h \left( \frac{\sqrt{u^2 + w \bar{w}}}{u} \right),~~~~\tau = \left( \frac{L_{AdS}^2}{2 i r_h} \right) \ln \left( \frac{w}{\bar{w}} \right),~~~~\phi = \left( \frac{L_{AdS}}{2 r_h} \right) \ln \left( u^2 + w \bar{w} \right),
\end{split}
\label{map BTZ to AdS3}
\ee
where $r_h$ is the radius of the horizon. It is obvious that the above transformations are simply obtained from the following ones \cite{Carlip:1994gc}
\bea
&& w = \sqrt{1- \frac{r_h^2}{r^2}} \; e^{\frac{r_h}{L_{AdS}} \left( \phi + \frac{i \tau}{L_{AdS}} \right)},
\cr && \cr 
&& \bar{w} = \sqrt{1- \frac{r_h^2}{r^2}} \; e^{\frac{r_h}{L_{AdS}} \left( \phi - \frac{i \tau}{L_{AdS}} \right)},
\cr && \cr
&& u = \left( \frac{r_h}{r} \right) e^{\frac{r_h \phi}{L_{AdS}}}.
\label{map AdS3 to BTZ}
\eea 
Then, we identify the pure $AdS_3$ mapped from the Ba$\tilde{\rm n}$ados geometry with the one mapped from the BTZ black hole.
Consequently, the Ba$\tilde{\rm n}$ados geometry is mapped to a BTZ black hole in global coordinates as
\bea \label{eq:BTZ-metric}
ds^2 = \left( \frac{r^2 - r_h^2}{L_{AdS}^2} \right) d \tau^2 + \left( \frac{L_{AdS}^2}{r^2 - r_h^2} \right) dr^2 + r^2 d\phi^2.
\eea 
Not that we denote the Euclidean time in the bulk spacetime with $\tau$ and show the Euclidean time on the boundary by $\tau_E$.
Thus, we obtain the map from the Ba$\tilde{\rm n}$ados geometry to the BTZ one as
\bea
\text{Ba$\tilde{\rm n}$ados geometry} \; (y,z, \bar{z}) \rightarrow AdS_3 \; (u, w, \bar{w}) \rightarrow BTZ \; (r, \tau, \phi).
\label{map-Banados-AdS-BTZ}
\eea 
The expressions for $r(y, z, \bar{z})$, $\tau (y, z, \bar{z})$ and $\phi (y, z, \bar{z})$ are given by \eqref{r-y-z-zb}, \eqref{tau-y-z-zb} and \eqref{phi-y-z-zb}, respectively. Now, we begin with the BTZ metric in (\ref{eq:BTZ-metric}), and then change the boundary coordinates from $(\tau, \phi)$ to $(z, \bar{z})$ which were exploited for the calculations in Section \ref{Sec: Gravity Dual}, while keeping $r$ as the radial coordinate. In other words, one can easily find (see Appendix \ref{Sec: Asymptotic Metric})
\be
\tau = \tau (y, z, \bar{z}) = \tau (r, z, \bar{z}), ~~~~
\phi = \phi (y, z, \bar{z}) = \phi (r, z, \bar{z}).
\ee
Then, the line element is given as a function of $r$, $z$ and $\bar{z}$. The details of the metric are reported in Appendix \ref{Sec: Asymptotic Metric}.
It should be pointed out that this method was also applied in ref. \cite{Mao:2025hkp}.
Moreover, a similar method was used in ref. \cite{Das:2024lra} for a model of driven CFT to study the shape of the horizon.
The asymptotic geometry near the boundary, i.e. $r\approx \infty$, is given by \eqref{asymptotic-metric-1} which we rewrite it here as
\bea
ds^2 \big{|}_{r \rightarrow \infty} = \left( \frac{L_{AdS}}{r} \right)^2 dr^2 + r^2 \Bigg{\lvert} \frac{(z^{\rm new}_{i, - \epsilon} - z^{\rm new}_{i, \epsilon})}{(z_X - z^{\rm new}_{i, - \epsilon}) (z_X- z^{\rm new}_{i, \epsilon})} \Bigg{\rvert}^2 \left( d\tau_{E}^2 + dX^2 \right).
\label{asymptotic-metric-2}
\eea 
Moreover, one has $\tau_{E}=\f{z+\bar{z}}{2}$, $X=\f{z-\bar{z}}{2i}$ and $z_X= i X$
\footnote{Here, we denote the spatial coordinate on the boundary by $X$ and keep $x$ for the location of the primary operator. Moreover, we calculated $\langle T(z) \rangle_i$ and $\langle \bar{T}(\bar{z}) \rangle_i$ at $z=z_X= iX$.}.
To obtain the asymptotically pure $AdS_3$ geometry in the Poincar\'e coordinates, we define a new radial coordinate as 
\bea \label{eq:map-from-r-to-uprime}
u^{\prime}= \left( \f{L_{AdS}}{r} \right) \Bigg{\lvert} \frac{(z^{\rm new}_{i, - \epsilon} - z^{\rm new}_{i, \epsilon})}{(z_X - z^{\rm new}_{i, - \epsilon}) (z_X- z^{\rm new}_{i, \epsilon})} \Bigg{\rvert}^{-1}= \left( \frac{L_{AdS}}{r} \right) \left( \frac{h^2_{\mathcal{O}}  }{\langle T(z_X) \rangle_i \langle \bar{T} (\bar{z}_X) \rangle_i} \right)^{\frac{1}{4}}.
\label{up-rp}
\eea 
Note that we applied \eqref{T-holo-antiholo-1-2} in the last equality. By using this coordinate, we obtain the asymptotic geometry, i.e., that for $u^{\prime} \rightarrow 0$, as the pure $AdS_3$ geometry in the Poincar\'e coordinates
\bea
ds^2 \big{|}_{u^{\prime} \rightarrow 0} = \left( \frac{L_{AdS}}{u'} \right)^2 \left( du^{\prime 2} + d\tau_{E}^2 + dX^2 \right).
\label{asymptotic-metric-up}
\eea 
Furthermore, after the analytic continuation, i.e. $\tau_{E} =it$, the metric on the asymptotic boundary is the Minkowski metric, $ds^2_{\rm bdy} = -dt^2 + dX^2$, as it was expected. 

\subsection{Shape of the Horizon \label{Sec:Shape-of-Horizon}}

\begin{figure}
\begin{center}
\includegraphics[scale=0.36]{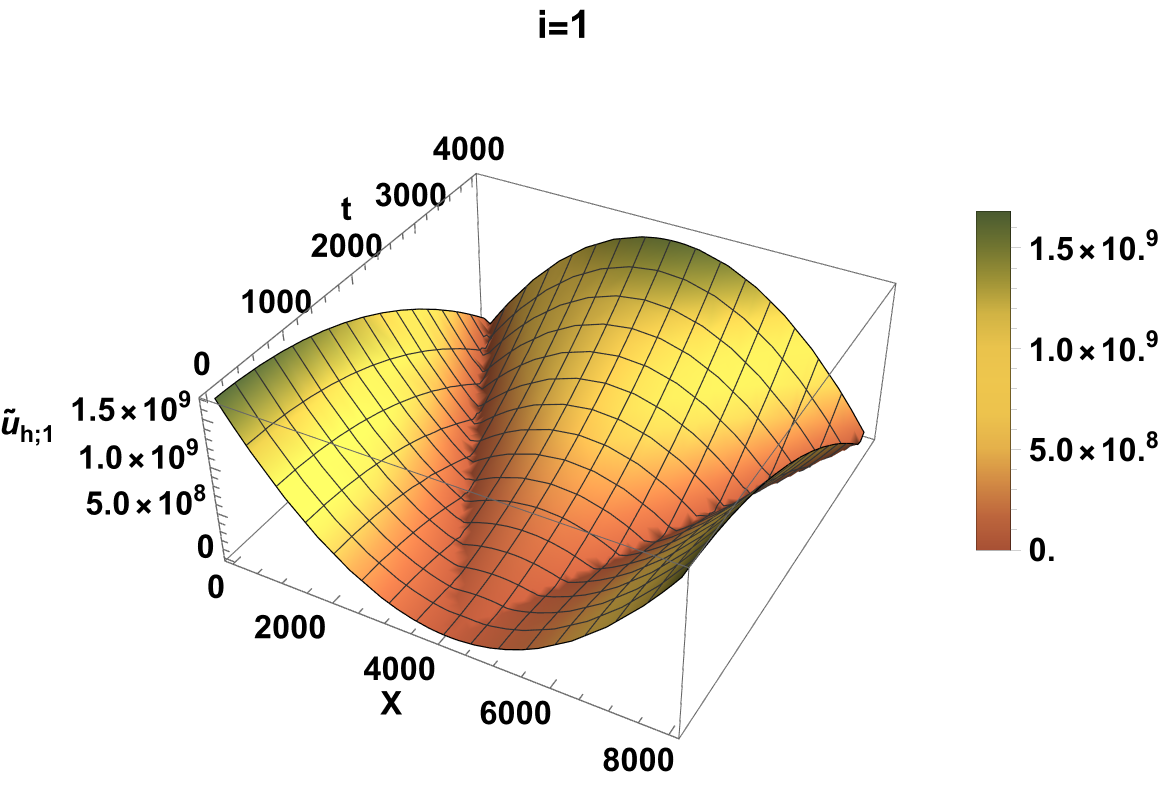}
\hspace{0.2cm}
\includegraphics[scale=0.36]{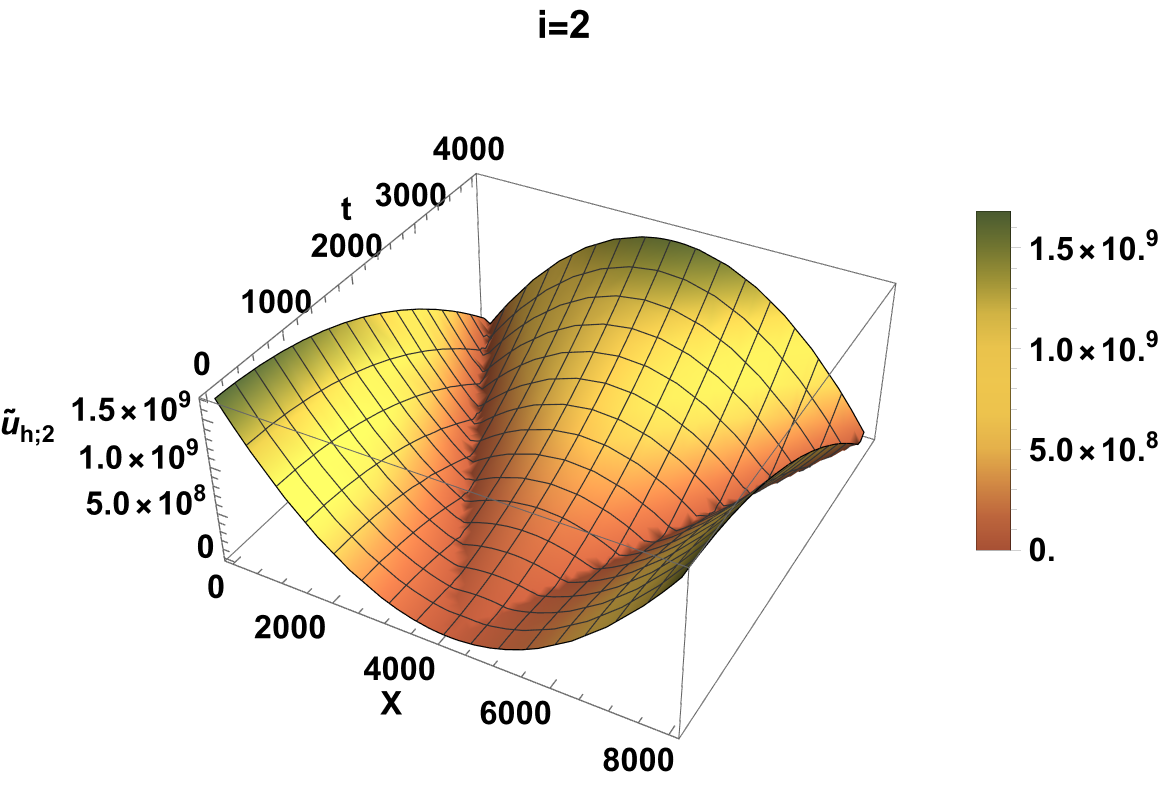}
\end{center}
\caption{The horizon radius $\tilde{u}_{h;i}$ as a function of $X$ and $t$. {\it Left}) For the non-unitary time evolution where $i=1$. {\it Right}) For the unitary time evolution where $i=2$. We set $x=4.1 \times 10^3$, $L_{AdS}=10^5$, $r_h = 10^3$, $a = 10^{-4}$ and $\epsilon = 10$.}
\label{fig: u-tilde-h-3D}
\end{figure}

In this section, we will report the spacetime-dependence of the horizon in the black brane geometry, i.e., the property, reflecting the dynamical one of the systems under consideration, of the gravity dual.
The coordinate transformation in \eqref{eq:map-from-r-to-uprime} maps $r_h$, the location of the horizon in $r$ coordinate, to the one in $u^{\prime}$ coordinate as
\bea \label{eq:L-o-HR-in-Poncare}
u^{\prime}_{h;i}= \left( \frac{L_{AdS}}{r_h} \right) \left( \frac{h^2_{\mathcal{O}}  }{\langle T(z_X) \rangle_i \langle \bar{T} (\bar{z}_X) \rangle_i} \right)^{\frac{1}{4}}.
\eea
After the analytic continuation, i.e. $\tau_{E} =it$, $u_{h;i}'$ in (\ref{eq:L-o-HR-in-Poncare}) becomes a function of $X$, $x$, $\epsilon$, and $t$. Now, we focus on the spacetime dependence of the horizon for $i=1$ and $2$. By plugging \eqref{z-new1-z-new2} into \eqref{eq:L-o-HR-in-Poncare}, we obtain the location of the horizon as
\be
\begin{split}
& u'_{h;1} = \f{L_{AdS}}{r_h}\cdot\frac{|X + e^{i a \epsilon} (t- x)| |X - e^{i a \epsilon} (t+ x)|}{2 \lvert 
\sin(a \epsilon) \rvert \sqrt{ \lvert x^2 - t^2 \rvert } },\\
&u'_{h;2} = \f{L_{AdS}}{r_h} \cdot\frac{|t +X - x e^{i a \epsilon}| |t - X + x e^{i a \epsilon}|}{2  \lvert x \sin(a \epsilon) \rvert },
\end{split}
\label{uhnew-1}
\ee 
or equivalently,
{\small
\be
\begin{split}
u^{\prime}_{h;1} &= \f{L_{AdS}}{r_h}\cdot \frac{\sqrt{(X + (t-x) \cos(a \epsilon) )^2 + (t-x)^2 \sin(a \epsilon)^2}\cdot\sqrt{(X - (t+x) \cos(a \epsilon) )^2 + (t+x)^2 \sin(a \epsilon)^2}}{2  \lvert \sin(a \epsilon) \rvert \sqrt{ \lvert x^2 - t^2 \rvert } },\\
u^{\prime}_{h;2} &= \f{L_{AdS}}{r_h}\frac{\sqrt{(t + X -x \cos(a \epsilon))^2 + x^2 \sin(a \epsilon)^2}\cdot\sqrt{(t - X + x \cos(a \epsilon))^2 + x^2 \sin(a \epsilon)^2}} {2 \;  \lvert x \sin(a \epsilon) \rvert }.
\end{split}
\label{uhnew-2}
\ee
}
To make the holographic entanglement entropy finite, we need to introduce a surface where the geodesic is anchored. 
Here, we assume that it is localized at $u^{\prime}_{UV}= 1$. 
Since we are interested in the entanglement dynamics in the energy-region where the energy scale under consideration is much smaller than one, at least, the radius of the black brane horizon should be much larger than the UV-cutoff scale, $u^{\prime}_{UV}$.
As in (\ref{uhnew-2}), the location of the horizon depends on the spacetime.
Therefore, the location of the black brane horizon, nearest to the boundary of the spacetime, should follow
\be
u^{\prime \; {\rm min}}_{h;i} \ge u^{\prime}_{UV}=1.
\label{eq:inequality-for-gravity}
\ee
Here, $u^{\prime \; {\rm min}}_{h;i}$ denotes the global minimum of $u^{\prime}_{h;i}$ and corresponds to the closest location, along the $u$ direction, of the horizon to the spacetime boundary. 
Note that in the coordinate $u^{\prime}$, the asymptotic boundary is located at $u^{\prime} \rightarrow 0$.
Moreover, the closest point of the horizon to the boundary necessarily satisfies two following equations,
{\small
\be
\begin{split}
& \frac{\partial u^{\prime}_{h;1} (X,t) }{\partial t} = 0, ~~\Rightarrow~~ t \Big[ (t^2 - X^2)^2 - 4 x^2 X^2 - X^4 + 4 x X^3 \cos(a \epsilon) \Big] = 0,\\
&\frac{\partial u^{\prime}_{h;1} (X,t) }{\partial X} = 0, ~~\Rightarrow~~ \sqrt{\lvert t^2 - x^2 \rvert} \Big[ 2 x^2 X + X^3 + x (t^2 - x^2 - 3 X^2) \cos (a \epsilon) 
+ (x^2 - t^2) X \cos (2 a \epsilon) \Big]  =0, \\
&\frac{\partial u^{\prime}_{h;2}(X,t)}{\partial t} = 0, ~~\Rightarrow~~ t \Big[ X^2 - t^2 - 2x X \cos(a \epsilon)+ x^2 \cos(2 a \epsilon) \Big] =0,\\
&\frac{\partial u^{\prime}_{h;2}(X,t)}{\partial X} = 0, ~~\Rightarrow~~ \left( X- x \cos(a \epsilon) \right) \left( X^2+ x^2 -t^2 - 2 x X \cos(a \epsilon) \right) = 0.
\end{split}\label{duh-dt-dX} 
\ee }
As we assumed that $a\epsilon \ll 1$ in the calculations of the entanglement entropy and mutual information, we will report $u^{\prime}_{h;i}$ and $u^{\prime \; {\rm min}}_{h;i}$ at the leading order of the small $\epsilon$-expansion.
Since at the leading order in this small $\epsilon$-expansion, the expectation values of the chiral and anti-chiral energy-momentum densities are given as (\ref{eq:time-EM-smallepsilon}), by substituting it into (\ref{eq:L-o-HR-in-Poncare}), we can simplify $u^{\prime}_{h;i}$ as
\be
u^{\prime}_{h;i}=\f{L_{AdS}h^{\f{1}{2}}_{\mathcal{O}}}{r_h}\cdot \f{(Z^0_{L,i}Z^0_{R,i})^{\f{1}{4h_{\mathcal{O}}}}}{\left(\mathcal{F}_{L}(t,x,X)\mathcal{F}_{R}(t,x,X)\right)^{\f{1}{4}}}.
\ee
Thus, the system-dependence of the horizon appears in the numerator as the partition function.
Since the behavior of the partition function is determined by whether the time evolution is unitary or non-unitary, the system-dependence of the horizon is also determined by it.
If we define the difference between them as $D_{h}=\f{u^{\prime}_{h;1}}{u^{\prime}_{h;2}}$, then at the leading order in the small $\epsilon$-expansion, $D_{h}$ is simplified as the ratio of the partition functions as
\be
D_{h}=\left(\f{Z^0_{L,1}Z^0_{R,1}}{Z^0_{L,2}Z^0_{R,2}}\right)^{\f{1}{4h_{\mathcal{O}}}}=\f{|x|}{\sqrt{|x^2-t^2|}}.
\ee
By using $D_h$, we can see how large the horizon, at $X$, for $i=1$ is compared to that for $i=2$. 
In the regime, $t\gg x>0$, $D_h$ is simplified as
\be\label{eq:ratio}
D_{h}\approx\f{x}{t}.
\ee
Furthermore, we define a system-independent horizon as 
\be
\tilde{u}_{h;i}=\f{u^{\prime}_{h;i}}{\left(Z^0_{L,i}Z^0_{R,i}\right)^{\f{1}{4h_{\mathcal{O}}}}}.
\ee
In Fig \ref{fig: u-tilde-h-3D}, we show $\tilde{u}_{h;i}$ as the function of $X$ and $t$. 
Since the only system-independent pieces of $u^{\prime}_{h;i}$ depend on $X$, from this figure, we can observe the outlined $X$ and $t$ dependence of the horizon. 
From Fig \ref{fig: u-tilde-h-3D}, we can observe that there exists only one spatial point, $X=x$, where the horizon becomes the closest to the boundary at $t=0$, while for $t>0$ there are two spatial points, where it does.
Furthermore, in this limit, we obtain the spacetime location where the horizon is nearest to the boundary as 
\be \label{eq:spacetime-for-max}
(t,X)=(0, x).
\ee
Then, the location of the closest horizon radius is given by
\bea \label{eq:closest-horizon}
u^{\prime \; {\rm min}}_{h;i} = \frac{L_{AdS} \; x a \epsilon }{2 r_h  },
\label{rh-psi-1-t=0}
\eea 
which shows that $u^{\prime \; {\rm min}}_{h;i}$ is independent of $i$.
Thus, at the time when the local operator is inserted into the system, in the gravity dual, the black brane horizon becomes closest to the boundary.
This is consistent with the observation, mentioned above, from Fig. \ref{fig: u-tilde-h-3D}.
Moreover, this is consistent with the dependence on the energy-momentum densities of the horizon, which is reported in (\ref{eq:L-o-HR-in-Poncare}).
As $\left \langle T(z_X) \right \rangle_i \cdot \langle\bar T(\bar{z}_X) \rangle_i$ appears in the denominator of $u^{\prime}_{h;i}$, if $\left \langle T(z_X) \right \rangle_i \cdot \langle\bar T(\bar{z}_X)\rangle_i$ becomes larger, the horizon becomes closer to the boundary.
As mentioned in Section \ref{sec:TE-EMT}, the time evolution of the  energy densities can be described by the propagation of the two excitations, one of which propagates to the left while the other propagates to the right, induced by the local operator.
Therefore, only if $t=0$ and $X=x$, the two excitations are at the spatial point $X$, so that 
$\left \langle T(z_X) \right \rangle_i \cdot \langle\bar T(\bar{z}_X)\rangle_i$ takes its maximum value.
Correspondingly, the horizon becomes closest to the boundary.

Subsequently, let us focus on the parameter-dependence of $u^{\prime \; {\rm min}}_{h;i}$, the closest horizon radius. 
First, we report on the $x$-dependence of $u^{\prime \; {\rm min}}_{h;i}$.
Equation (\ref{eq:closest-horizon}) shows that if we move $x$, the insertion point of the local operator, closer to the origin, $u^{\prime \; {\rm min}}_{h;i}$ becomes closer to the spacetime boundary.
This may be due to the spatial dependence of the Rindler Hamiltonian, the one determining the Euclidean time evolution.
As the Hamiltonian density of the Rindler Hamiltonian becomes smaller if we move closer to the origin, the Euclidean time evolution of the Rindler Hamiltonian cannot tame the high-energy modes induced by the contraction of the local operator inserted around the origin.
Then, we present the $\epsilon$-dependence of $u^{\prime \; {\rm min}}_{h;i}$.
If we take $\epsilon$ to be smaller, then the closest horizon radius becomes closer to the spacetime boundary.
One possible interpretation of $\epsilon$-dependence of the closest horizon radius is that as the Euclidean time evolution with smaller $\epsilon$ more weakly suppresses the high-energy modes induced by the contraction of the local operator, the closest part of the horizon penetrates close to the boundary.

Then, we revisit the relation between $u^{\prime \; {\rm min}}_{h;i}$ and the location of the surface where the geodesic ends, which is reported in (\ref{eq:inequality-for-gravity}).
By exploiting a relation, $\frac{r_h}{L_{AdS}} = \sqrt{\frac{24 h_{\mathcal{O}}}{c} -1}$, we obtain the constraint on the parameters and $x$ as 
\be
1 \le \f{x a \epsilon}{2\sqrt{\f{24h_{\mathcal{O}}}{c}-1}}.
\label{eq:inequality-for-eUV}
\ee
When we take the parameters and the location of the operator such that they satisfy \eqref{eq:inequality-for-eUV}, then the gravity dual should describe the entanglement dynamics in the energy-region where the energy scale under consideration is much smaller than one.

\begin{figure}
\begin{center}
\includegraphics[scale=0.36]{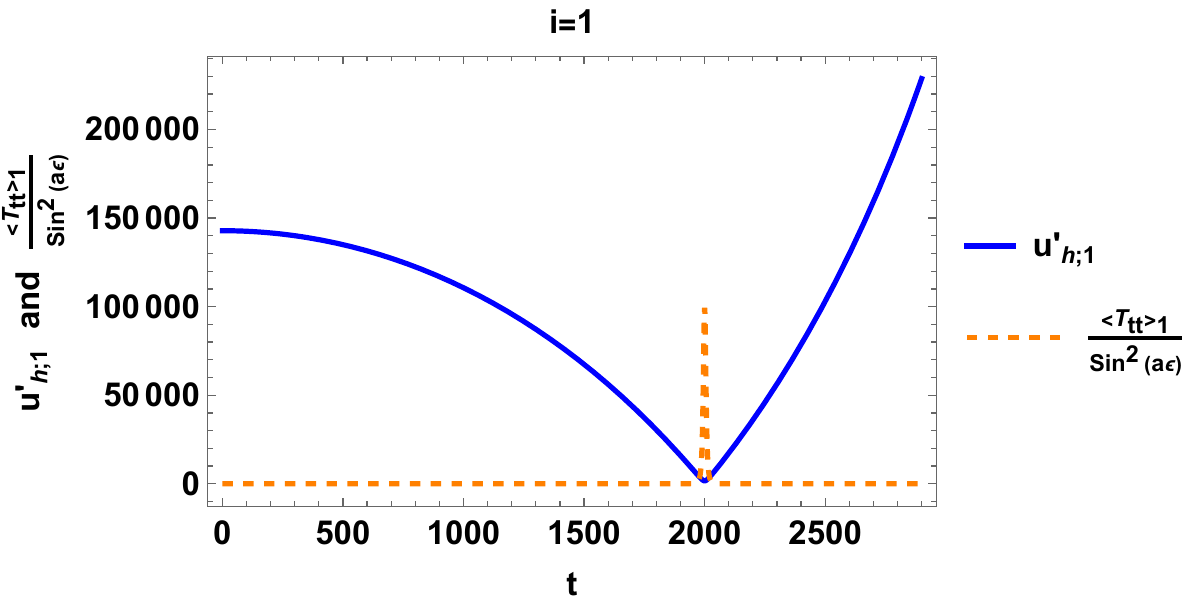}
\hspace{0.2cm}
\includegraphics[scale=0.36]{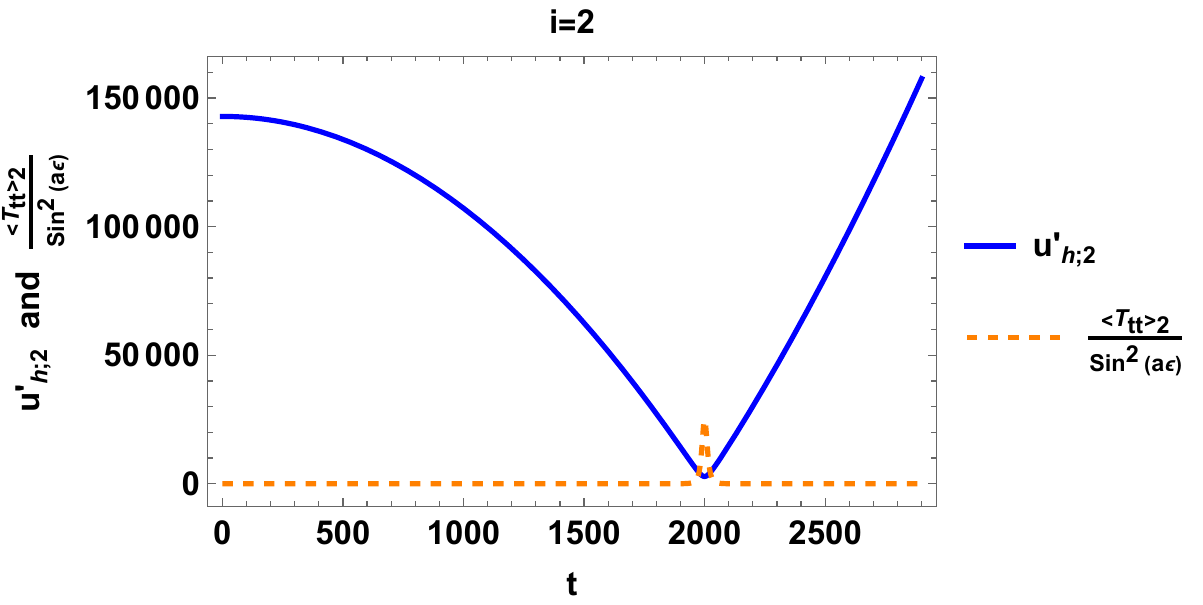}
\end{center}
\caption{The horizon radius $u^{\prime}_{h;i}$ and $\frac{\langle T_{tt} \rangle_i}{\sin^2 (a \epsilon)}$ as a function of $t$. {\it Left}) For the non-unitary time evolution where $i=1$. {\it Right}) For the unitary time evolution where $i=2$. We set $x=4 \times 10^3$, $X=2 \times 10^3$, $L_{AdS}=10^5$, $r_h = 7 \times 10^4$, $a = 10^{-4}$, $\epsilon = 50$ and $c=1000$. Moreover, $h_{\mathcal{O}} = \frac{c}{24} \left( 1+ \frac{r_h^2}{L_{AdS}^2 } \right)$. It is observed that the minima of the horizon coincide with the maxima of $\langle T_{tt} \rangle_i$.}
\label{fig: uph-Ttt}
\end{figure}

Moreover, we closely look at the behavior of the horizon.
From \eqref{eq:L-o-HR-in-Poncare}, one can see that the time dependence of $u^{\prime}_{h;i}$ is determined by the dependence of $\langle T(z_X) \rangle_i$ and $\langle \bar{T}(\bar{z}_X) \rangle_i$. 
Therefore, the minima of the horizon radius correspond to the maxima of the chiral and anti-chiral parts of the energy-momentum tensor, and hence, to the maxima of the energy density $\langle T_{tt} \rangle_i$.
In Fig. \ref{fig: uph-Ttt}, we show $u^{\prime}_{h;i}$ and $\langle T_{tt} \rangle_i$ as a function of $t$ for fixed values of $x$ and $X$. 
We observe that the spatial location, along $X$, where the horizon has peaks, coincides with the one where $\langle T_{tt} \rangle_i$ has peaks.
As explained in Sections \ref{sec:TE-EMT} and \ref{Sec:quasiperticle-picture}, insertion of the primary operator creates two  excitations which move to the left and right. The left moving and right moving excitations create a maximum in $\langle T(z_X) \rangle_i$ and $\langle \bar{T}(\bar{z}_X) \rangle_i$, respectively. Therefore, $\langle T_{tt} \rangle_i$ has two maxima and each of them corresponds to one of the excitations. Having said this, one might conclude that the two minima that are observed in the horizon radius are related to the propagation of the two quasiparticles on the boundary. It should be pointed out that the observation that a local operator quench can create two minima in the horizon radius was also reported recently in ref. \cite{Mao:2025hkp}
\footnote{It should be emphasized that the authors in ref. \cite{Mao:2025hkp}, considered a local operator quench for a 2d holographic CFT on a cylinder. They applied the $H_{\rm SSD}$ and $H_{2-{\rm SSD}}$ Hamiltonians for the Euclidean time evolution, i.e. UV regularization. Therefore, our setup is different from theirs.}.

We close this section with a discussion on the late-time behavior of the horizon in the region far from its peaks. 
By substituting \eqref{T-holo-antiholo-late-times} into \eqref{eq:L-o-HR-in-Poncare}, assuming $X$ is far from the points where the horizon radius has minima, and considering its behavior in the late-time region, we obtain the behavior of it as $u^{\prime}_{h;1} \propto t$ and $u^{\prime}_{h;2} \propto t^2$.
Thus, the location of the horizon far from the peaks becomes far from the spacetime boundary according to the time evolution.
Moreover, in the spatial region far from the peaks, the horizon for $i=2$ moves far away from the boundary faster than that for $i=1$. 
This is consistent with (\ref{eq:ratio}).

\section{Discussions and Future Directions}
\label{Sec: Discussions}

In Section \ref{Sec:intro}, we have already reported the summary of our findings. 
Therefore, we will close this paper with the discussions of our findings and comment on the future directions.
\begin{itemize}
\item[] {\it Beyond $\log{\left(1/a\epsilon\right)}$:} In this paper, we investigated the entanglement dynamics induced by the local operator with the composite time evolution operator constructed of the Euclidean and Lorentzian ones. 
Note that the Euclidean time evolution is determined by $H_R$, while the Lorentzian one is determined by $H_0$.
In the parameter region, $a\epsilon \ll 1$, we expanded the entanglement entropy and mutual information with respect to $a\epsilon \ll 1$.
The time dependence of the entanglement entropy for the single interval shows that in a certain time region, the leading order of the entanglement entropy becomes at the order of $\log(1/a\epsilon)$, while in the other time regions, that of the entanglement entropy becomes at the order of one, i.e., it becomes the entanglement entropy of the vacuum state.
We found that the quasiparticle picture can qualitatively describe the behavior of the entanglement entropy for the single interval when it becomes of the order of $\log(1/a\epsilon)$.
In contrast, the time evolution of the mutual information for some configurations, e.g., Figs. \ref{fig: HMI-configurations}(b), \ref{fig: HMI-configurations}(c), \ref{fig: HMI-configurations}(d), shows that the mutual information can be at the order of one for the entire time evolution (for the details, see Appendix \ref{sec:R-TE-MI} ).  
In those configurations, in a certain time region, the value of the mutual information can deviate from the vacuum one, while in some time region, it is the same as the vacuum one.
It should be emphasized that this behavior was originally pointed out by \cite{Asplund:2013zba}.
Moreover, this time evolution of the mutual information at the order of one is beyond our quasiparticle picture.
Therefore, it would be interesting to investigate how this time evolution at the order of one is induced from the gravity side.
For example, consider the time evolution from the vacuum state with the insertion of the heavy local operator, i.e., $h_{\mathcal{O}}>c/24$, in the case of (c) in Fig. \ref{fig: HMI-configurations}.
When the mutual information deviates from the vacuum one, a single peak of the black brane horizon is in the region between two spatial subregions.
This suggests that the time evolution of the mutual information at the order of one reflects the property of the dual gravity.
To investigate the mechanism inducing the time evolution of the mutual information at the order of one may lead us to a deep understanding of the holographic CFTs.
\item[]{\it Time-ordering effect:}
We investigated the time ordering effect of the composite time evolution operator constructed of the Euclidean time evolution operator determined by $H_{R}$ and the Lorentzian one determined by $H_0$. Then, we found that for $i=2$, the unitary time evolution induces the late-time logarithmic growth of 
the entanglement entropy for the semi-infinite interval, while for $i=1$, the non-unitary time evolution induces the late-time time-independent behavior of that.
One future problem is to investigate if the same mechanism can describe the time ordering effect induced by more general Hamiltonians on the entanglement dynamics.
For example, one simple generalization of our setup is to replace $H_R$ with 
\be
H_{G.R.}=a\int^{\infty}_{-\infty}|x|h(x) \; dx.
\ee
In this generalization, we can investigate the entanglement dynamics even for $x<0$ because $|x|h(x)$, the energy density of $H_{G.R.}$, is positive.
However, since the energy density at $x=0$ is still zero, the Euclidean time evolution induced by $H_{G.R.}$ may not play as the regulator at $x=0$.
Therefore, it may be interesting to investigate the time ordering effect when we replace $H_{G.R.}$ with $H_{\theta}$ which is defined as
\be \label{eq:generalized-regulator}
H_{\theta}=H_0+\tanh{\theta} \; H_{G.R.},
\ee
where $\theta$ is real. 
Note that for $\theta=0$, $H_{\theta}$ reduces to the uniform Hamiltonian, $H_0$.

\item[]{\it Measurement-induced phase-transition:} 
As discussed in Section \ref{eq:ETE-PSPM}, the Euclidean time evolution can be interpreted as a post-selecting measurement on the ancilla. 
Currently, considerable studies on the dynamics induced by the measurements \cite{2018PhRvB..98t5136L,2019PhRvB.100m4306L,2019PhRvX...9c1009S,2019PhRvB..99v4307C,2019PhRvB.100f4204S,2020PhRvX..10d1020G,2020PhRvB.101j4302J,2020arXiv200312721L,2021PhRvL.126f0501I,2025PhRvB.112j4322L} elucidated the novel phase-transition induced by the measurement, so-called measurement-induced phase-transition. 
Therefore, it would be interesting to generalize our system to the Floquet protocol as
\be
\ket{\Psi}\propto \underset{n}{\underbrace{e^{-\epsilon H_R}e^{-it H_0} \cdots e^{-\epsilon H_R}e^{-it H_0}}}\mathcal{O}(x)\ket{0},
\ee
where $n$ is an integer, and then investigate whether the measurement can induce the phase-transition.
\end{itemize}

\section*{Acknowledgments}

We would like to thank Chen Bai, Jia Tian and Yu-Xuan Zhang for very helpful discussions. We are also very grateful to Tom Hartman for his very illuminating comments. The work of M.N. is supported by funds from the University of Chinese Academy of Sciences (UCAS) and
the Kavli Institute for Theoretical Sciences (KITS). A. M. and F.O. are also supported by the funds from the UCAS and KITS. Moreover, A. M. is supported in part by MEXT KAKENHI Grant-in-Aid for Transformative Research Areas A “Extreme Universe” No. 21H05184.

\appendix

\section{Time Evolution of the Primary Operators}
\label{Sec: Time Evolution of The Primary Operator}

In this appendix, we find the time evolution of the primary operators in the Heisenberg picture. To do so, we first write the Hamiltonian $H_0$ of the CFT and the Rindler Hamiltonian $H_R$ in terms of the Virasoro generators. We work in the Euclidean signature, i.e. $\tau_E=i t$, and choose the coordinates on the plane as $z= \tau_E + ix$ and $\bar{z} = \tau_E - i x$. Then, by applying
\bea
T_{\tau_E \tau_E} = T(z) + \bar{T}(\bar{z}),
\eea 
one can rewrite the Hamiltonian of the CFT as
\bea
H_0 &=& \frac{1}{2 \pi} \int_{- \infty}^{\infty} dx T_{\tau_E \tau_E} 
\cr && \cr
&=& \frac{1}{2 \pi i} \int_{- i \infty}^{i\infty} dz T(z) + \frac{1}{2 \pi i} \int_{-i \infty}^{i \infty} d \bar{z} \bar{T}(\bar{z})
\cr && \cr
&=& \frac{1}{2 \pi i} \oint dz T(z) + \frac{1}{2 \pi i} \oint d \bar{z} \bar{T}(\bar{z}).
\label{H0-stress-tensor}
\eea 
Note that we slightly shift the vertical contour in the positive real direction and close it counterclockwise.
Next, by using the definition of the Virasoro generators as
\bea
L_n = \frac{1}{2 \pi i} \oint dz z^{n+1} T(z), \;\;\;\;\;\;\;\; \bar{L}_n = \frac{1}{2 \pi i} \oint d \bar{z} \bar{z}^{n+1} \bar{T}(\bar{z}),
\eea 
one obtains
\bea
H_0 = L_{-1} + \bar{L}_{-1}.
\label{H0-Virasoro}
\eea 
On the other hand, one can rewrite the Rindler Hamiltonian in \eqref{H0-HR} as
\bea
H_R &=& \frac{a}{2 \pi} \int_{- \infty}^{\infty} dx x T_{\tau_E \tau_E} 
\cr && \cr
&=& - \frac{a}{2 \pi} \int_{- i \infty}^{i\infty} dz z T(z) + \frac{a}{2 \pi} \int_{-i \infty}^{i \infty} d \bar{z} \bar{z} \bar{T}(\bar{z})
\cr && \cr
&=& - i a \left( \frac{1}{2 \pi i} \oint dz z T(z) - \frac{1}{2 \pi i} \oint d \bar{z} \bar{z} \bar{T}(\bar{z}) \right).
\label{HR-stress-tensor}
\eea 
From which, one has
\bea
H_R = \alpha \left( L_0 - \bar{L}_0 \right),
\label{HR-Virasoro}
\eea 
where we defined 
\bea
\alpha= -i a.
\label{alpha-a}
\eea 
Next, by applying \eqref{H0-Virasoro} and \eqref{HR-Virasoro}, one obtains
\bea
[H_R, H_0] \neq 0.
\eea 
In the following, we obtain the time evolutions of the primary operators.

\subsection{OPE Method}
\label{Sec: Method 1}

In this section, we use the OPE of the energy-momentum tensor and the primary operator to find the behavior of the primary operator under the time evolutions by $H_0$ and $H_R$. It should be pointed out that this method was applied in ref. \cite{Bai:2024azk} for the time evolution by a q-M\"obius Hamiltonian. By applying \eqref{H0-stress-tensor} and the OPE between the energy-momentum tensor and the primary operator $\mathcal{O}(z, \bar{z})$ \cite{YellowBook} 
\bea
T(z') \mathcal{O}(z, \bar{z}) = \frac{h_{\mathcal{O}}}{(z' - z)^2} \mathcal{O}(z, \bar{z}) + \frac{1}{(z' - z)} \partial_z \mathcal{O}(z, \bar{z}) + \cdots,
\label{OPE-T-O}
\eea 
one arrives at
\bea
[H_0, \mathcal{O}(z, \bar{z})] 
&=& \frac{1}{2 \pi i} \oint dz' \Big[  \frac{h_{\mathcal{O}}}{(z' - z)^2} \mathcal{O}(z, \bar{z}) + \frac{1}{(z' - z)} \partial_z \mathcal{O}(z, \bar{z}) \Big] + \text{anti-chiral}
\cr && \cr
&=& \left( \partial_z + \partial_{\bar{z}} \right) \mathcal{O}(z, \bar{z}).
\label{H0-O-1}
\eea 
Here, 
$\mathcal{R}$ is the radial ordering operator. Moreover, one obtains
\bea
[ H_0, [H_0, \mathcal{O}(z, \bar{z})] ] 
&=& \frac{1}{2 \pi i} \oint dz' \Big[ \frac{2 h_{\mathcal{O}}}{(z' - z)^3} \mathcal{O}(z, \bar{z}) + \frac{(h_{\mathcal{O}} +1)}{(z' - z)^2} \partial_z \mathcal{O}(z, \bar{z}) + \frac{1}{(z' - z)} \partial_z^2 \mathcal{O}(z, \bar{z}) \Big] 
\cr && \cr 
&+& \frac{1}{2 \pi i} \oint dz' \Big[ \frac{h_{\mathcal{O}}}{(z' - z)^2} \partial_{\bar{z}} \mathcal{O}(z, \bar{z}) + \frac{1}{(z' - z)} \partial_{\bar{z}} \partial_z \mathcal{O}(z, \bar{z}) \Big] + \text{anti-chiral}
\cr && \cr
& = & \left( \partial_z + \partial_{\bar{z}} \right)^2 \mathcal{O}(z, \bar{z}).
\label{H0-O-2}
\eea 
Next, by applying the Baker-Campbell-Hausdorff (BCH) formula
\bea
e^A B e^{-A} = B + [A, B] + \frac{1}{2!} [ A, [A,B] ] + \frac{1}{3!} [A, [A, [A,B]]] + \cdots,
\label{BCH-formula}
\eea 
as well as \eqref{H0-O-1} and \eqref{H0-O-2}, one simply finds
\bea
e^{\tau_E H_0} \mathcal{O}(z, \bar{z}) e^{- \tau_E H_0} &=& \mathcal{O}(z, \bar{z}) + \tau_E \left( \partial_z + \partial_{\bar{z}} \right) \mathcal{O}(z, \bar{z}) + \frac{1}{2} \tau_E^2 \left( \partial_z + \bar{\partial}_{\bar{z}} \right)^2 \mathcal{O}(z, \bar{z}) + \cdots
\cr && \cr
&=& e^{ \tau_E \left( \partial_z + \partial_{\bar{z}} \right)} \mathcal{O}(z, \bar{z})
=  \mathcal{O}(z + \tau_E, \bar{z} + \tau_E).
\label{time-evolution-H0}
\eea 
Note that in the last line, we applied the following identity
\bea
e^{ \tau_E \partial_z} f(z) = f(z+ \tau_E).
\label{translation}
\eea 
Therefore, $H_0$ generates time translation, as it was expected. On the other hand, for the Rindler Hamiltonian by using \eqref{HR-stress-tensor} and \eqref{OPE-T-O}, one obtains
\bea
[H_R, \mathcal{O}(z, \bar{z})] &=& 
\frac{\alpha}{2 \pi i} \oint dz' z' \Big[  \frac{h_{\mathcal{O}}}{(z' - z)^2} \mathcal{O}(z, \bar{z}) + \frac{1}{(z' - z)} \partial_z \mathcal{O}(z, \bar{z}) \Big] - \text{anti-chiral}
\cr && \cr
&=& \alpha \bigg[ \frac{1}{z^{h_{\mathcal{O}}-1}} \vec{\partial}_z (z^{h_{\mathcal{O}}}) - \frac{1}{\bar{z}^{\bar{h}_{\mathcal{O}}-1}} \vec{\partial}_{\bar{z}} (\bar{z}^{\bar{h}_{\mathcal{O}}} ) \bigg] \mathcal{O}(z, \bar{z}).
\label{HR-O-1}
\eea 
In the above expression, $\vec{\partial}_z$ and $\vec{\partial}_{\bar{z}}$ are acting on all of the functions on the right hand side. Moreover, by applying \eqref{HR-O-1}, one obtains
\bea
[ H_R, [H_R, \mathcal{O}(z, \bar{z})] ] 
&=& \frac{1}{2 \pi i} \oint dz' z' \Bigg[ \alpha^2 h_{\mathcal{O}} \Big[ \frac{h_{\mathcal{O}}}{(z'-z)^2} \mathcal{O}(z, \bar{z}) 
+ \frac{1}{(z'-z)} \partial_z \mathcal{O}(z, \bar{z}) \Big]
\cr && \cr
&+& \alpha^2 z \Big[ \frac{2 h_{\mathcal{O}}}{(z' - z)^3} \mathcal{O}(z, \bar{z}) + \frac{(h_{\mathcal{O}} +1)}{(z' - z)^2} \partial_z \mathcal{O}(z, \bar{z}) + \frac{1}{(z' - z)} \partial_z^2 \mathcal{O}(z, \bar{z}) \Big]
\cr && \cr
&-& \alpha^2 \bar{h}_{\mathcal{O}} \Big[ \frac{h_{\mathcal{O}}}{(z'-z)^2} \mathcal{O}(z, \bar{z}) 
+ \frac{1}{(z'-z)} \partial_z \mathcal{O}(z, \bar{z}) \Big]
\cr && \cr
&-& \alpha^2 \bar{z} \Big[ \frac{h_{\mathcal{O}}}{(z'-z)^2} \partial_{\bar{z}} \mathcal{O}(z, \bar{z}) 
+ \frac{1}{(z'-z)}  \partial_{\bar{z}} \partial_z \mathcal{O}(z, \bar{z}) \Big]
\Bigg] + \text{anti-chiral}
\cr && \cr
&=& \alpha^2 \Bigg[ \frac{1}{z^{h_{\mathcal{O}}-1}} \vec{\partial}_z (z^{h_{\mathcal{O}}}) - \frac{1}{\bar{z}^{\bar{h}_{\mathcal{O}}-1}} \vec{\partial}_{\bar{z}} (\bar{z}^{\bar{h}_{\mathcal{O}}} ) \Bigg]^2 \mathcal{O}(z, \bar{z}).
\label{HR-O-2}
\eea 
Next, by exploiting \eqref{BCH-formula}, \eqref{HR-O-1} and \eqref{HR-O-2}, one has
\bea
e^{\tau_E H_R} \mathcal{O}(z, \bar{z}) e^{- \tau_E H_R} &=& 
e^{ \frac{\alpha \tau_E}{z^{h_{\mathcal{O}}-1}} \vec{\partial}_z (z^{h_{\mathcal{O}}}) } e^{- \frac{\alpha \tau_E}{\bar{z}^{\bar{h}_{\mathcal{O}}- 1} } \vec{\partial}_{\bar{z}} (\bar{z}^{\bar{h}_{\mathcal{O}} }) } \mathcal{O}(z, \bar{z})
\cr && \cr
&=& \frac{1}{z^{h_{\mathcal{O}}} \bar{z}^{\bar{h}_{\mathcal{O}}} } e^{ \alpha \tau_E z \vec{\partial}_z} e^{- \alpha \tau_E \bar{z} \vec{\partial}_{\bar{z}} } \left( z^{h_{\mathcal{O}}} \bar{z}^{\bar{h}_{\mathcal{O}}} \mathcal{O} (z, \bar{z}) \right).
\label{time-evolution-HR-z}
\eea 
On the other hand, by changing the coordinates to $z= e^{y}$, one has \cite{Bai:2024azk}
\bea
z \partial_z = \partial_y.
\label{z-y}
\eea 
Then, by using \eqref{translation} and \eqref{z-y}, one can rewrite \eqref{time-evolution-HR-z} as
\bea
e^{\tau_E H_R} \mathcal{O}(z, \bar{z}) e^{- \tau_E H_R} &=& \frac{1}{e^{y h_{\mathcal{O}} } e^{\bar{y} \bar{h}_{\mathcal{O}}} } e^{ \alpha \tau_E \vec{\partial}_y} 
e^{- \alpha \tau_E \vec{\partial}_{\bar{y}} } \left( e^{y h_{\mathcal{O}}} e^{\bar{y} \bar{h}_{\mathcal{O}} } \mathcal{O} \left( e^y , e^{\bar{y}} \right) \right)
\cr && \cr 
&=& e^{ \alpha \tau_E (h_{\mathcal{O}} - \bar{h}_{\mathcal{O}}) } \mathcal{O} \left( e^{\alpha \tau_E} z , e^{- \alpha \tau_E } \bar{z} \right).
\label{time-evolution-HR-y}
\eea 
Therefore, the effect of the time evolution under $H_R$ is similar to a rotation on the complex coordinates $(z, \bar{z})$.

\subsection{$\mathcal{O}^{H}_{n,1}(z, \bar{z})$ and $\mathcal{O}^{H}_{n,2}(z, \bar{z})$}
\label{Sec: O-H-n-1-2}

After finding the behavior of the primary operator $\mathcal{O}(z, \bar{z})$ under the time evolutions by the Hamiltonians $H_0$ and $H_R$ in the last section, we can easily simplify the expressions for the Heisenberg operators $\mathcal{O}^{H}_{n,1}(z, \bar{z})$ and $\mathcal{O}^{H}_{n,2}(z, \bar{z})$ which are defined in \eqref{O1-Heisneberg} and \eqref{O2-Heisneberg}. We first start with
\bea
\mathcal{O}^{H}_{n,1}(z, \bar{z}) = e^{- \epsilon H_R} e^{- H_0 \tau_E} \mathcal{O}_{n}(z, \bar{z}) e^{H_0 \tau_E} e^{\epsilon H_R}.
\eea 
By applying \eqref{time-evolution-H0} and \eqref{time-evolution-HR-z}, one obtains
\bea
\mathcal{O}^{H}_{n,1}(z, \bar{z}) = e^{n \alpha \epsilon (\bar{h}_\mathcal{O} - h_{\mathcal{O}})} \mathcal{O}_{n} \left( e^{- \alpha \epsilon} (z - \tau_E) , e^{\alpha \epsilon} (\bar{z} - \tau_E) \right).
\label{time evolution-O-n-1}
\eea 
Then, by defining
\bea
z_{1,\epsilon}^{\rm new} = e^{- \alpha \epsilon} (z - \tau_E), \;\;\;\;\;\;\; \bar{z}_{1, \epsilon}^{\rm new} = e^{ \alpha \epsilon} (\bar{z} - \tau_E),
\label{z1-new}
\eea 
one can rewrite \eqref{time evolution-O-n-1} as \cite{Wen_2018,Goto:2021sqx}
\bea
\mathcal{O}^{H}_{n,1}(z, \bar{z}) = \left( \frac{\partial z^{\rm new}_{1, \epsilon}}{\partial z} \right)^{n h_{\mathcal{O}}}
\left( \frac{\partial \bar{z}^{\rm new}_{1, \epsilon}}{\partial \bar{z} } \right)^{n \bar{h}_{\mathcal{O}}} \mathcal{O}_{n}(z^{\rm new}_{1, \epsilon}, \bar{z}^{\rm new}_{1, \epsilon} ).
\eea 
Here, $n h_{\mathcal{O}}$ and $n \bar{h}_{\mathcal{O}}$ are the conformal dimensions of $\mathcal{O}^{H}_{n,1}(z, \bar{z})$. On the other hand, for
\bea
\mathcal{O}^{H}_{n,2}(z, \bar{z}) = e^{- H_0 \tau_E} e^{- \epsilon H_R} \mathcal{O}_{n}(z, \bar{z}) e^{\epsilon H_R} e^{H_0 \tau_E},
\eea 
by applying \eqref{time-evolution-H0} and \eqref{time-evolution-HR-z}, one has
\bea
\mathcal{O}^{H}_{n,2}(z, \bar{z}) = e^{n \alpha \epsilon (\bar{h}_\mathcal{O} - h_{\mathcal{O}})} \mathcal{O}_{n} \left( e^{- \alpha \epsilon} z - \tau_E , e^{\alpha \epsilon} \bar{z} - \tau_E \right).
\label{time evolution-O-n-2}
\eea 
Then, by defining
\bea
z_{2,\epsilon}^{\rm new} = e^{- \alpha \epsilon} z - \tau_E, \;\;\;\;\;\;\; \bar{z}_{2, \epsilon}^{\rm new} = e^{ \alpha \epsilon} \bar{z} - \tau_E,
\label{z2-new}
\eea 
one can rewrite \eqref{time evolution-O-n-2} as
\bea
\mathcal{O}^{H}_{n,2}(z, \bar{z}) = \left( \frac{\partial z^{\rm new}_{2, \epsilon}}{\partial z} \right)^{n h_{\mathcal{O}}}
\left( \frac{\partial \bar{z}^{\rm new}_{2, \epsilon}}{\partial \bar{z} } \right)^{n \bar{h}_{\mathcal{O}}} \mathcal{O}_{n}(z^{\rm new}_{2, \epsilon}, \bar{z}^{\rm new}_{2, \epsilon} ).
\eea 
Putting everything together, one can write
\bea
\mathcal{O}^{H}_{n,i}(z, \bar{z}) = \left( \frac{\partial z^{\rm new}_{i, \epsilon}}{\partial z} \right)^{n h_{\mathcal{O}}}
\left( \frac{\partial \bar{z}^{\rm new}_{i, \epsilon}}{\partial \bar{z} } \right)^{n \bar{h}_{\mathcal{O}}} \mathcal{O}_{n}(z^{\rm new}_{i, \epsilon}, \bar{z}^{\rm new}_{i, \epsilon} ).
\label{time evolution-conformal transform-O}
\eea 
Moreover, it is straightforward to verify that
\bea
\mathcal{O}^{H, \dagger}_{n,i}(z, \bar{z}) = \left( \frac{\partial z^{\rm new}_{i, - \epsilon}}{\partial z} \right)^{n h_{\mathcal{O}}}
\left( \frac{\partial \bar{z}^{\rm new}_{i, - \epsilon}}{\partial \bar{z} } \right)^{n \bar{h}_{\mathcal{O}}} \mathcal{O}^{\dagger}_{n}(z^{\rm new}_{i, - \epsilon}, \bar{z}^{\rm new}_{i, - \epsilon} ).
\label{time evolution-conformal transform-O-dagger}
\eea
At the end, by applying the definition of $\alpha$ in \eqref{alpha-a}, we can rewrite \eqref{z1-new} and \eqref{z2-new} as
\bea
&& z_{1,\epsilon}^{\rm new} = e^{i a \epsilon} (z - \tau_E), \;\;\;\;\;\;\; \bar{z}_{1, \epsilon}^{\rm new} = e^{- i a \epsilon} (\bar{z} - \tau_E),
\label{z1-new-a}
\\
&& z_{2,\epsilon}^{\rm new} = e^{i a \epsilon} z - \tau_E, \;\;\;\;\;\;\;\;\;\; \bar{z}_{2, \epsilon}^{\rm new} = e^{- i a \epsilon} \bar{z} - \tau_E.
\label{z2-new-a}
\eea 

\section{Entanglement Entropy for the Insertion into $A$
\label{Sec:Insetion-into-A}}

In this section, we will explore the time dependence of the entanglement entropy when the local operator is inserted into a subsystem $A \in [x_1, x_2]$
\footnote{Here, we assume that $2x< x_1 + x_2$, or equivalently $x- x_1 < x_2 -x$.}.
In this case, there are two OPE channels that contribute to the entanglement entropy. Moreover, 
the choice of the OPE channel corresponds to the choice of the branch cut. In our procedure, we take the minimal value of the OPE channels as the entanglement entropy. 
By the same reasoning which was explained in Section \ref{Sec:EE-for-finite-interval}, for $i=1$, and for the first channel, one chooses the cross ratios as
\bea
(\eta_i, \bar{\eta}_i) \rightarrow
\begin{cases}
(e^{2 \pi i} \eta_i, e^{-2 \pi i} \bar{\eta}_i), &~~ 0 < t < x -x_1, \\
(\eta_i, e^{-2 \pi i} \bar{\eta}_i), &~~  x- x_1 < t < x_2- x, \\
(\eta_i, \bar{\eta}_i), &~~ t>  x_2- x.
\end{cases}
\label{cross-ratio-first-channel}
\eea 
It leads to the following expression for the entanglement entropy
\bea 
S_{A;1}= S_A^{\rm Vac}+
\begin{cases}
\frac{c}{6} \log \Big[ \frac{\kappa^2_{\mathcal{O}} ((x-x_1)^2- t^2) ((x_2-x)^2-t^2)}{a^2 \epsilon^2 l^2 (x^2-t^2)} \Big], &~~ 0 < t < x -x_1,
\vspace{1mm}
 \\
\frac{c}{6} \log \Big[ \frac{\kappa_{\mathcal{O}} (t+x -x_1) (x_2- x -t)}{a \epsilon l (t+x) }\Big], &~~  x-x_1 < t < x_2- x, \\
0, &~~ t>  x_2- x,
\end{cases}
\label{SA-1-finite-interval-inside-ER-first-channel}
\eea 
where $l=x_2-x_1$. Similarly, for the second channel, one chooses the cross ratios as
\bea
(\eta_i, \bar{\eta}_i) \rightarrow
\begin{cases}
(\eta_i, \bar{\eta}_i), &~~ 0 < t < x -x_1, \\
(e^{-2 \pi i} \eta_i, \bar{\eta}_i), &~~  x- x_1 < t < x_2- x, \\
(e^{-2 \pi i} \eta_i, e^{2 \pi i} \bar{\eta}_i), &~~ t>  x_2- x.
\end{cases}
\label{cross-ratio-second-channel}
\eea 
Then, one arrives at
\bea 
S_{A;1}= S_A^{\rm Vac}+
\begin{cases}
0, &~~ 0 < t < x -x_1, \\ 
\frac{c}{6} \log \Big[ \frac{\kappa_{\mathcal{O}} (t-x +x_1) (t-x+x_2)}{a \epsilon l (x-t) }\Big], &~~  x-x_1 < t < x_2- x, 
\vspace{1mm}
\\
\frac{c}{6} \log \Big[ \frac{\kappa^2_{\mathcal{O}} (t^2 - (x-x_1)^2) (t^2- (x_2-x)^2)}{a^2 \epsilon^2 l^2 (x^2-t^2)} \Big], &~~ t>  x_2- x.
\end{cases}
\label{SA-1-finite-interval-inside-ER-second-channel}
\eea 
Next, by choosing the minimum value between \eqref{SA-1-finite-interval-inside-ER-first-channel} and \eqref{SA-1-finite-interval-inside-ER-second-channel}, one obtains the time dependence of the entanglement entropy as
\bea 
S_{A;1}= S_A^{\rm Vac}+
\begin{cases}
0, &~~ 0 < t < x -x_1, \\
\frac{c}{6} \log \Big[\frac{\kappa_{\mathcal{O}} (t-x +x_1) (t- x +x_2) }{a \epsilon l (x- t) } \Big], &~~  x- x_1 < t < \sqrt{\frac{x (x- x_1) (x_2 -x)}{x_1 + x_2 -x}}, \\
\frac{c}{6} \log \Big[\frac{\kappa_{\mathcal{O}} (t+x -x_1) (x_2- x -t)}{a \epsilon l (t+x) }\Big], &~~  \sqrt{\frac{x (x- x_1) (x_2 -x)}{x_1 + x_2 -x}} < t < x_2- x, \\
0, &~~ t>  x_2- x.
\end{cases}
\label{SA-1-finite-interval-inside-ER}
\eea
In this case, the phase transition between the two channels happens at $t= \sqrt{\frac{x (x- x_1) (x_2 -x)}{x_1 + x_2 -x}}$. 

On the other hand, for $i=2$, the cross ratios in the first and second channels are again given by \eqref{cross-ratio-first-channel} and \eqref{cross-ratio-second-channel}. It is straightforward to show that for the first channel one has
\bea 
S_{A;2}= S_A^{\rm Vac}+
\begin{cases}
\frac{c}{6} \log \Big[ \frac{\kappa^2_{\mathcal{O}} ((x-x_1)^2- t^2) ((x_2-x)^2-t^2)}{a^2 \epsilon^2 l^2 x^2} \Big], &~~ 0 < t < x -x_1,
\vspace{1mm}
 \\
\frac{c}{6} \log \Big[ \frac{\kappa_{\mathcal{O}} (t+x -x_1) (x_2- x -t)}{a \epsilon l x }\Big], &~~  x-x_1 < t < x_2- x, \\
0, &~~ t>  x_2- x,
\end{cases}
\label{SA-2-finite-interval-inside-ER-first-channel}
\eea 
and for the second channel one obtains
\bea 
S_{A;2}= S_A^{\rm Vac}+
\begin{cases}
0, &~~ 0 < t < x -x_1, \\ 
\frac{c}{6} \log \Big[ \frac{\kappa_{\mathcal{O}} (t-x +x_1) (t-x+x_2)}{a \epsilon l x }\Big], &~~  x-x_1 < t < x_2- x,
\vspace{1mm}
 \\
\frac{c}{6} \log \Big[ \frac{\kappa^2_{\mathcal{O}} (t^2 - (x-x_1)^2) (t^2- (x_2-x)^2)}{a^2 \epsilon^2 l^2 x^2} \Big], &~~ t>  x_2- x.
\end{cases}
\label{SA-2-finite-interval-inside-ER-second-channel}
\eea 
Next, by choosing the minimum value of the expressions in \eqref{SA-2-finite-interval-inside-ER-first-channel} and \eqref{SA-2-finite-interval-inside-ER-second-channel}, one arrives at 
\bea
S_{A;2}= S_A^{\rm Vac} +
\begin{cases}
0, &~~ 0 < t < x- x_1, \\
\frac{c}{6} \log \Big[\frac{\kappa_{\mathcal{O}} (t-x +x_1) (t- x +x_2) }{a \epsilon l x } \Big], &~~  x- x_1 < t < \sqrt{(x- x_1) (x_2 -x)}, 
\vspace{1mm}
\\
\frac{c}{6} \log \Big[\frac{\kappa_{\mathcal{O}} (t+x -x_1) (x_2- x -t)}{a \epsilon l x }\Big], &~~  \sqrt{(x- x_1) (x_2 -x)} < t < x_2- x, \\
0, &~~ t>  x_2- x.
\end{cases}
\label{SA-2-finite-interval-inside-ER}
\eea 
In this case, the phase transition between the two channels occurs at $t= \sqrt{(x- x_1) (x_2 -x)}$.

\section{Asymptotic Metric}
\label{Sec: Asymptotic Metric}

In this section, we derive the asymptotic metric in \eqref{asymptotic-metric-2}. As explained in Section \ref{Sec:Shape-of-Horizon}, by plugging \eqref{Banados map} into \eqref{map BTZ to AdS3} and applying \eqref{boundary-map-BTZ}, one can find the coordinates $(r, \tau, \phi)$ of the BTZ geometry as functions of the coordinates $(y, z, \bar{z})$ of the Ba$\tilde{\rm n}$ados geometry as
\bea
\label{r-y-z-zb}
&& r(y, z, \bar{z})^2 = 
- \left( \frac{r_h}{y \; \alpha_{\mathcal{O}} } \right)^2 \Bigg{\lvert} \frac{(z - z^{\rm new}_{i, - \epsilon}) (z- z^{\rm new}_{i, \epsilon})}{(z^{\rm new}_{i, - \epsilon} - z^{\rm new}_{i, \epsilon})} \Bigg{\rvert}^2
\\
&& \;\;\;\;\;\;\;\;\;\;\;\;\;\;\;\;\;\;\;\; \times 
\Bigg[ 1 + \frac{y^2 \left( \big{\lvert} 2z - z^{\rm new}_{i, - \epsilon} - z^{\rm new}_{i, \epsilon} \big{\rvert}^2 - \alpha_{\mathcal{O}}^2 \big{\lvert} z^{\rm new}_{i, - \epsilon} - z^{\rm new}_{i, \epsilon} \big{\rvert}^2 \right)}{2 \big{\lvert} z - z^{\rm new}_{i, - \epsilon} \big{\rvert}^2 \big{\lvert} z- z^{\rm new}_{i, \epsilon} \big{\rvert}^2}
\cr && \cr
&& \;\;\;\;\;\;\;\;\;\;\;\;\;\;\;\;\;\;\;\;\;\;\;\;\; + \frac{y^4 \Big{\lvert} \left(  2z- z^{\rm new}_{i, - \epsilon} - z^{\rm new}_{i, \epsilon} \right)^2 - \alpha_{\mathcal{O}}^2 \left( z^{\rm new}_{i, - \epsilon} - z^{\rm new}_{i, \epsilon} \right)^2 \Big{\rvert}^2 }{16 \big{\lvert} z - z^{\rm new}_{i, - \epsilon} \big{\rvert}^4 \big{\lvert} z- z^{\rm new}_{i, \epsilon} \big{\rvert}^4} \Bigg],
\nonumber
\eea 

\bea
\label{tau-y-z-zb}
&& \tau (y , z, \bar{z}) =
\left( \frac{L_{AdS}^2}{2 i r_h} \right) \log \left[ \Bigg{\lvert} \frac{z - z^{\rm new}_{i, - \epsilon}}{z- z^{\rm new}_{i, \epsilon}} \Bigg{\rvert}^{2 \alpha_{\mathcal{O}}} \right]
\\
&&\;\;\;\;\;\;\;\;\;\;\;\;\;\;\;\;\;\; +
\left( \frac{L_{AdS}^2}{2 i r_h} \right) \log \left[ \frac{1 +
y^2 \left( \frac{(2z + z^{\rm new}_{i, - \epsilon} (\alpha_{\mathcal{O}}-1)- z^{\rm new}_{i, \epsilon} (1+ \alpha_{\mathcal{O}}) )
(2 \bar{z} + \bar{z}^{\rm new}_{i, - \epsilon} (\alpha_{\mathcal{O}}- 1)- \bar{z}^{\rm new}_{i, \epsilon} (1+ \alpha_{\mathcal{O}}) )}{4 \big{\lvert} z - z^{\rm new}_{i, - \epsilon} \big{\rvert}^2 \big{\lvert} z- z^{\rm new}_{i, \epsilon} \big{\rvert}^2} \right)}
{1 +
y^2 \left( \frac{(2z - z^{\rm new}_{i, - \epsilon} (1+ \alpha_{\mathcal{O}})+ z^{\rm new}_{i, \epsilon} (\alpha_{\mathcal{O}}- 1) )
(2 \bar{z} - \bar{z}^{\rm new}_{i, - \epsilon} (1+ \alpha_{\mathcal{O}})+ \bar{z}^{\rm new}_{i, \epsilon} (\alpha_{\mathcal{O}}- 1) )}{4 \big{\lvert} z - z^{\rm new}_{i, - \epsilon} \big{\rvert}^2 \big{\lvert} z- z^{\rm new}_{i, \epsilon} \big{\rvert}^2} \right)} \right],
\nonumber
\eea 

\bea
\label{phi-y-z-zb}
&&\phi (y, z, \bar{z}) = 
\left( \frac{L_{AdS}}{2 r_h} \right) \log \Bigg[ \frac{(z - z^{\rm new}_{i, - \epsilon}) (\bar{z} - \bar{z}^{\rm new}_{i, \epsilon}) }{(z - z^{\rm new}_{i, \epsilon}) (\bar{z} - \bar{z}^{\rm new}_{i, - \epsilon})} \Bigg]^{\alpha_{\mathcal{O}} } 
\\
\!\!\!\!\!\!\!\!\! && + \left( \frac{L_{AdS}}{2 r_h} \right) \log \left[ \frac{1 + \frac{y^2 \left( \big{\lvert} 2z - z^{\rm new}_{i, - \epsilon} - z^{\rm new}_{i, \epsilon} \big{\rvert}^2 - \alpha_{\mathcal{O}}^2 \big{\lvert} z^{\rm new}_{i, - \epsilon} - z^{\rm new}_{i, \epsilon} \big{\rvert}^2 \right)}{2 \big{\lvert} z - z^{\rm new}_{i, - \epsilon} \big{\rvert}^2 \big{\lvert} z- z^{\rm new}_{i, \epsilon} \big{\rvert}^2}
+ \frac{y^4 \big{\lvert} \left(  2z- z^{\rm new}_{i, - \epsilon} - z^{\rm new}_{i, \epsilon} \right)^2 - \alpha_{\mathcal{O}}^2 \left( z^{\rm new}_{i, - \epsilon} - z^{\rm new}_{i, \epsilon} \right)^2 \big{\rvert}^2 }{16 \big{\lvert} z - z^{\rm new}_{i, - \epsilon} \big{\rvert}^4 \big{\lvert} z- z^{\rm new}_{i, \epsilon} \big{\rvert}^4}
}{1+ \frac{y^2 \big{\lvert} 2z- z^{\rm new}_{i, - \epsilon} (1+ \alpha_{\mathcal{O}}) + z^{\rm new}_{i, \epsilon} (\alpha_{\mathcal{O}} - 1) \big{\rvert}^2}{4 \big{\lvert} z - z^{\rm new}_{i, - \epsilon} \big{\rvert}^2 \big{\lvert} z- z^{\rm new}_{i, \epsilon} \big{\rvert}^2} }
\right].
\nonumber 
\eea 
From the above transformations, one can easily obtain the following expressions to the leading order in $y \approx 0$
\footnote{Note that in the limit $r \rightarrow \infty$ which we are interested in here, one has $y \rightarrow 0$.}
\bea
&& r(y, z, \bar{z}) = \left( \frac{L_{AdS}}{y} \right) \Bigg{\lvert} \frac{(z - z^{\rm new}_{i, - \epsilon}) (z- z^{\rm new}_{i, \epsilon})}{(z^{\rm new}_{i, - \epsilon} - z^{\rm new}_{i, \epsilon})} \Bigg{\rvert} + \mathcal{O}(y) + \cdots,
\label{r-y-z-zb-asymptotic}
\\ 
&& \tau(y , z, \bar{z}) = - L_{AdS} \log \Bigg{\lvert} \frac{z - z^{\rm new}_{i, - \epsilon}}{z- z^{\rm new}_{i, \epsilon}} \Bigg{\rvert} + \mathcal{O}(y^2) + \cdots,
\label{tau-y-z-zb-asymptotic}
\\
&& \phi(y, z, \bar{z}) = - \left( \frac{i}{2} \right) \log \Bigg[ \frac{(z - z^{\rm new}_{i, - \epsilon}) (\bar{z} - \bar{z}^{\rm new}_{i, \epsilon}) }{(z - z^{\rm new}_{i, \epsilon}) (\bar{z} - \bar{z}^{\rm new}_{i, - \epsilon})} \Bigg] + \mathcal{O}(y^2) + \cdots.
\label{phi-y-z-zb-asymptotic}
\eea 
It should be pointed out that we applied \eqref{rh-alpha} to obtain the above expressions. Then, from \eqref{r-y-z-zb}, one can simply find $y$ as a function of $r, z$ and $\bar{z}$
\bea
y(r, z, \bar{z}) &=& \frac{2 \sqrt{2} \big| z- z^{\rm new}_{i, - \epsilon} \big| \big| z- z^{\rm new}_{i, \epsilon} \big| \big| z^{\rm new}_{i, - \epsilon} - z^{\rm new}_{i, \epsilon} \big|} {L_{AdS} \Big{\lvert} \left( 2z - z^{\rm new}_{i, - \epsilon} - z^{\rm new}_{i, \epsilon} \right)^2 - \alpha_{\mathcal{O}}^2 \left( z^{\rm new}_{i, - \epsilon} - z^{\rm new}_{i, \epsilon} \right)^2 \Big{\rvert}}
\cr && \cr
&& \times
\Bigg[ r^2 - \frac{L_{AdS}^2 \big| 2z - z^{\rm new}_{i, - \epsilon} - z^{\rm new}_{i, \epsilon} \big|^2 - \alpha_{\mathcal{O}}^2 \big| z^{\rm new}_{i, - \epsilon} - z^{\rm new}_{i, \epsilon} \big|^2}{2 \big| z^{\rm new}_{i, - \epsilon} - z^{\rm new}_{i, \epsilon} \big|^2}
\cr && \cr 
&&- \bigg[ \left( r^2 - \frac{L_{AdS}^2 \big| 2z - z^{\rm new}_{i, - \epsilon} - z^{\rm new}_{i, \epsilon} \big|^2 - \alpha_{\mathcal{O}}^2 \big| z^{\rm new}_{i, - \epsilon} - z^{new}_{i, \epsilon} \big|^2}{2 \big| z^{\rm new}_{i, - \epsilon} - z^{\rm new}_{i, \epsilon} \big|^2} \right)^2 
\cr && \cr
&& - \frac{L_{AdS}^4 \Big{\lvert} \left( 2z - z^{\rm new}_{i, - \epsilon} - z^{\rm new}_{i, \epsilon} \right)^2 - \alpha_{\mathcal{O}}^2 \left( z^{\rm new}_{i, - \epsilon} - z^{\rm new}_{i, \epsilon} \right)^2 \Big{\rvert}^2}{4 \big| z^{\rm new}_{i, - \epsilon} - z^{\rm new}_{i, \epsilon} \big|^4}
\bigg]^{\frac{1}{2}}
\Bigg]^{\frac{1}{2}}.
\label{y-r-z-zb}
\eea 
Note that to the leading order in the limit $r \rightarrow \infty$, it reduces to 
\bea
y(r, z, \bar{z}) = \left( \frac{L_{AdS}}{r} \right) \Bigg{\lvert} \frac{(z - z^{\rm new}_{i, - \epsilon}) (z - z^{\rm new}_{i, \epsilon})}{(z^{\rm new}_{i, - \epsilon} - z^{\rm new}_{i, \epsilon})} \Bigg{\rvert} + \cdots.
\label{y-r-z-zb-asymptotic}
\eea 
Next, by replacing \eqref{y-r-z-zb} into \eqref{tau-y-z-zb} and \eqref{phi-y-z-zb}, one can find $\tau$ and $\phi$ in terms of $(r, z, \bar{z})$. Since the corresponding expressions are complicated, we do not write them here. To the leading order in the limit $r \rightarrow \infty$, $\tau(r, z, \bar{z})$ and $\phi(r, z, \bar{z})$ are given by \eqref{tau-y-z-zb-asymptotic} and \eqref{phi-y-z-zb-asymptotic}, respectively.
Then, by plugging the expressions for $\tau(r, z, \bar{z})$ and $\phi(r, z, \bar{z})$ into the BTZ metric \eqref{eq:BTZ-metric}, one can write the metric as a function of $(r, z, \bar{z})$. The full expression for the metric is very complicated and writing it here is not very illuminating. It is straightforward to show that 
the asymptotic metric is given by
\bea
ds^2 \mid_{r \rightarrow \infty} &=& \left( \frac{L_{AdS}}{r} \right)^2 dr^2 + r^2 \Bigg{\lvert} \frac{(z^{\rm new}_{i, - \epsilon} - z^{\rm new}_{i, \epsilon})}{(z - z^{\rm new}_{i, - \epsilon}) (z - z^{\rm new}_{i, \epsilon})} \Bigg{\rvert}^2 dz d\bar{z}
\cr && \cr
&=& \left( \frac{L_{AdS}}{r} \right)^2 dr^2 + r^2 \Bigg{\lvert} \frac{(z^{\rm new}_{i, - \epsilon} - z^{\rm new}_{i, \epsilon})}{(z - z^{\rm new}_{i, - \epsilon}) (z - z^{\rm new}_{i, \epsilon})} \Bigg{\rvert}^2 \left( d\tau_{E}^2 + dX^2 \right),
\label{asymptotic-metric-1}
\eea 
where $\tau_{E}=\f{z+\bar{z}}{2}$ and $X=\f{z-\bar{z}}{2i}$.

\section{Report on Time Evolution of Mutual Information \label{sec:R-TE-MI}}

In this section, we will report the rest of the time evolution of the holographic mutual information for configurations (b), (c) and (d) in Fig. \ref{fig: HMI-configurations}.

\subsection{Asymmetric Configuration}
\label{Sec: Operator Between the Subsystems A and B: Asymmetric Configuration}

First, we consider the asymmetric configuration where the operator is inserted between $A$ and $B$ and is closer to the subsystem $A$
(see Fig. \ref{fig: HMI-configurations}(b)). 
We take the locations of the subsystems and local operator to be $0 < x_1 < x_2 < x < x_3 < x_4$.
One can show that 
\bea
b-l < d+l-b < b < d+2l-b,
\eea 
where $b$ denotes the distance between $x$ and $x_1$ as
\bea
x-x_1 = b.
\label{b-asymmetric}
\eea 
Then, from \eqref{SA-i-finite-interval-outside-ER-LHS}, \eqref{SA-i-finite-interval-outside-ER-RHS}, \eqref{SA-1-finite-interval-inside-ER} and \eqref{SA-2-finite-interval-inside-ER}, one can find the connected and disconnected pieces of the entanglement entropies.
One can verify there is a phase transition at $d= d_i$ between $S_{{\rm dis};i}$ and $S_{{\rm con};i}$ in the time interval $t \in ( d+l-b, b)$. At this distance, the following equations are satisfied
\bea
&& (2l+d)^2 d^2 = \frac{ \kappa^2_{\mathcal{O}} \;  l^2 (b-t) (t- (b-l)) (t- (d+l-b)) (d+2l-b-t)}{a^2 \epsilon^2 (x^2 - t^2)}, ~~~ \text{for} ~~~~ i=1,
\cr && \cr
&& (2l+d)^2 d^2 = \frac{ \kappa^2_{\mathcal{O}} \;  l^2  (b-t) (t- (b-l)) (t- (d+l-b)) (d+2l-b-t)}{a^2 \epsilon^2 x^2}, ~~~ \text{for} ~~~~ i=2,
\nonumber
\\
\eea 
when $t \in ( d+l-b, b)$. Then, one can find the holographic mutual information. However, these expressions are complicated, and we do not report them here.
Moreover, one can verify that in some time intervals the holographic mutual information is at the order of $\mathcal{O}(\log{\left(1/a \epsilon\right)})$. In contrast, in other time intervals it is at the order of $\mathcal{O}(1)$. Therefore, its behavior cannot be interpreted by the quasiparticle picture as we discussed in Section \ref{Sec:quasiperticle-picture}. 

\begin{figure}
\begin{center}
\includegraphics[scale=0.33]{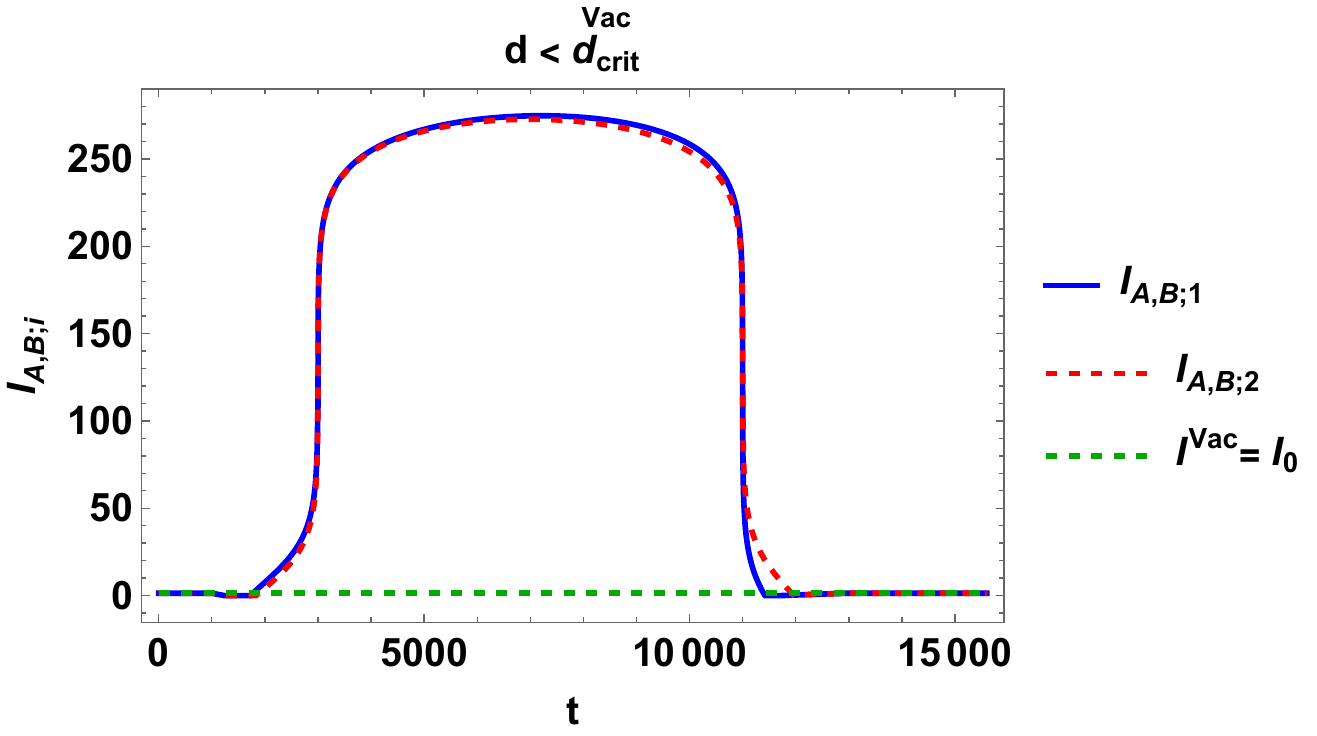}
\includegraphics[scale=0.33]{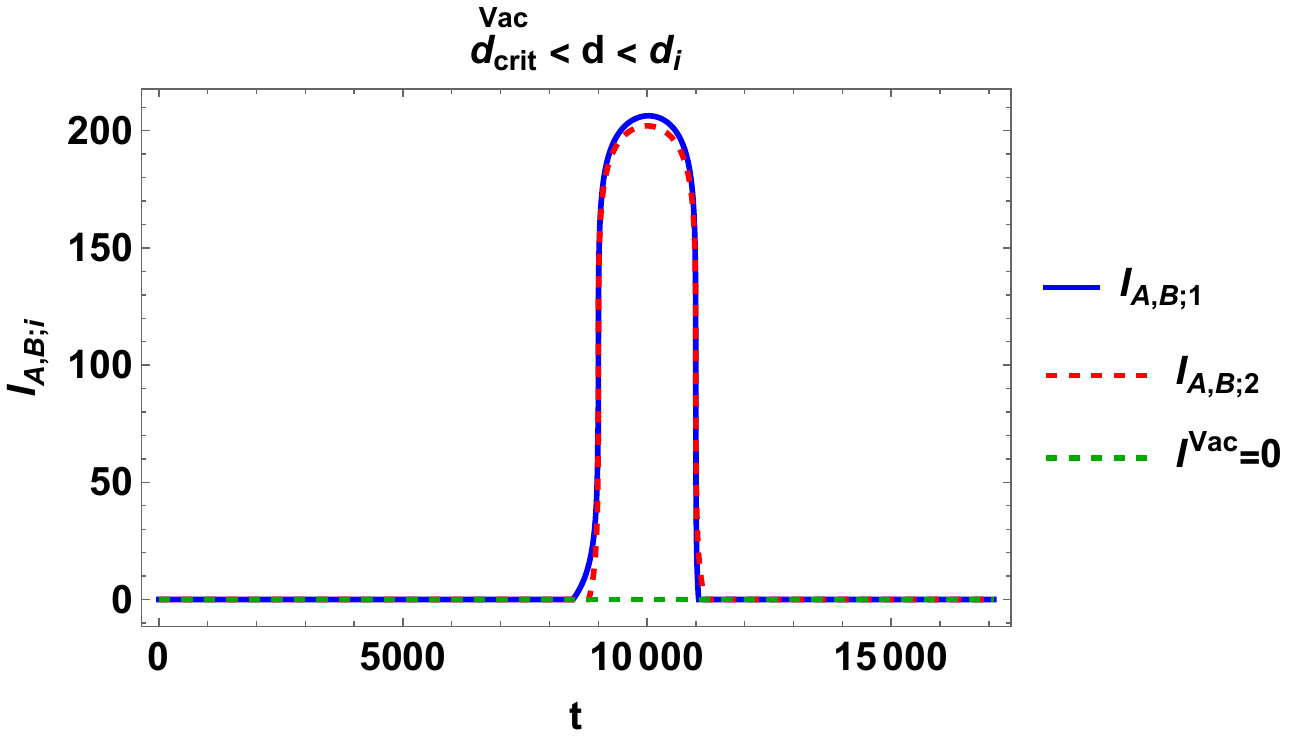}
\end{center}
\caption{  
$I_{A,B;i}$ as a function of time for the asymmetric configuration where the primary operator is inserted between the subsystems $A$ and $B$ and closer to $A$.
{\it Left)} For $d < d_{\rm crit}^{\rm Vac}$, $x_3= 2.4 \times 10^4$ and $x_4= 3.4 \times 10^4$. 
{\it Right)} For $d_{\rm crit}^{\rm Vac} < d < d_i$, $x_3= 3 \times 10^4$ and $x_4= 4 \times 10^4$. 
Moreover, we set $x= 2.1 \times10^4$, $x_1 = 10^4$, $x_2= 2 \times 10^4$, $a= 10^{-4}$, $\epsilon = 10$, $c= 100$ and $h_{\mathcal{O}} = \bar{h}_{\mathcal{O}}= 3 h_{0}$ where $h_{0} = \frac{c}{24}$. 
}
\label{fig: HMI-Asymmetric}
\end{figure}

To show the behavior of $I_{A,B;i}$, we plot it as a function of $t$ in Fig. \ref{fig: HMI-Asymmetric}.
In the left panel of Fig. \ref{fig: HMI-Asymmetric}, we chose the parameters such that $d< d_{\rm crit}^{\rm Vac}$. In this case, at early and late times, the holographic mutual information is equal to $I_{A,B}^{\rm Vac}= I_0$. Moreover, in the intermediate times, it is smaller or larger than that of the vacuum state. 
On the other hand, in the right panel of Fig. \ref{fig: HMI-Asymmetric}, we chose the parameters such that 
$ d_{\rm crit}^{\rm Vac} < d < d_i$.
In this case, at early and late times, $I_{A,B;i}$ is equal to $I_{A,B}^{\rm Vac}= 0$. Moreover, in the intermediate times, it is larger than that of the vacuum state. 

\subsection{Operator on the Left Hand Side of the Subsystems $A$ and $B$}
\label{Sec: Operator on the Left Hand Side of the Regions A and B}

In this section, we consider the case where the operator is inserted on the left hand side of the subsystems $A$ and $B$ (see Fig. \ref{fig: HMI-configurations}(c)). Therefore, one has $0 < x < x_1 < x_2 < x_3 < x_4$.
For convenience, we define 
\footnote{Here, we abused the notation for $b$. The definition of $b$ in \eqref{b-asymmetric} and \eqref{b-LHS} are different by a minus sign.}
\be
x_1 - x =b.
\label{b-LHS}
\ee
Then, one can apply \eqref{SA-i-finite-interval-outside-ER-LHS} and obtain the connected and disconnected pieces of the entanglement entropies.
Next, one can compare $S_{{\rm dis};i}$ with $S_{{\rm con};i}$ for $t \in \left( b+l, b+l+d \right)$. It is straightforward to verify that there is a phase transition at $d= \tilde{d}_i$ between $S_{{\rm dis};i}$ and $S_{{\rm con};i}$ when the following equations are satisfied
\bea
&& l^4 = \frac{ (2l+d) d \; \kappa_{\mathcal{O}}^2 (t-b) (b+2l+d -t) (t-b-l) (b+l+d -t)}{a^2 \epsilon^2 (t+x)^2}, ~~~ \text{for}~~~ i=1,
\cr && \cr
&& l^4 = \frac{ (2l+d) d \; \kappa_{\mathcal{O}}^2 (t-b) (b+2l+d -t) (t-b-l) (b+l+d -t)}{a^2 \epsilon^2 x^2}, ~~~ \text{for}~~~ i=2.\;\;
\eea 
Then, one can show that for $d< \tilde{d}_i$,
the holographic mutual information is given by 
\bea
I_{A,B;1} = I_0 + 
\begin{cases}
0 , &~ 0 < t < b, \\
\frac{c}{6} \log \Big[ \frac{(2l+d) (b+l-t)}{l (b+2l+d-t) }
\Big], &~  b < t < b+l,
\vspace{1mm} \\
\frac{c}{6} \log \Big[ \frac{ a^2 \epsilon^2 d (2l + d) (t+x)^2}{\kappa_{\mathcal{O}}^2 (t-b) (b+2l+d -t) (t-b-l) (b+l+d -t)}
\Big], &~ b+l < t <  b+l+d,
\vspace{1mm}\\
\frac{c}{6} \log \Big[ \frac{ (2l+d) (t-b-l-d)}{l (t-b)}
\Big], &~ b+l+d < t <  b+2l+d,
\\
0, &~ t >  b+2l+d,
\end{cases}
\label{I-1-outside-ER-LHS-A-d-smaller-d-tilde}
\eea
and
\bea
I_{A,B;2} = I_0 + 
\begin{cases}
0 , &~ 0 < t < b, \\
\frac{c}{6} \log \Big[ \frac{(2l+d) (b+l-t)}{l (b+2l+d-t) }
\Big], &~  b < t < b+l,
\vspace{1mm} \\
\frac{c}{6} \log \Big[ \frac{a^2 \epsilon^2 d (2l + d) x^2}{\kappa_{\mathcal{O}}^2 (t-b) (b+2l+d -t) (t-b-l) (b+l+d -t)}{}
\Big], &~ b+l < t <  b+l+d,
\vspace{1mm} \\
\frac{c}{6} \log \Big[ \frac{ (2l+d) (t-b-l-d)}{l (t-b)}
\Big], &~ b+l+d < t <  b+2l+d,
\\
0, &~ t >  b+2l+d.
\end{cases}
\label{I-2-outside-ER-LHS-A-d-smaller-d-tilde}
\eea
In this case, $I_{A,B;1}$ and $I_{A,B;2}$ are only different from each other when $t \in (b+l, b+l+d)$.

On the other hand, for $\tilde{d}_i <d < d_{\rm crit}^{\rm Vac} $, one obtains
\bea
I_{A,B;1} = I_{A,B;2} =
\begin{cases}
I_0 , &~ 0 < t < b, \\
\frac{c}{6} \log \Big[ \frac{(2l+d) (b+l-t)}{l (b+2l+d-t) }
\Big], &~  b < t < \tilde{t}_1, \\
0, &~ \tilde{t}_1 < t <  \tilde{t}_2,
\\
\frac{c}{6} \log \Big[ \frac{ (2l+d) (t-b-l-d)}{l (t-b)}
\Big], &~ \tilde{t}_2 < t <  b+2l+d,
\\
I_0, &~ t >  b+2l+d,
\end{cases}
\label{I-1-outside-ER-LHS-A-d-larger-d-tilde}
\eea
where $\tilde{t}_{1,2}$ are defined as 
\bea
&& \tilde{t}_1 = b+l+ \frac{(2l+d) d^2}{d^2+dl-l^2} < b+l,
\cr && \cr 
&& \tilde{t}_2 = b - \frac{l^3}{d^2+dl-l^2} > b+l+d.
\label{t-tilde-1-t-tilde-2}
\eea 
Therefore, $I_{A,B;1}$ and $I_{A,B;2}$ are the same in this case. 
Moreover, in \eqref{I-1-outside-ER-LHS-A-d-smaller-d-tilde}, \eqref{I-2-outside-ER-LHS-A-d-smaller-d-tilde} and \eqref{I-1-outside-ER-LHS-A-d-larger-d-tilde}, $I_{A,B;i}$ can be at the order of one in the small $\epsilon$-expansion. Thus, its behavior cannot be explained by the quasiparticle picture as we discussed in Section \ref{Sec:quasiperticle-picture}.
Furthermore, for $d> d_{\rm crit}^{\rm vac}$, one has $S_{{\rm dis};i} < S_{{\rm con};i}$ for all times. Therefore, $I_{A,B;i}$ is always zero.

\begin{figure}
\begin{center}
\includegraphics[scale=0.33]{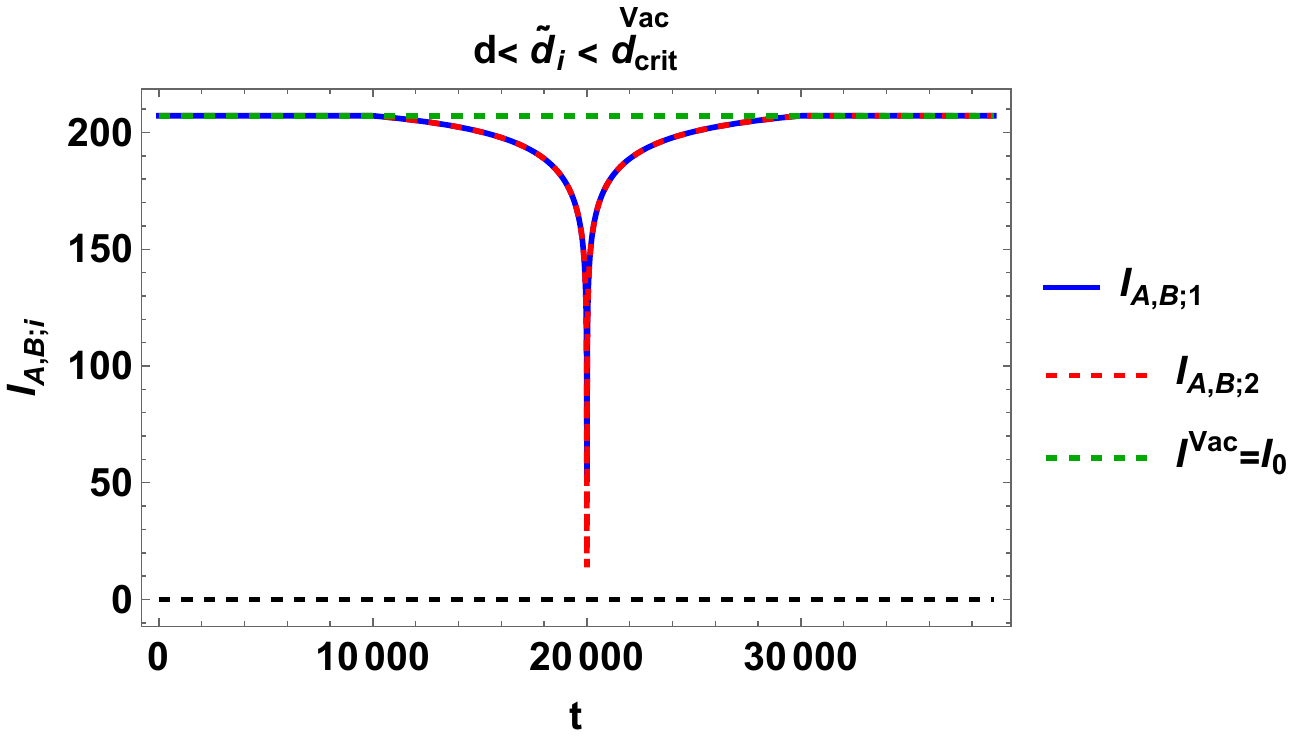}
\includegraphics[scale=0.33]{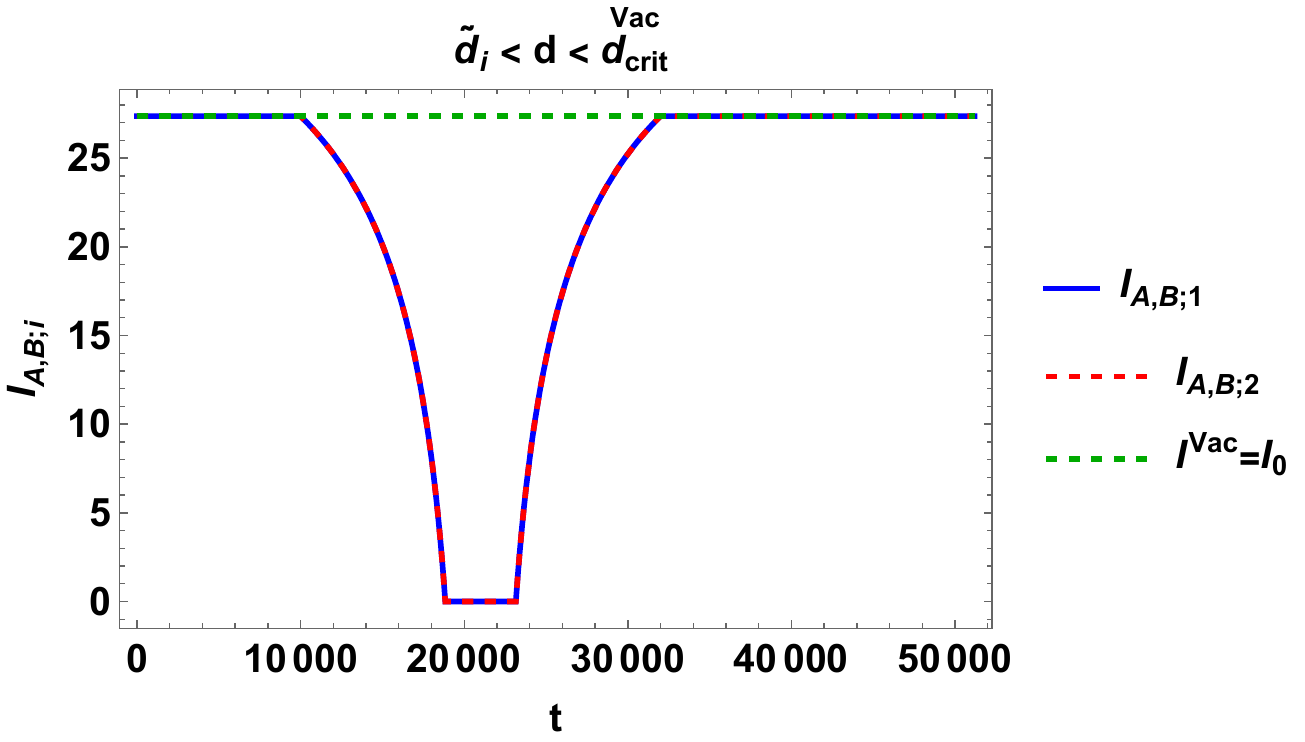}
\end{center}
\caption{  
$I_{A,B;i}$ as a function of time when the primary operator is inserted outside of the subsystems $A$ and $B$ and on their left hand sides.
{\it Left)} For $d < \tilde{d}_i < d_{\rm crit}^{\rm Vac}$, $x_3= 30010$ and $x_4= 40010$. 
{\it Right)} For $\tilde{d}_i < d < d_{\rm crit}^{\rm Vac}$, $x_3= 3.2 \times 10^4$ and $x_4= 4.2 \times 10^4$.
Here, we set $x=10^4$, $x_1= 2 \times 10^4$, $x_2= 3 \times 10^4$, $a= 10^{-4}$, $\epsilon = 10$, $c= 100$ and $h_{\mathcal{O}} = \bar{h}_{\mathcal{O}} = 3 h_{0}$.
In this case, the early and late time values of $I_{A,B;i}$ are non-zero and equal to $I_0$. 
}
\label{fig: HMI-LHS}
\end{figure}

To show the behavior of  $I_{A,B;i}$, we plot it as a function of $t$ in Fig. \ref{fig: HMI-LHS}. 
In these figures, which are for $d< d_{\rm crit}^{\rm Vac}$, the early and late time values of the holographic mutual information are equal to each other and given by $I_0$. Moreover, in the left panel of Fig. \ref{fig: HMI-LHS}, which is for $d < \tilde{d}_i$, the holographic mutual information is non-zero for all times. Furthermore, $I_{A,B;i=1,2}$ are only different in the time interval $t \in (b+l, b+l+d)$. In contrast, in the right panel of Fig. \ref{fig: HMI-LHS} which is for $\tilde{d}_i < d < d_{\rm crit}^{\rm Vac}$, the holographic mutual information is zero in the time interval $t \in \left( \tilde{t}_1, \tilde{t}_2 \right)$. 

\subsection{Operator Inside the Subsystem $A$}
\label{Sec: Operator Inside the Region A}

In this section, we consider the case where the operator is inserted inside the subsystem A (see Fig. \ref{fig: HMI-configurations}(d)). 
In this case, one has $0 < x_1 < x < x_2 < x_3 < x_4$. Moreover, the operator is inserted inside $[x_1 , x_2]$ and $[x_1, x_4]$. It is also located on the left hand side of $[x_2, x_3]$ and $[x_3, x_4]$. 
For simplicity, we define $b$ the same as \eqref{b-asymmetric}.
Furthermore, we assume
\footnote{It should be pointed out that we assume that for $i=1$, we have
$\sqrt{\frac{x b (d+2l-b)}{d+2(l-b)+x}} < l-b$. On the other hand, for $i=2$, we assume that $\sqrt{b (d+2l-b)} < l-b$. 
Otherwise, the time intervals in $S_{\rm con;i}$ should be changed. }
\bea
x- x_1 < x_2 - x, \;\;\;\;\;\; \text{or equivalently} \;\;\;\;\;\; b < l-b.
\eea
Then, we apply \eqref{SA-i-finite-interval-outside-ER-LHS} and \eqref{SA-1-finite-interval-inside-ER} to obtain the connected and disconnected pieces of the entanglement entropies.
By comparing $S_{{\rm dis};i}$ with $S_{{\rm con};i}$ for $t \in \left( l- b, d+ l -b \right)$, one can verify that there is a phase transition at $d= d'_i$ between $S_{{\rm dis};i}$ and $S_{{\rm con};i}$ when the following equations are satisfied
\bea
&& l^4 = \frac{ (2l+d) d \; \kappa_{\mathcal{O}}^2 (t+b) (d + 2l -b -t) (t- (l-b) ) (d+ l -b -t)}{a^2 \epsilon^2 (t+x)^2}, ~~~ \text{for} ~~~ i=1,
\cr && \cr
&& l^4 = \frac{ (2l+d) d \; \kappa_{\mathcal{O}}^2 (t+b) (d + 2l -b -t) (t- (l-b) ) (d+ l -b -t)}{a^2 \epsilon^2 x^2}, ~~~ \text{for} ~~~ i=2.\;\;\;\;\;
\eea 
Next, one can calculate the holographic mutual information. However, the expressions are complicated and we do not report them here. It should be pointed out that $I_{A,B;i}$ can be at the order of one in the small $\epsilon$-expansion. Therefore, its behavior cannot be described by the quasiparticle picture as we discussed in Section \ref{Sec:quasiperticle-picture}.

\begin{figure}
\begin{center}
\includegraphics[scale=0.33]{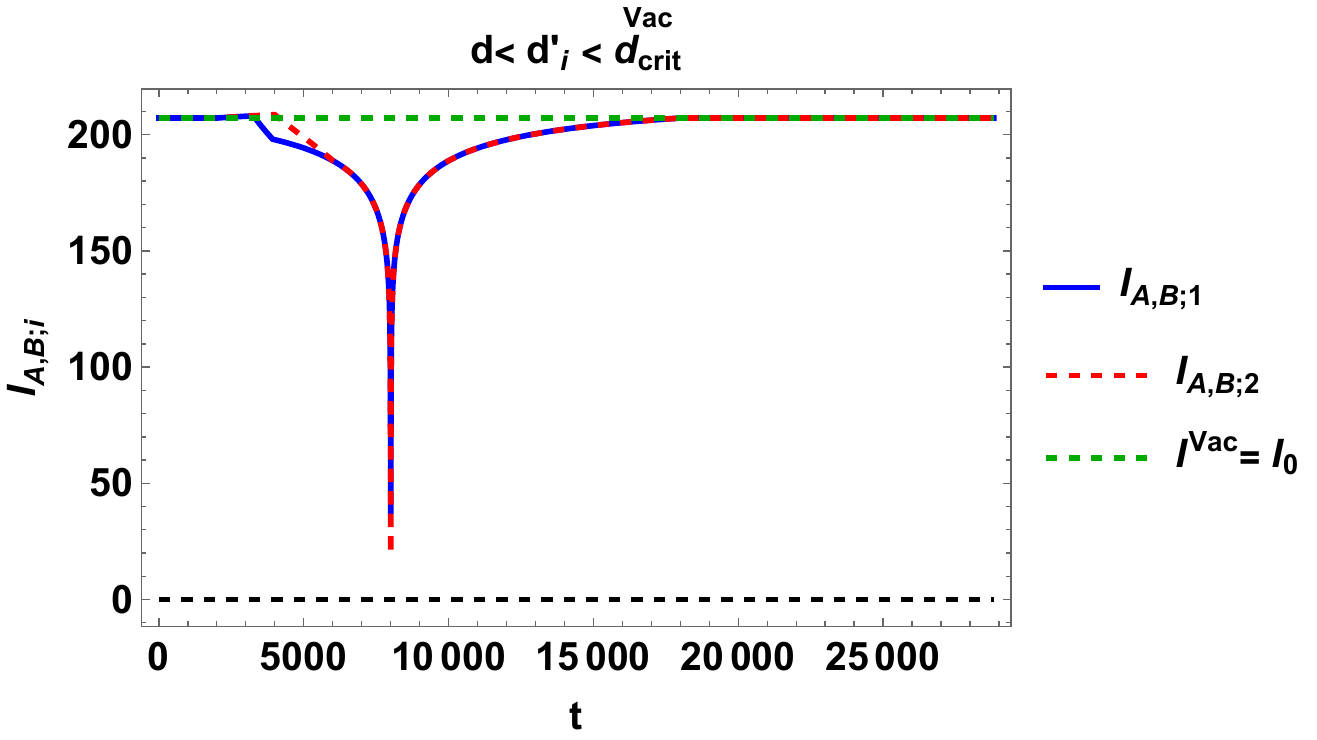}
\includegraphics[scale=0.33]{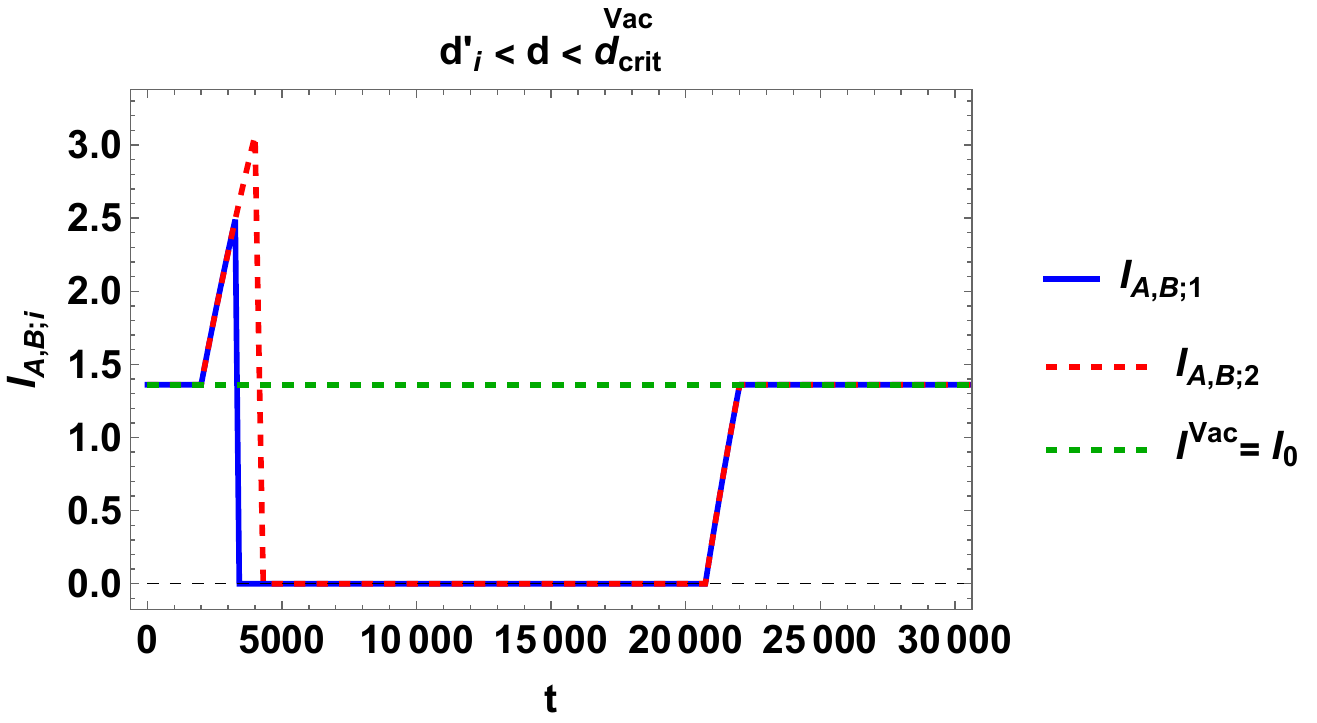}
\end{center}
\caption{
$I_{A,B;i}$ as a function of time when the primary operator is inserted inside the subsystem $A$. 
{\it Left)} For $d < d'_i < d_{\rm crit}^{\rm Vac}$, $x= 1.2 \times 10^4$, $x_3= 20010$ and $x_4= 30010$.
{\it Right)} For $d'_i < d < d_{\rm crit}^{\rm Vac}$, $x_3= 2.4 \times 10^4$ and $x_4= 3.4 \times 10^4$.
Here, we set $x_1= 10^4$, $x_2= 2 \times 10^4$, $a= 10^{-4}$, $\epsilon = 10$, $c= 100$ and $h_{\mathcal{O}}= \bar{h}_{\mathcal{O}} = 3 h_{0}$. 
}
\label{fig: HMI-inside}
\end{figure}

To show the behavior of the holographic mutual information, we plot it in Fig. \ref{fig: HMI-inside}. 
In the left panel, we chose the parameters such that $d < d'_i < d_{\rm crit}^{\rm Vac}$. On the other hand, in the right panel, one has $d'_i < d < d_{\rm crit}^{\rm Vac}$.
From Fig. \ref{fig: HMI-inside}, one observes that the early and late time values of the holographic mutual information are equal to each other and given by $I_{A,B}^{\rm Vac}= I_0$.
Moreover, in the intermediate times, it can be smaller or larger than $I_{A,B}^{\rm Vac}= I_0$. Furthermore, one can show that by increasing the distance between the two subsystems, $I_{A,B;i}$ is zero for all times.

\bibliographystyle{ieeetr}
\bibliography{reference.bib}

\end{document}